\documentclass[11pt]{article}
\usepackage[utf8]{inputenc}
\usepackage[margin=1in]{geometry}
\usepackage{amsmath,amssymb,amsthm}
\usepackage{graphicx,subfig}
\usepackage{placeins}

\usepackage[numbers]{natbib}
\usepackage{xcolor}

\newtheorem{Def}{Definition}

\title{Network resampling for estimating uncertainty}
\author{Qianhua Shan and Elizaveta Levina \\ Department of Statistics, University of Michigan \\  qshan@umich.edu,  elevina@umich.edu}

\begin{document}

\maketitle

\begin{abstract}

    With network data becoming ubiquitous in many applications, many models and algorithms for network analysis have been proposed.    Yet methods for providing uncertainty estimates in addition to point estimates of network parameters are much less common.    While bootstrap and other resampling procedures have been an effective general tool for estimating uncertainty from i.i.d. samples, adapting them to networks is highly nontrivial.   In this work, we study three different network resampling procedures for uncertainty estimation, and propose a general algorithm to construct confidence intervals for network parameters through network resampling.    We also propose an algorithm for selecting the sampling fraction, which has a substantial effect on performance.   We find that, unsurprisingly, no one procedure is empirically best for all tasks, but that selecting an appropriate sampling fraction substantially improves performance in many cases.  We illustrate this on simulated networks and on Facebook data.   
\end{abstract}

\section{Introduction}
\label{sec:intro}
With network data becoming common in many fields, from gene interactions to communications networks,  much research has been done on fitting various models to networks and estimating parameters of interest.     In contrast, few methods are available for evaluating uncertainty associated with these estimates, which is key to statistical inference.     In classical settings, when asymptotic analysis for obtaining error estimates is intractable,  various resampling schemes have been successfully developed for estimating uncertainty, such as the jackknife \cite{shao_1989_general}, the bootstrap \cite{efron_1986_bootstrap},  and subsampling or $m$-out-of-$n$ bootstrap \cite{bickel_1997_resampling,politis_subsampling_1999}.
Most of the classical resampling methods were developed for i.i.d.\ samples and are not easily transferred to the network setting, where nodes are connected by edges and have a complex dependence structure.    Earlier work on bootstrap for dependent data focused on settings with a natural ordering, such as time series \cite{politis_1994_stationary} or spatial random fields \cite{bickel_2006_covariance}.  

More recently, a few analogues of bootstrap have been proposed specifically for network data.  One general approach is to extract some features of the network that can be reasonably treated as i.i.d., and resample those following classical bootstrap.  For instance, under a latent space model assumptio  latent node positions are assumed to be drawn i.i.d.\ from some unknown distribution, and the network model is based on these positions.   Then as long as the latent positions can be estimated from a given data, they can be resampled in the usual bootstrap fashion to generate new bootstrap network samples, as proposed in \cite{levin_bootstrapping_2019},.      Another network feature that easily lends itself to this kind of approach is subgraph counts, because they are calculated from small network ``patches'' that can be extracted and then treated as i.i.d.\ for resampling purposes.    This approach has been applied to estimating uncertainty in the mean degree of the graph   \cite{thompson_using_2016}, and extended to  bootstrapping the distribution of node degrees  \cite{gel_bootstrap_2017, green_2017_bootstrapping}.    Related work \cite{bhattacharyya_subsampling_2015} proposed to estimate the variance of network count statistics by bootstrapping subgraphs isomorphic to the pattern of interest.   This approach works well for network quantities  that can be computed from small patches, and does not require the latent space model assumption, but it does not generalize easily to other more global inference tasks. 


A more general approach analogous to the jackknife was proposed in \cite{lin_2020_on},  who analyzed a leave-one-node-out procedure for network data. This procedure can be applied to most network statistics, and the authors show that the variance estimates are conservative in expectation under some regularity conditions. However, for a size $N$ network, this procedure requires to compute the statistics of interest on $N$  networks with $N-1$ nodes each, which  can be computationally expensive.

Another class of methods works by subsampling the network, which we can think of as analogous to $m$-out-of-$n$ bootstrap, e.g., \cite{bickel_1981_asymptotic}.  Inference based on a partially observed or subsampled network was first considered  for computational reasons, when it is either not computationally feasible to conduct inference on the whole network, or only part of the network was observed in the first place.     Most commonly, subsampling is implemented through node sampling, where we first select a fraction of all nodes at random and then observe the induced subgraph, but there are alternative subsampling strategies, to be discussed below.
Examples in this line of work include  subsampling under a sparse graphon model \cite{lunde_2019_subsampling},  estimating the number of connected components in a unknown parent graph from a subsampled subgraph \cite{frank_estimation_1978,klusowski_estimating_2020}, and estimating the degree distribution of a network from a subgraph generated by node sampling or a  random walk \cite{zhang_2015_estimating}.        Similar methods have also been employed for cross-validation on networks, with subsampling either nodes \cite{chen_network_2018} or edges \cite{li_network_2020} multiple times.  


Our goal in this paper is to study general subsampling schemes  that can be used to estimate uncertainty in most network statistics with minimal model assumptions.    We will study three different subsampling procedures, discussed in detail in the next section:  node sampling, row sampling, and node pair sampling.   Our goal is to understand how the choice of subsampling procedure affects performance for different inference tasks, and how each subsampling method performs under different models.   One of our  main findings is that choosing the fraction of the network to sample has significant implications for performance.  To this end, we propose a data-driven procedure to select the resampling fraction and show it performs close to optimal.


The paper is organized as follows: In Section \ref{sec:methods},  we present the proposed resampling procedures and the data-driven method to choose the subsampling fraction. In Section \ref{sec:results}, we compare numerical performance of different subsampling methods for different inference tasks.   Section \ref{sec:data} presents an application of our methods to triangle density estimation in Facebook networks, and Section \ref{sec:disc} concludes with discussion.  

\section{Network subsampling methods}
\label{sec:methods}

We start with fixing notation.  Let  $G=(V,E)$ be the observed network, with $V$ the node set and $E$ the edge set, and $n = |V|$ nodes.    We represent $G$  using its $n\times n$ adjacency matrix $A$, where $A_{ij}=1$ if there exists an edge from node $i$ to node $j$, i.e.,  $(i,j)\in E$,  and $A_{ij}=0$ otherwise.   For the purposes of this paper, we focus on binary undirected networks.     Generalization to weighted undirected networks should be straightforward, as the resampling mechanisms would not change.   Directed networks will, generally speaking, require a different approach, as for the undirected setting a single row of the adjacency matrix contains all the information about the corresponding node, whereas for directed matrices this would not be the case.

In classical statistics, we have a well-established standard bootstrap procedure \cite{efron_1994_introduction}: given an i.i.d.\ sample $\mathcal{X} = \{X_1,\dots,X_n\}$ from an underlying distribution $F$, we can construct a confidence set for some statistics of interest $T(F)$ by estimating the sample distribution of $\hat{T}_n(\mathcal{X})$, and, typically, using its quantiles to construct a confidence interval for $T(F)$.   
 By drawing new samples $\mathcal{X}_1,\dots,\mathcal{X}_B$ with replacement from the set $\mathbf{X}$, or in other words sampling i.i.d.\ from the empirical distribution $\hat{F}_n$, we can obtain the empirical distribution of $\hat{T}_n(\mathcal{X}_b)$, $b = 1, \dots, B$, to approximate  the distribution of $\hat{T}_n(\mathcal{X})$.

 Thinking about applying this procedure to a network immediately reveals a number of questions:   what are the units of resampling, nodes or edges?  What does it mean to sample a node or an edge with replacement?   Sampling without replacement seems more amendable to working with resulting induced graphs, and thus we turn to the $m$-out-of-$n$ bootstrap, typically done without replacement, which provides asymptotically consistent estimation not only for i.i.d.\ but also for weakly dependent stationary samples \cite{politis_subsampling_1999}.   Instead of drawing samples of size $n$ with replacement, we will now draw new samples of size $m<n$ without replacement.

 For networks, we argue that the the best unit to resample and compare across procedures is {\em node pairs} (whether connected by an edge or not).    Resampling algorithms can be constructed in different ways and operate either on nodes or on node pairs, but in the end it is the number of node pairs that determines the amount of useful  information we have available.     Resampling node pairs with replacement does not really work as it is not clear how to incorporate duplicated node pairs.   Thus we will focus on network resampling without replacement,  similar to the $m$-out-of-$n$ bootstrap approach.   The fraction of resampled node pairs, denoted by $q$, turns out to be an important tuning parameter, and we will propose a data-driven way of choosing it.    First, we describe three different resampling procedures we study.

\subsection{Subsampling procedures}
\label{subsec:subsampling}

Next, we describe in detail the three subsampling procedures we study.  While there are other subsampling strategies and some may have advantages in specific applications, we focus on these three because they are popular and general.

\begin{figure}[htbp]
\centering
\subfloat[Node sampling\label{fig:1a}]{\includegraphics[width=0.3\textwidth]{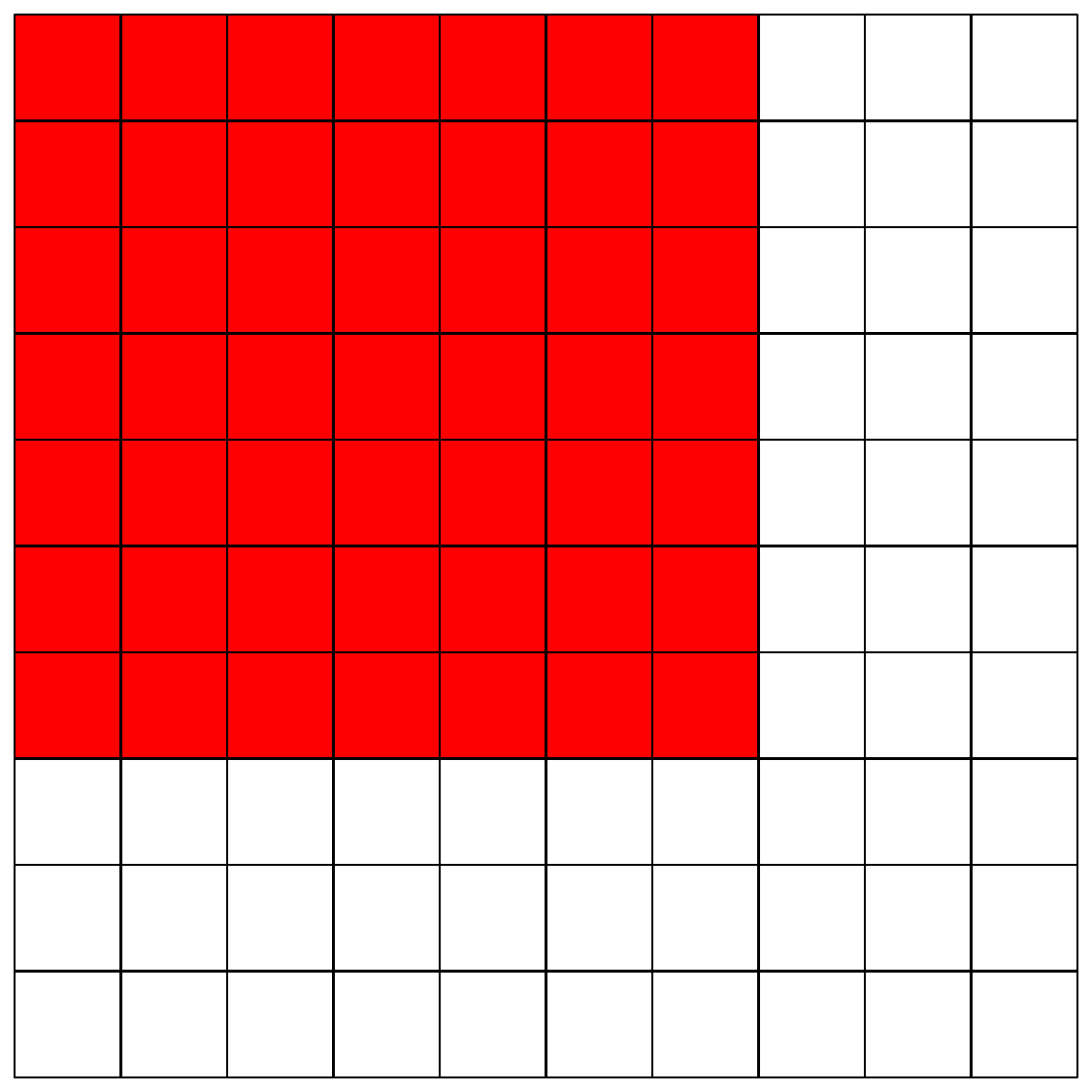}}\hfill
\subfloat[Row sampling\label{fig:1b}] {\includegraphics[width=0.3\textwidth]{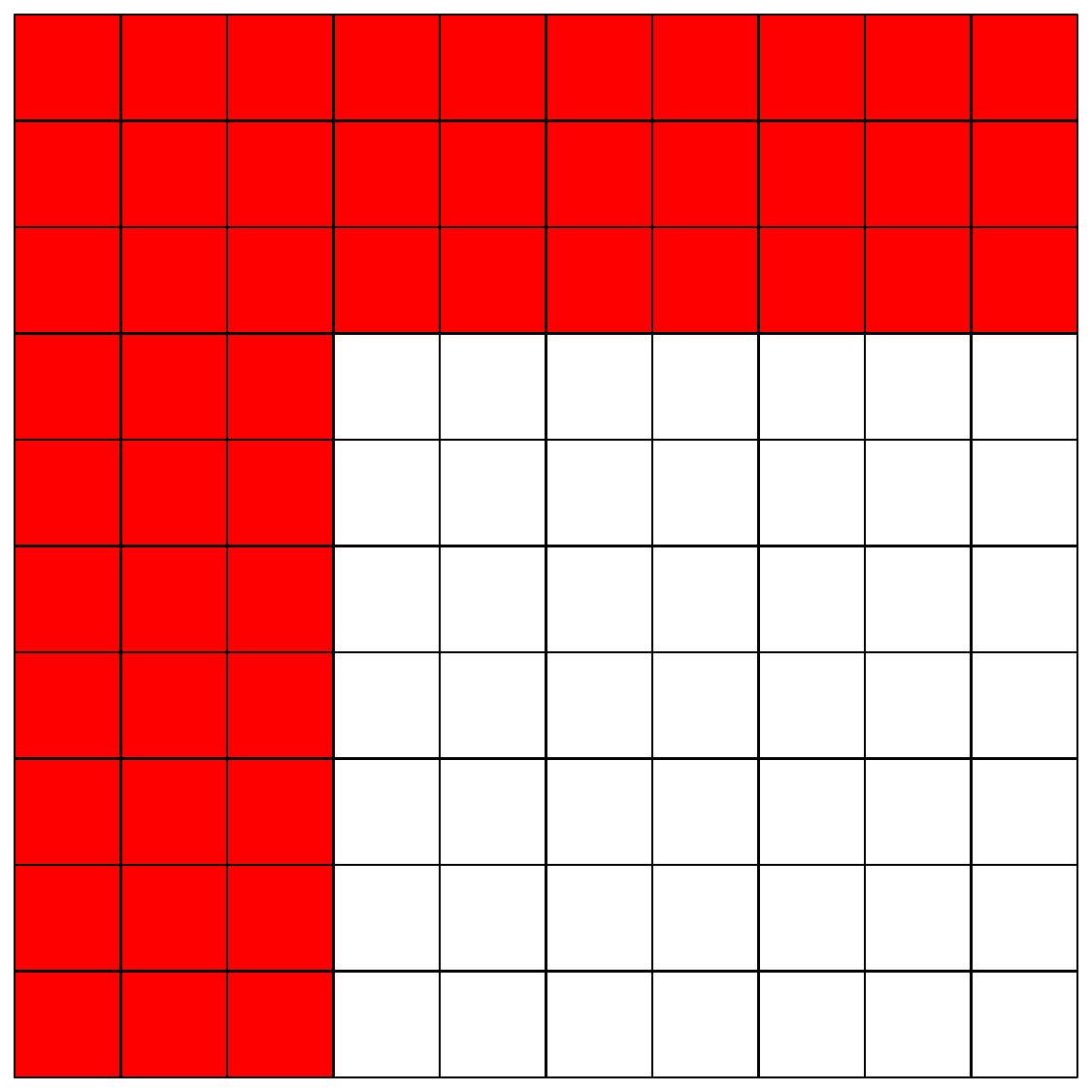}}\hfill
\subfloat[Node pair sampling\label{fig:1c}]{\includegraphics[width=0.3\textwidth]{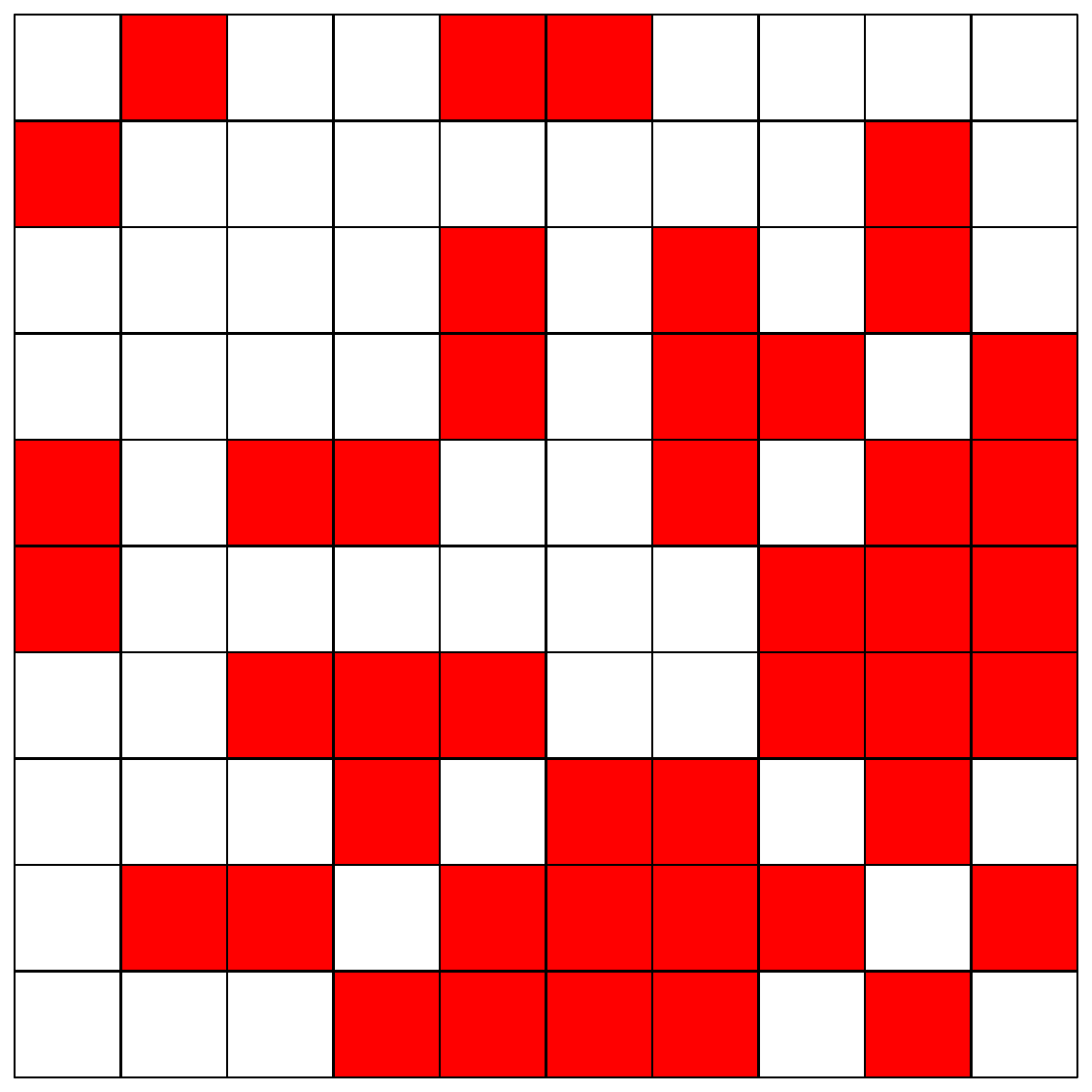}}
\caption{Different subsampling methods.   In all cases, the fraction of node pairs sampled is $q \approx 0.5$ (exact $q=0.5$ is not always possible with only 10 nodes). } \label{sampling}
\end{figure}

    \subsubsection*{Subsampling by node sampling}

    This is perhaps the most common subsampling scheme, also known as $p$-sampling or induced subgraph sampling.   A subset of nodes $\tilde V$ is first sampled from the node set $V$ independently at random, through $n$ independent Bernoulli trials with success probability $p$.   We then observe the subgraph induced by these nodes, which consists of all edges between the nodes in $\tilde V$, that is, $\{(i,j) \in E : i \in \tilde V, j \in \tilde V\}$.    The expected fraction of node pairs sampled under this scheme is $q = p^2$ of the entries in the adjacency matrix.

    Sampling nodes is arguably the most intuitive way to subsample a network, as it is analogous to first selecting the objects of interest and then observing the relationships among them,  a common way of collecting network data in the real world.   The intuition behind node sampling is that the induced network should inherit the global structure of its parent network, such as communities.    A variation where inference is on node-level statistics, such as node degree, will work well if these statistics are subsampled directly with the nodes themselves, rather than recomputed from the induced subgraph \cite{kolaczyk_sampling_2009}.   At the same time, since each subsampled network will be on a different set of nodes, it is less clear how well this method can work for more general inferential tasks.     Recently,  node sampling was shown to consistently estimate the distribution of some  network statistics such as triangle density \cite{lunde_2019_subsampling}, under a sparse graphon model satisfying some regularity conditions.    But this result only holds asymptotically, and the question of how to choose $p$ for finite samples was not considered.

\subsubsection*{Subsampling by row sampling}
This subsampling method was first proposed for the cross-validation on networks in \cite{chen_network_2018}.   In row sampling, one first chooses a subset of nodes $\Tilde{V}$ by independent Bernoulli trials with success probability $p$; then edges from nodes in $\Tilde{V}$ and {\em all} nodes in $V$ are observed.  This is equivalent to sampling whole rows from the adjacency matrix.    This way of sampling is related to star sampling, a common model for social network data collected through surveys, where first a set of nodes is sampled, and then each of the sampled nodes reports all of its connections, especially relevant in situations where we may not know all of the nodes in $V$.  We assume that both 1s and 0s are reported accurately, meaning that a 0 represents that an edge is absent rather than the status of that connection is unknown.    For undirected graphs with symmetric adjacency matrices, row sampling is equivalent to masking a square submatrix from the original adjacency matrix, that is, $q = 1 - (1-p)^2$ proportion of all node pairs observed.

\subsubsection*{Subsampling by node pair sampling}

Finally, a network can be subsampled by directly selecting a subset of node pairs to observe.  This sampling method was used in \cite{li_network_2020} for cross-validation on networks, and shown to be superior to row sampling in most cross-validation tasks considered. A subset of node pairs $(i,j)$ are selected from $V\times V$ through independent Bernoulli trials with success probability $p$, resulting in $q=p$ proportion of entries in the adjacency matrix observed.   Again, zeros are treated as truly absent edges, and for undirected networks the sampling is adjusted to draw only from entries with $i < j$ and fill in the symmetric matrix  accordingly.
In  \cite{li_network_2020}, noisy matrix completion was applied first to create a low rank approximation to a full matrix before proceeding with cross-validation, and in principle, we have a choice of whether to use the subsampled matrix or its completed version for downstream tasks.  


\subsection{Constructing confidence intervals by resampling}
\label{subsec:bootstrap}
Before we proceed to the data-driven method of choosing the settings for resampling, we summarize the algorithm we employ for constructing confidence intervals for network statistics based on a given resampling method with a given sampling fraction $q$.     This is a general bootstrap-style algorithm that can be applied to any network statistic $T$ computable on a subgraph.

\noindent {\bf Algorithm 1: constructing a bootstrap confidence interval. }   \\
Input:  a graph $G$, a  network statistic $T(G)$, a subsampling method, a fraction $q$, and the number of bootstrap samples $B$.  
\begin{enumerate}
\item For $b = 1, \dots, B$ 
    \begin{enumerate}
        \item Randomly subsample subgraph $G_b$, using the chosen method and fraction $q$.  
        \item Calculate the estimate $\hat{T}_b = \hat{T}(G_b)$.
    \end{enumerate}
\item Sort the $B$ estimates,  $\hat{T}_{(1)} \le \dots \le \hat{T}_{(B)}$, and construct a $100(1-\alpha)\%$ confidence interval for $T(G)$ as $[\hat{T}_{(l)},\hat{T}_{(u)}]$, where $l = \lfloor \frac{\alpha}{2} B \rfloor $ and $u = \lceil \frac{1-\alpha}{2}  B \rceil$.  
\end{enumerate}
We give this version of bootstrap as the most commonly used one, but any other version of bootstrap based on the empirical cumulative distribution function (symmetrized quantiles, etc) can be equally well applied.

\subsection{Choosing the subsampling fraction $q$}
For any method of network subsampling, we have to choose the sampling fraction;  for consistency across different methods, we focus on the choice of the node pair fraction $q$, which can be converted to $p$ for node or row sampling.    In the classical bootstrap analogue, the $m$-out-of-$n$ bootstrap, consistency results have been obtained for  $m\rightarrow \infty$ and $m/n\rightarrow 0$ as $n\rightarrow \infty$.    These do not offer us much practical guidance for how to choose $m$ for a given finite sample size $n$.   The situation is similar for network bootstrap, with no practical rules for choosing $q$ implied by consistency analysis.

The iterated bootstrap  has been commonly used to calibrate bootstrap results since it was proposed  and developed in the 1980s \cite{hall_1986_bootstrap, bera_1987_prepivoting, hall_1988_bootstrap}.
An especially relevant algorithm for our purposes is the double bootstrap approach, originally proposed to improve coverage of bootstrap intervals \cite{martin_bootstrap_1990}.   A similar double bootstrap approach was proposed by  \cite{lee_class_1999} for $m$-out-of-$n$ bootstrap, to correct a discrepancy between the nominal and the actual coverage for bootstrap confidence intervals due to a different sample size $m$.   

A generic double bootstrap algorithm proceeds as follows.    Given the original i.i.d.\ sample  $\mathcal{X} = \{X_1,\dots,X_n\}$  and an estimator $\hat\theta$ based on $\mathcal{X}$, consider a size $n$ bootstrap sample  $\mathcal{X}^*$, drawn from $\mathcal{X}$ with replacement, and the corresponding estimator $\hat{\theta}^*$.      If the distribution of $\hat{\theta}$ can be approximated by the distribution of $\hat{\theta}^*$ conditional on $\mathcal{X}$, a confidence interval of nominal level $1-\alpha$ can be constructed as $I(\alpha) = [\hat{t}_{\alpha/2},\hat{t}_{1-(\alpha/2)}]$, with $\hat{t}_{\alpha}$ defined as $\sup\{t:\mathbf{P}(\hat{\theta}^*\leq t\mid \mathcal{X})\leq \alpha\}$.    This confidence interval typically has correct coverage up to $O(n^{-1})$, and double bootstrap proposes to improve this bound by  using  $I(\beta_{\alpha})$ instead, where $\beta_{\alpha}$ is defined as the solution to $\mathbf{P}(\theta \in I(\beta_{\alpha}\mid\mathcal{X})) = 1-\alpha$ and can be approximated by $\hat{\beta}_{\alpha}$, the solution to $\mathbf{P}(\hat{\theta} \in I(\hat{\beta}_{\alpha}\mid\mathcal{X}^*)\mid\mathcal{X}) = 1-\alpha$.   
In practice, this is implemented by sampling multiple size $n$  $\mathcal{X}^{**}$ draws from each $\mathcal{X}^{*}$ with replacement, and estimating  $\mathbf{P}(\hat{\theta} \in I(\beta_j\mid\mathcal{X}^*)\mid\mathcal{X})$  as the proportion of confidence intervals based on $(1-\beta_j)$ percentiles of $\mathcal{X}^{**}$  that cover $\hat{\theta}$ for a set of candidate $\beta_j$'s.   
We then choose $\hat{\beta}_{\alpha} = \beta_j$ such that $\mathbf{P}(\hat{\theta} \in I(\beta_j\mid\mathcal{X}^*)\mid\mathcal{X}) $ is the closest to $1-\alpha$ among all candidate $\beta_j$'s. 

Since different choice of subsampling fraction $q$ will lead to different level to coverage, here we propose an analogous double bootstrap method for network resampling. Compared to the i.i.d. case, instead of using size $n$ resamples sampled with replacement, we will generate resamples using network subsampling methods described above and instead of choosing an appropriate $\alpha$, we aim to choose an appropriate $q$. The algorithm is detailed as follows:

\noindent {\bf Algorithm 2: choosing the sampling fraction $q$. }   \\

Input:  a graph $G$, a  network statistic $T(G)$, a subsampling method, a set of candidate fractions  $q_1, \dots, q_J$, and the number of bootstrap samples $B$.  

 For $j = 1, \dots, J$, 
\begin{enumerate}
        \item For $b = 1,\dots,B$,
        \begin{enumerate}
            \item Split all nodes randomly into two sets $\mathcal{V}_1^b$, $\mathcal{V}_2^b$. 
            \item On the subgraph induced by $\mathcal{V}_1^b$, calculate the estimate $\hat{T}_{b,j}^{(1)}$. 
            \item On the subgraph induced by $\mathcal{V}_2^b$, run Algorithm 1 with $q = q_j$ to construct a confidence interval $R_{b,j} $ for $T_{b,j}^{(2)}$ 
        \end{enumerate}
        \item Calculate empirical coverage rate $\pi_j = \frac{1}{B}\sum_{b=1}^{B}\mathbf{1}\{\hat{T}_{b,j}\in R_{b,j}\}$.
    \end{enumerate}
 
Output:   $\hat j = \arg\min (\pi_j-(1-\alpha))^2$.  

One key difference from the i.i.d.\ case is that we estimate empirical coverage  by checking if the confidence interval constructed on each subsample covers $\hat{T}_{b,j}^{(1)}$, 
not $\hat{T}(G)$, the estimate on the original observed graph $G$. This is because with network subsampling, we cannot preserve the original sample size $n$, and if we are going to compare network statistics on two subsamples, we need to match the graph sizes (in this case $n/2$).

\section{Empirical results}
\label{sec:results}
We evaluate the three different subsampling techniques and our method for choosing the subsampling fraction on four distinct tasks.  Three focus on estimating uncertainty, in  (1) the normalized triangle density, a global network summary statistic;   (2)   the number of communities under the stochastic block model, a network model selection parameter; and  (3) coefficients estimated in regression with network cohesion, regression parameters estimated with the use of network information.     Task (4) is to choose a tuning parameter for regression with network cohesion with a lasso penalty, and evaluate subsampling as a tuning method in this setting.

The networks for all the tasks will be generated from the  stochastic block models (SBM).   To briefly review, the SBM assigns each node $i$ to one of $K$ communities; the node labels $c_i \in \{1, \dots, K\}$, for $i = 1, \dots, n$ are sometimes viewed as randomly generated from a multinomial distribution with given probabilities for each community, but for the purposes of this evaluation we treat them as fixed, assigning a fixed number of nodes $n_k$, $k = 1, \dots K$, to each community $k$.    Edges  are then generated as independent Bernoulli variables, with  probabilities of success determined by the communities of their incident nodes,  $P(A_{ij}=1)=B_{c_i c_j}$, where $B$ is  a $K \times K$ symmetric probability matrix.      We will investigate the effects of varying the number of communities, the number of nodes in each community, the expected overall edge density $\rho$, and the ratio of probabilities of within- and between-community edges $t$, to be defined below.   
    
    \subsection{Normalized Triangle Density}

	Count statistics are functions of counts of subgraphs in network, and are commonly viewed as key statistical summaries of  a network \cite{bhattacharyya_subsampling_2015, bickel_method_2011}, analogous to moments for i.i.d.\  data.     Here we will examine a simple and commonly studied statistic,  the density of triangles.  A triangle subgraph is three vertices all connected to each other.  Triangle density is frequently studied because it is related to the concept of transitivity, a key property of social networks where two nodes that have a ``friend'' in common are more likely to be connected themselves.    We will apply resampling  to construct confidence intervals for the normalized triangle density for a given network.     


        \begin{Def}[edge density]
	Let $G$ be a graph on $n$ vertices with adjacency matrix $A$. The edge density is defined as
	$$\rho = \dfrac{1}{{n\choose 2}}\sum_{i<j
	\subset\mathcal{V}}A_{ij} . $$
	\end{Def}
	
	\begin{Def}[normalized triangle density]
	Let $G$ be a graph on $n$ vertices with adjacency matrix $A$. The normalized triangle density is defined as
	$$T = \rho^{-3}\frac{1}{{n\choose 3}}\sum_{i<j<k
	\subset\mathcal{V}}A_{ij}A_{jk}A_{ik} . $$

	\end{Def}

	 We generate the networks from the SBM models with number of communities $K$ set to either 1 (no communities, the Erd\"os-R\'enyi graph) or 3, with equal community sizes.  We consider two values of $n$, 300 and 600.    The matrix of probabilities $B$ for the SBM is defined by, for all $k, l = 1, \dots, K$, 
	 $$
B_{kl} = \begin{cases}   \rho \gamma_{1}& k = l ,  \\
 \rho \gamma_{2} & k \neq l . \end{cases}
$$
The difficulty of community detection in this setting is  controlled by the edge density $\rho$ and the ratio $t = \gamma_{1}/\gamma_{2}$.  We set $t=5$, and vary 
the expected edge density $\rho$  from 0.01 to 0.1, with higher $\rho$ corresponding to more information available for community detection.   
	
\begin{figure}[ht!]
 \begin{minipage}{0.15\linewidth}  
 \centering
 \begin{align*}
 n & =  300 \\  
 \rho & =  0.01
 \end{align*}
 \end{minipage}  
  \begin{minipage}{0.4\linewidth}  
   \centering
  \includegraphics[scale=0.35]{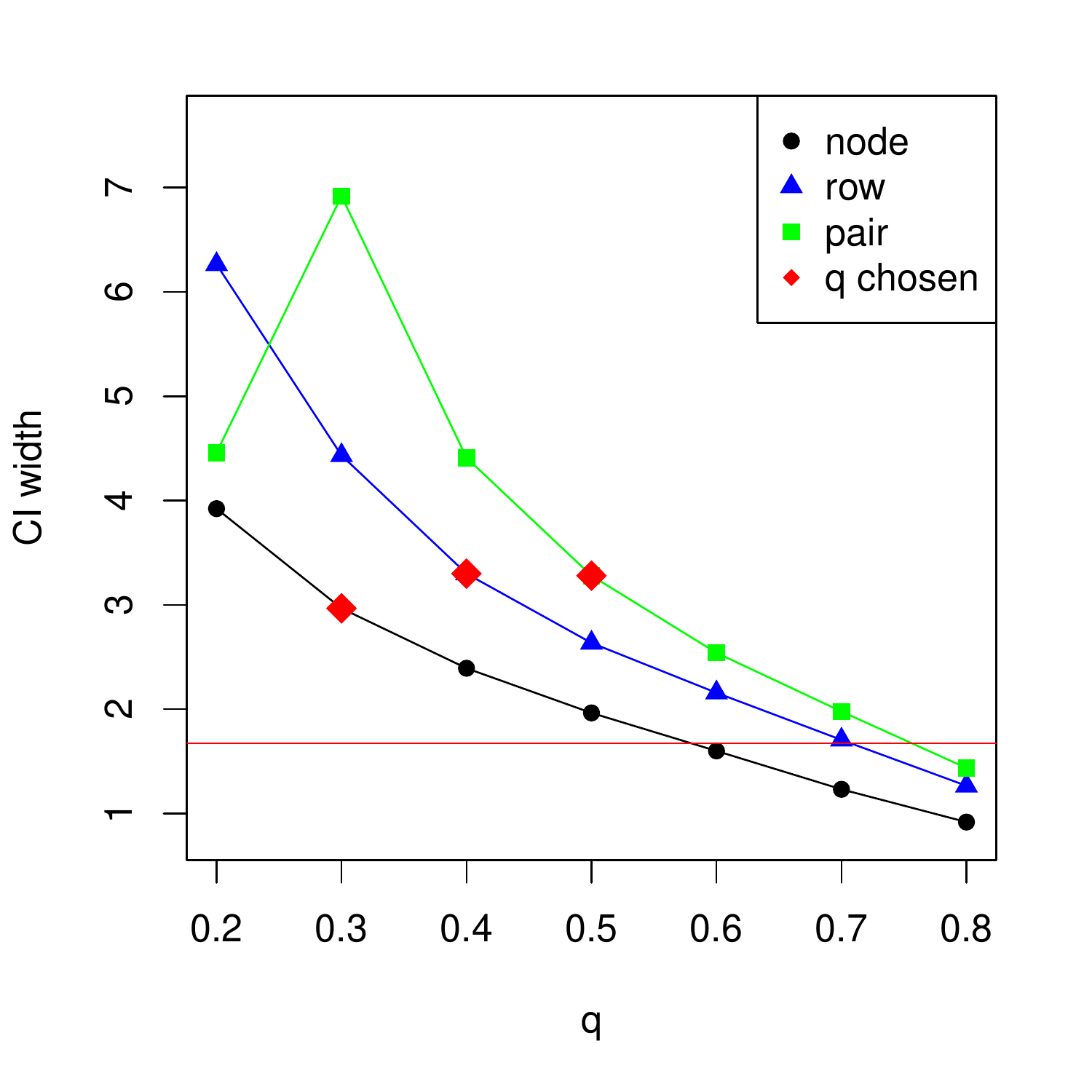} 
   \end{minipage}  
  \begin{minipage}{0.4\linewidth}  
   \centering
   \includegraphics[scale=0.35]{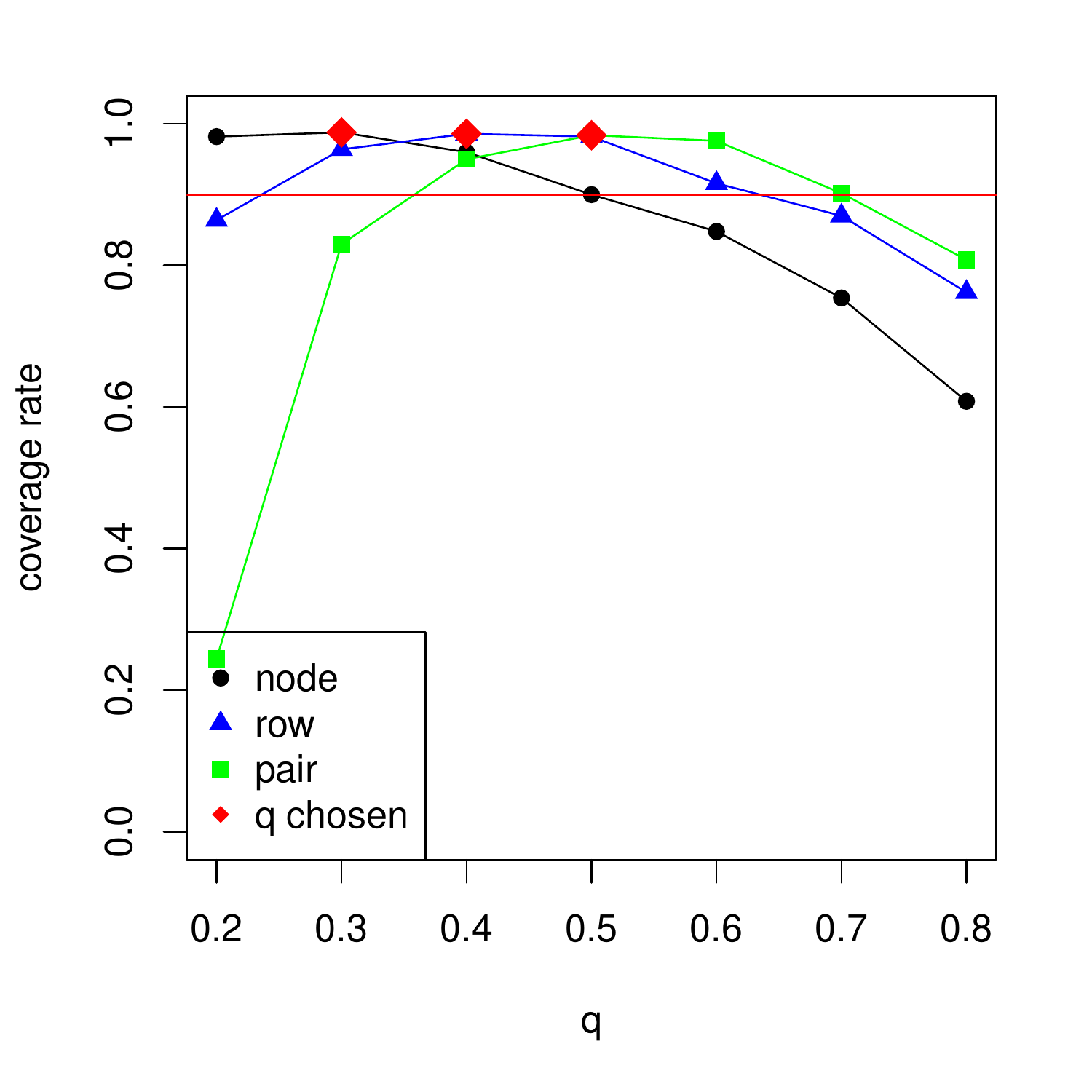}  
   \end{minipage} \\
   
 \begin{minipage}{0.15\linewidth}  
 \centering
 \begin{align*}
 n & =  300 \\  
 \rho & =  0.05
 \end{align*}
 \end{minipage}  
  \begin{minipage}{0.4\linewidth}  
   \centering
  \includegraphics[scale=0.35]{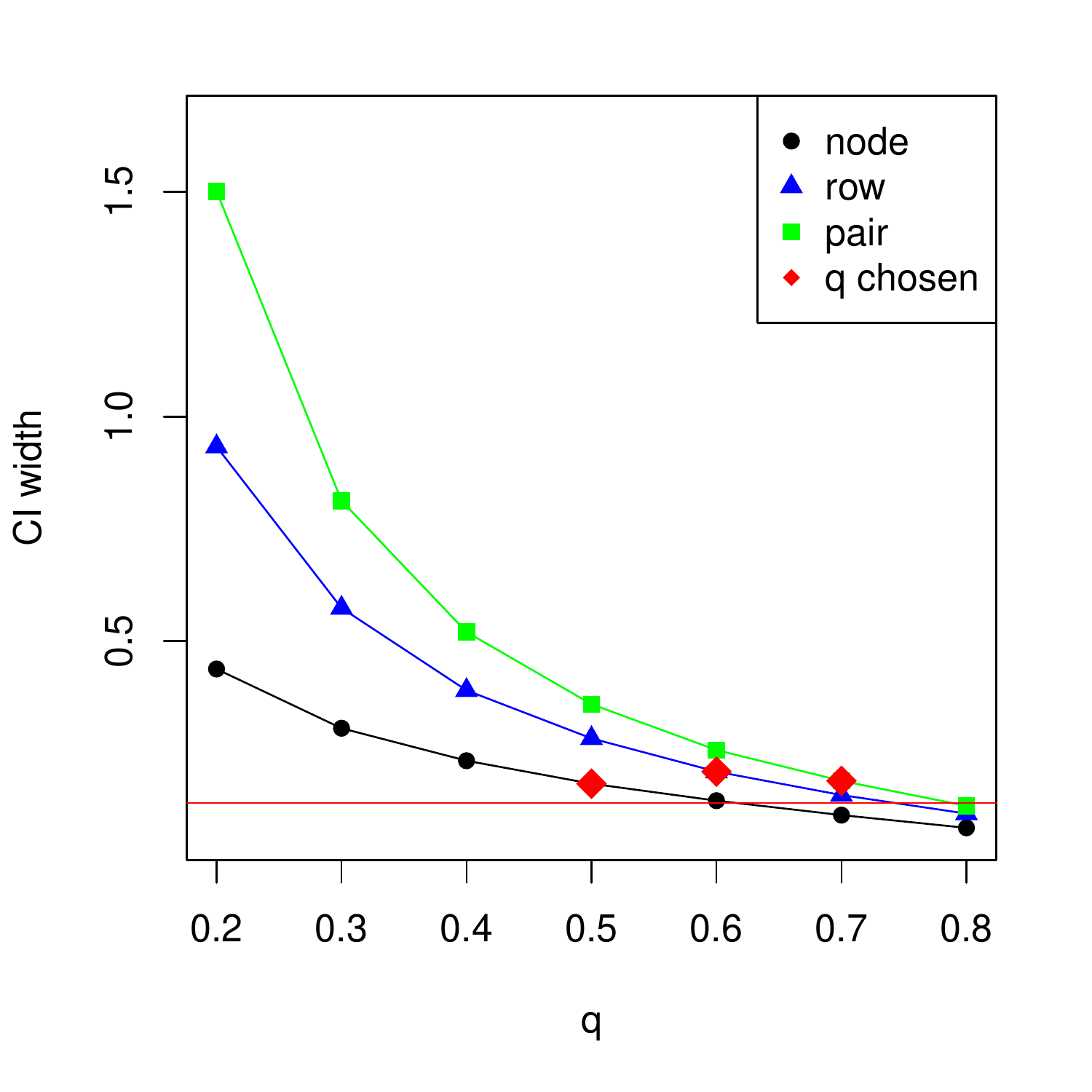} 
   \end{minipage}  
  \begin{minipage}{0.4\linewidth}  
   \centering
   \includegraphics[scale=0.35]{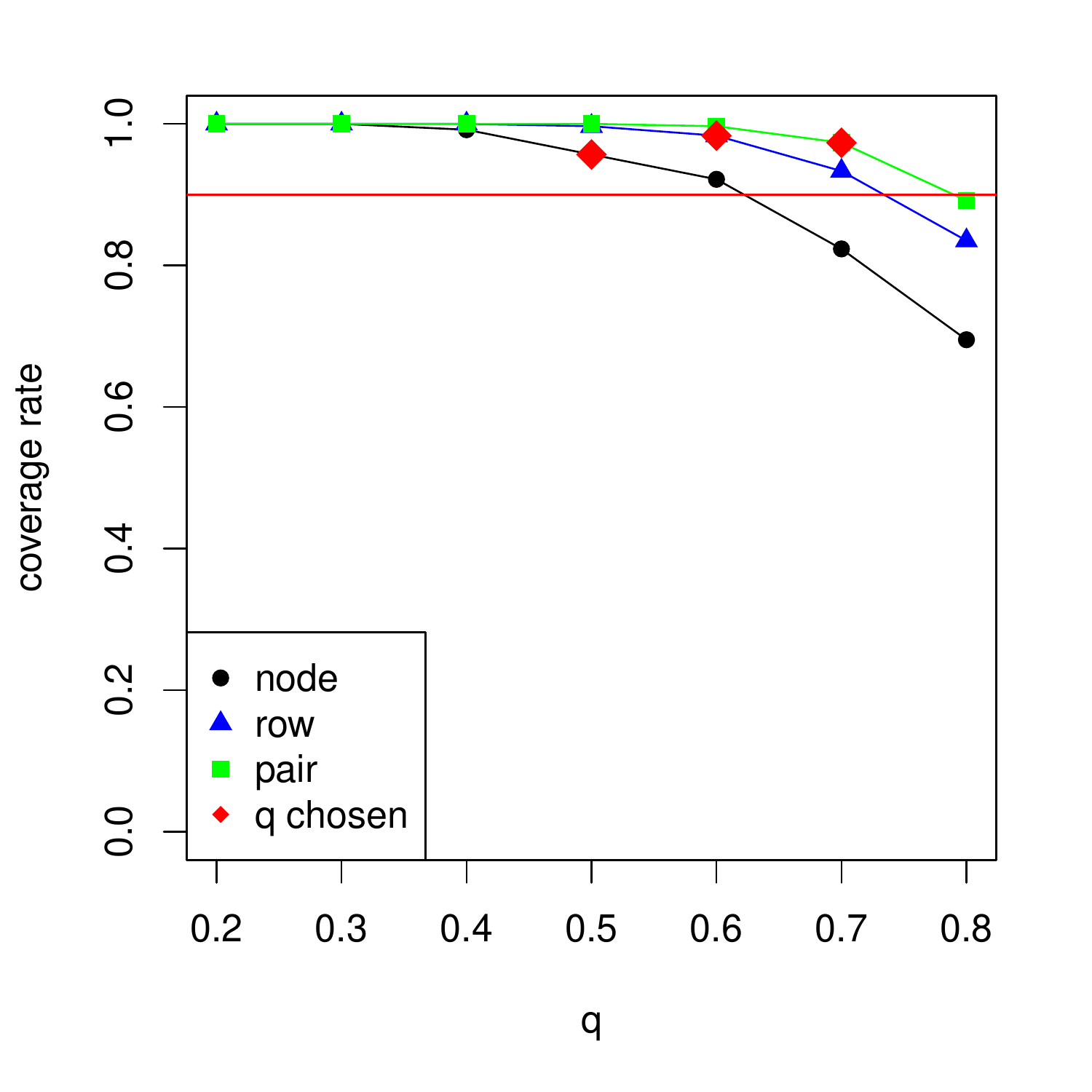}  
   \end{minipage} \\

 \begin{minipage}{0.15\linewidth}  
 \centering
 \begin{align*}
 n & =  300 \\  
 \rho & =  0.10
 \end{align*}
 \end{minipage}  
  \begin{minipage}{0.4\linewidth}  
   \centering
  \includegraphics[scale=0.35]{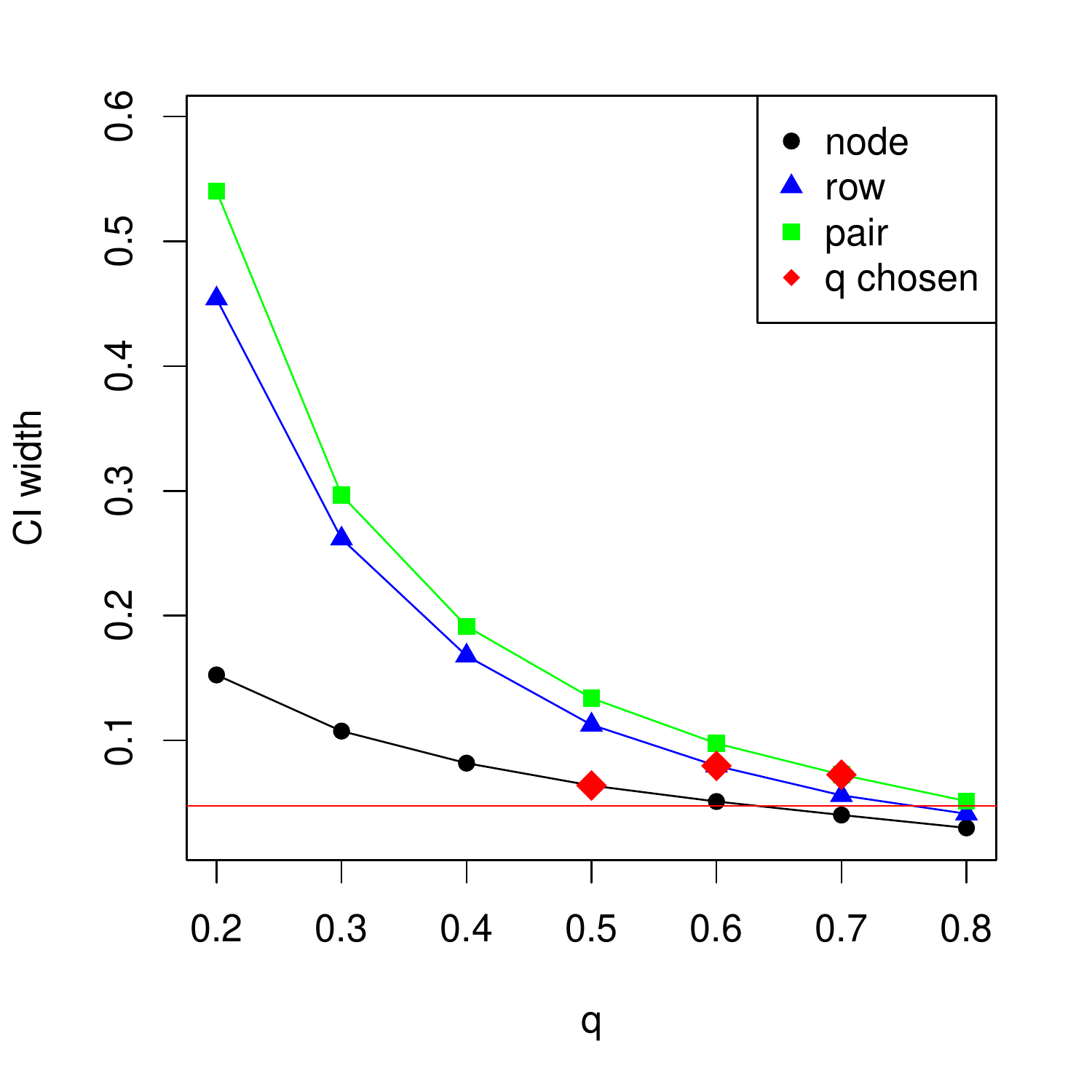} 
   \end{minipage}  
  \begin{minipage}{0.4\linewidth}  
   \centering
   \includegraphics[scale=0.35]{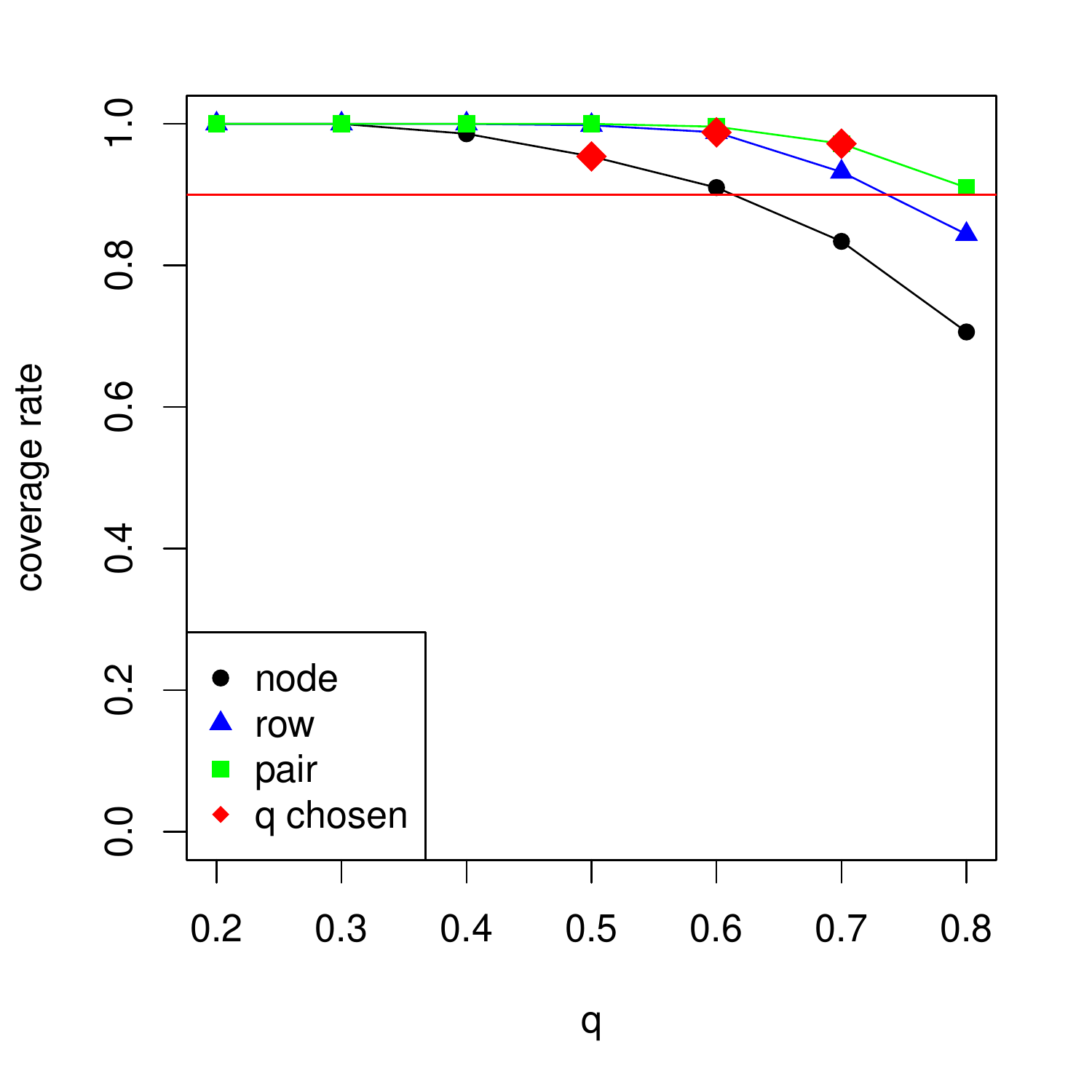}  
   \end{minipage}  \\

 \begin{minipage}{0.15\linewidth}  
 \centering
 \begin{align*}
 n & =  600 \\  
 \rho & =  0.05
 \end{align*}
 \end{minipage}  
  \begin{minipage}{0.4\linewidth}  
   \centering
  \includegraphics[scale=0.35]{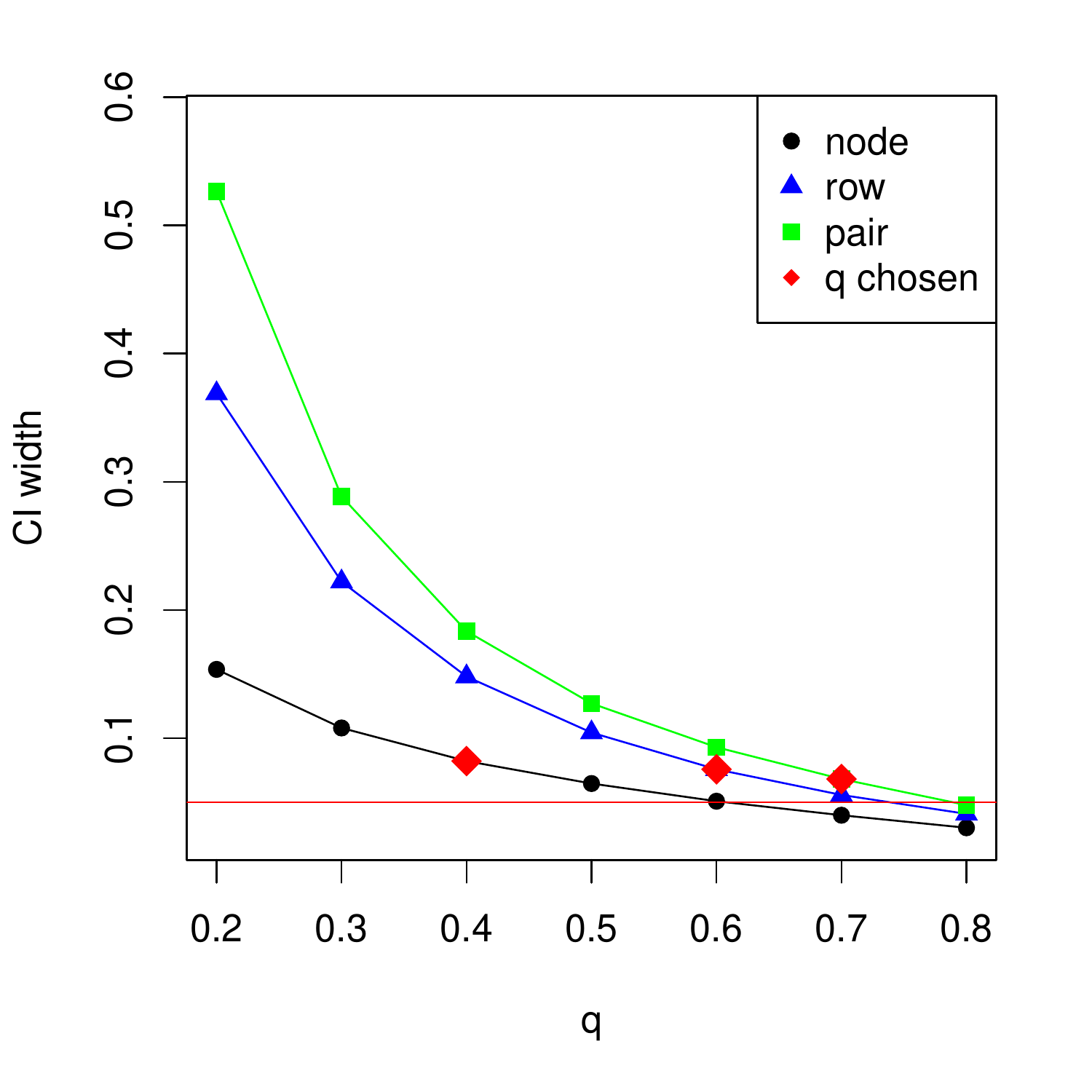} 
   \end{minipage}  
  \begin{minipage}{0.4\linewidth}  
   \centering
   \includegraphics[scale=0.35]{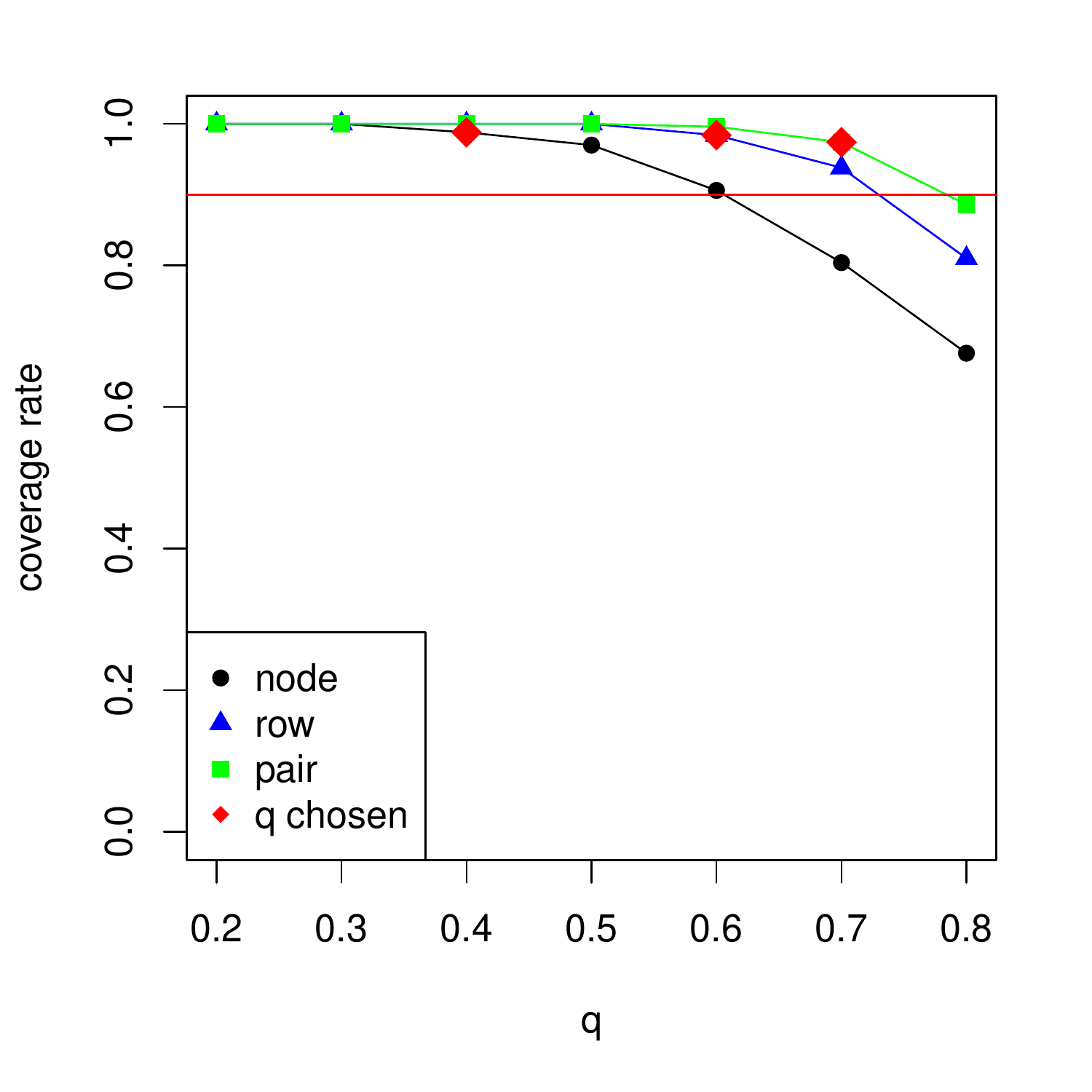}  
   \end{minipage} 	
	    \caption{Confidence intervals width  (left) and coverage (right) for the three different subsampling schemes on the Erd\"os-R\'enyi graphs with various values of $n$ and edge density $\rho$.   The value of $q$ chosen by our algorithm is shown as a red dot.}
	     \label{Triden1}
    \end{figure}

\begin{figure}[ht!]
 \begin{minipage}{0.15\linewidth}  
 \centering
 \begin{align*}
 n & =  300 \\  
 \rho & =  0.01
 \end{align*}
 \end{minipage}  
  \begin{minipage}{0.4\linewidth}  
   \centering
  \includegraphics[scale=0.35]{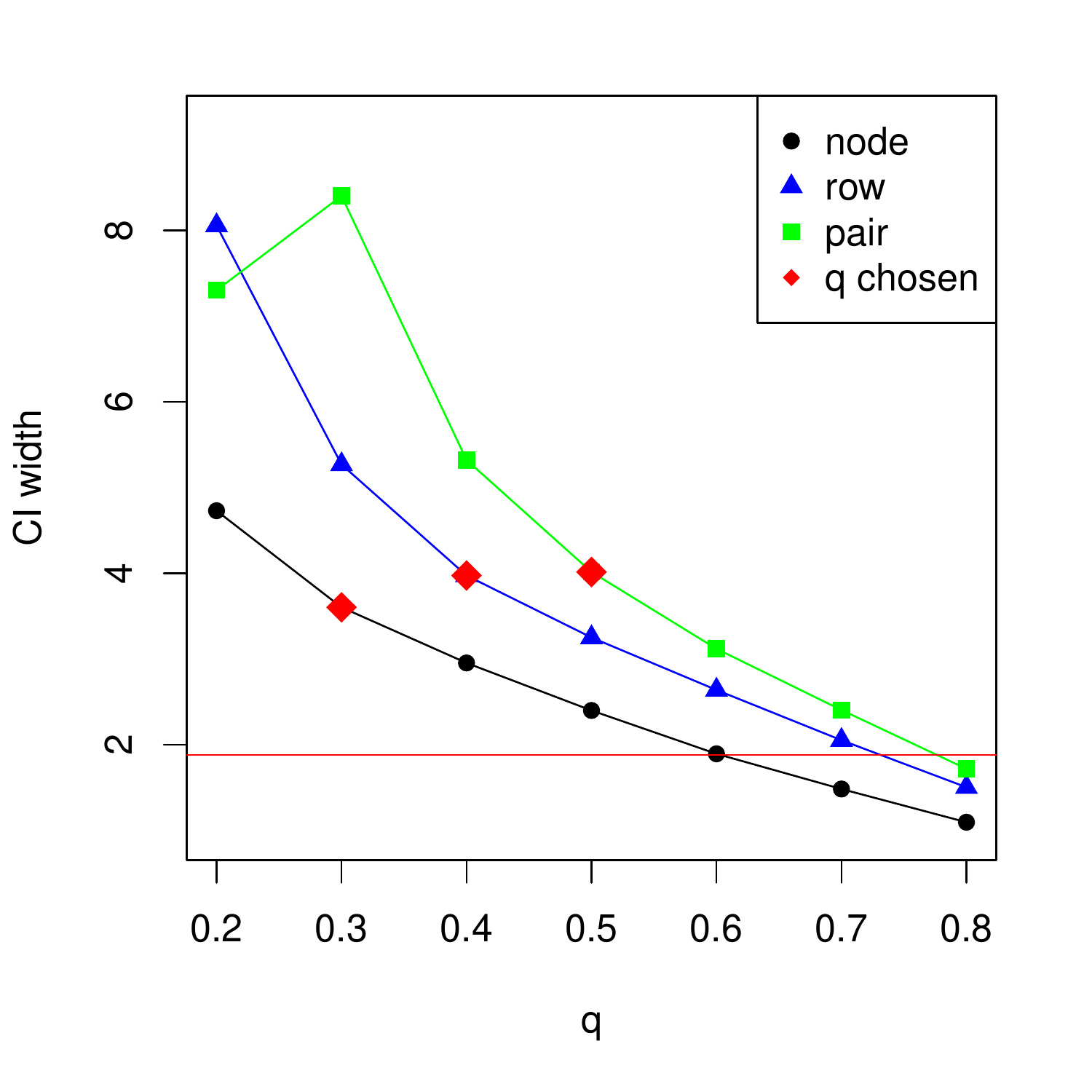} 
   \end{minipage}  
  \begin{minipage}{0.4\linewidth}  
   \centering
   \includegraphics[scale=0.35]{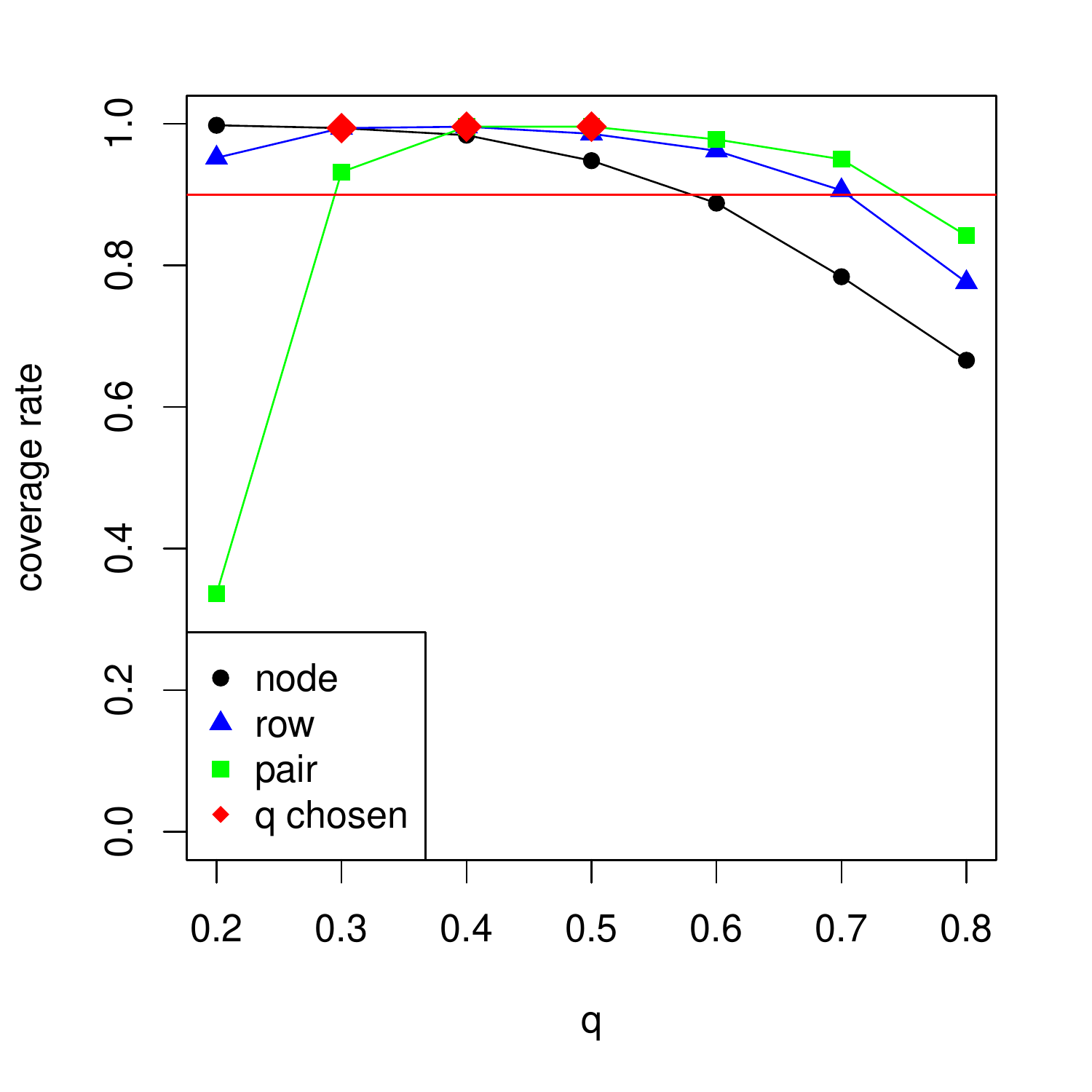}  
   \end{minipage} \\
   
 \begin{minipage}{0.15\linewidth}  
 \centering
 \begin{align*}
 n & =  300 \\  
 \rho & =  0.05
 \end{align*}
 \end{minipage}  
  \begin{minipage}{0.4\linewidth}  
   \centering
  \includegraphics[scale=0.35]{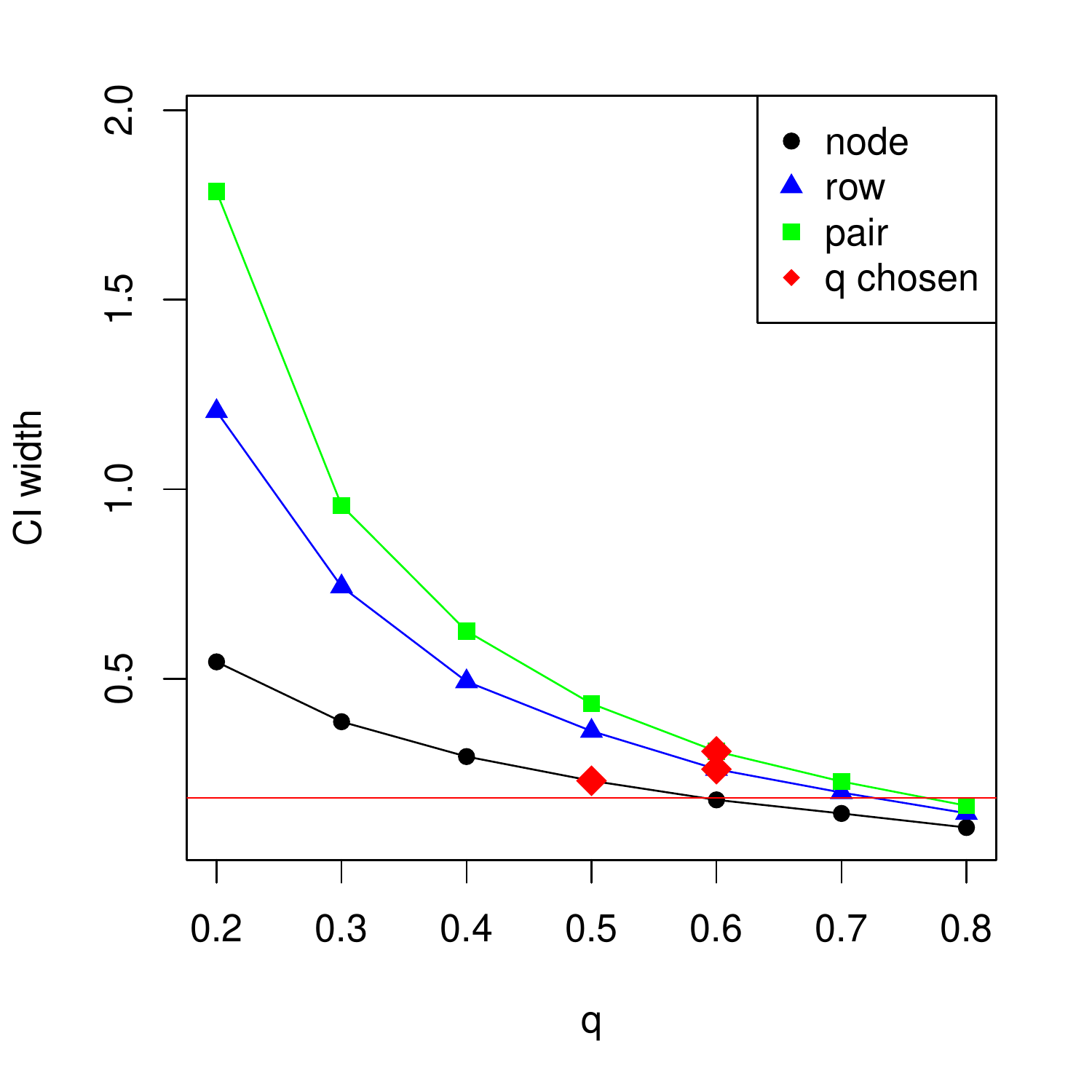} 
   \end{minipage}  
  \begin{minipage}{0.4\linewidth}  
   \centering
   \includegraphics[scale=0.35]{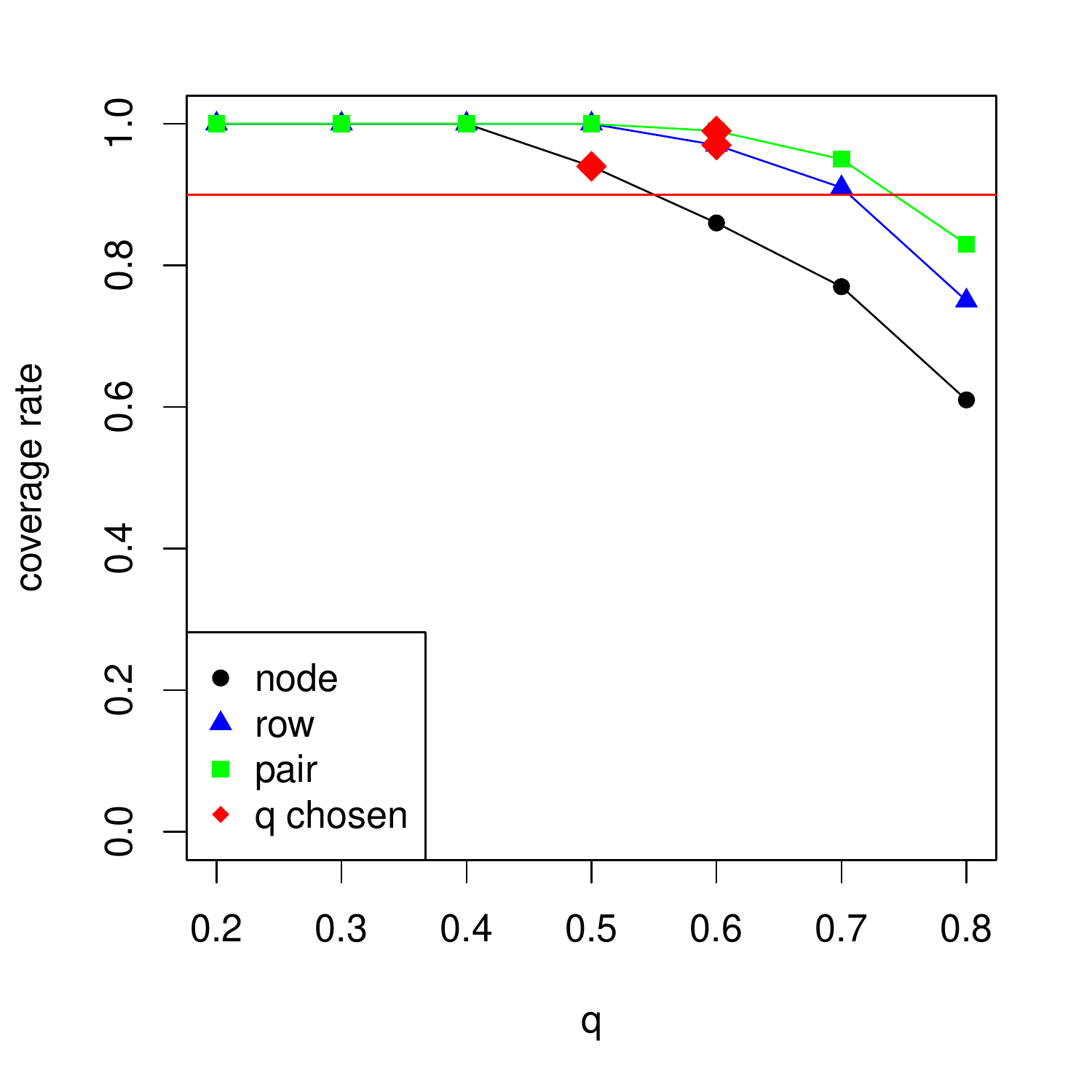}  
   \end{minipage} \\

 \begin{minipage}{0.15\linewidth}  
 \centering
 \begin{align*}
 n & =  300 \\  
 \rho & =  0.10
 \end{align*}
 \end{minipage}  
  \begin{minipage}{0.4\linewidth}  
   \centering
  \includegraphics[scale=0.35]{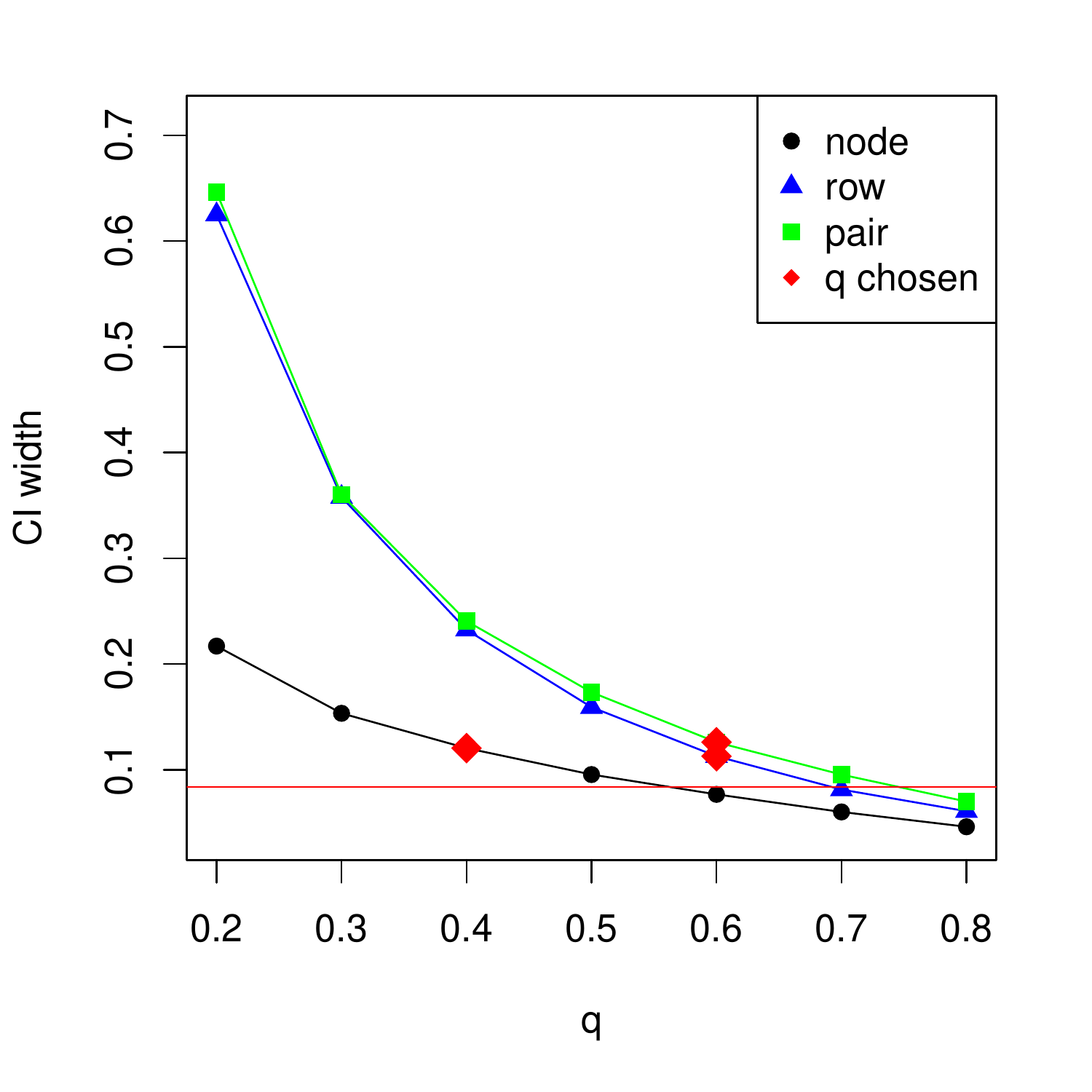} 
   \end{minipage}  
  \begin{minipage}{0.4\linewidth}  
   \centering
   \includegraphics[scale=0.35]{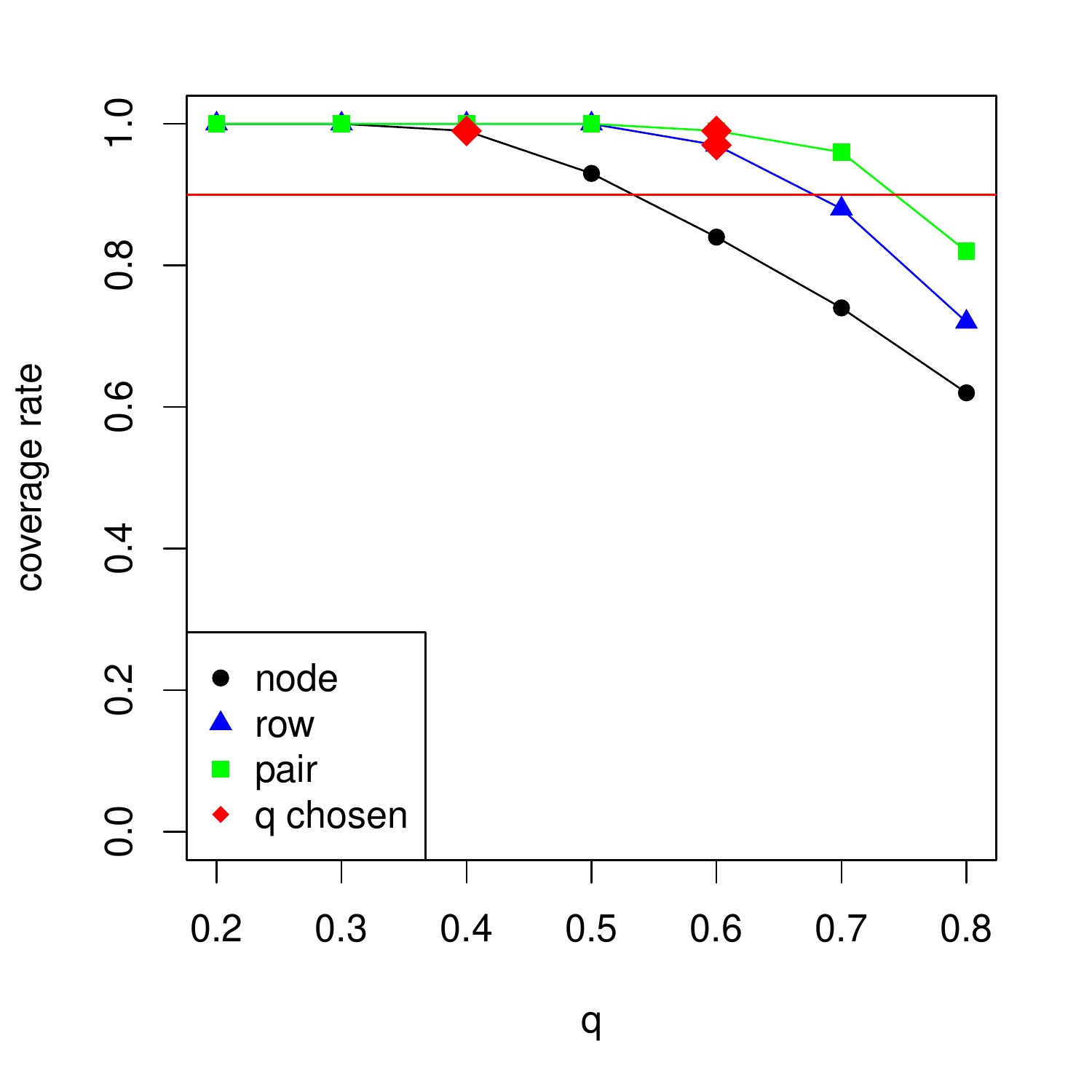}  
   \end{minipage}  \\

 \begin{minipage}{0.15\linewidth}  
 \centering
 \begin{align*}
 n & =  600 \\  
 \rho & =  0.05
 \end{align*}
 \end{minipage}  
  \begin{minipage}{0.4\linewidth}  
   \centering
  \includegraphics[scale=0.35]{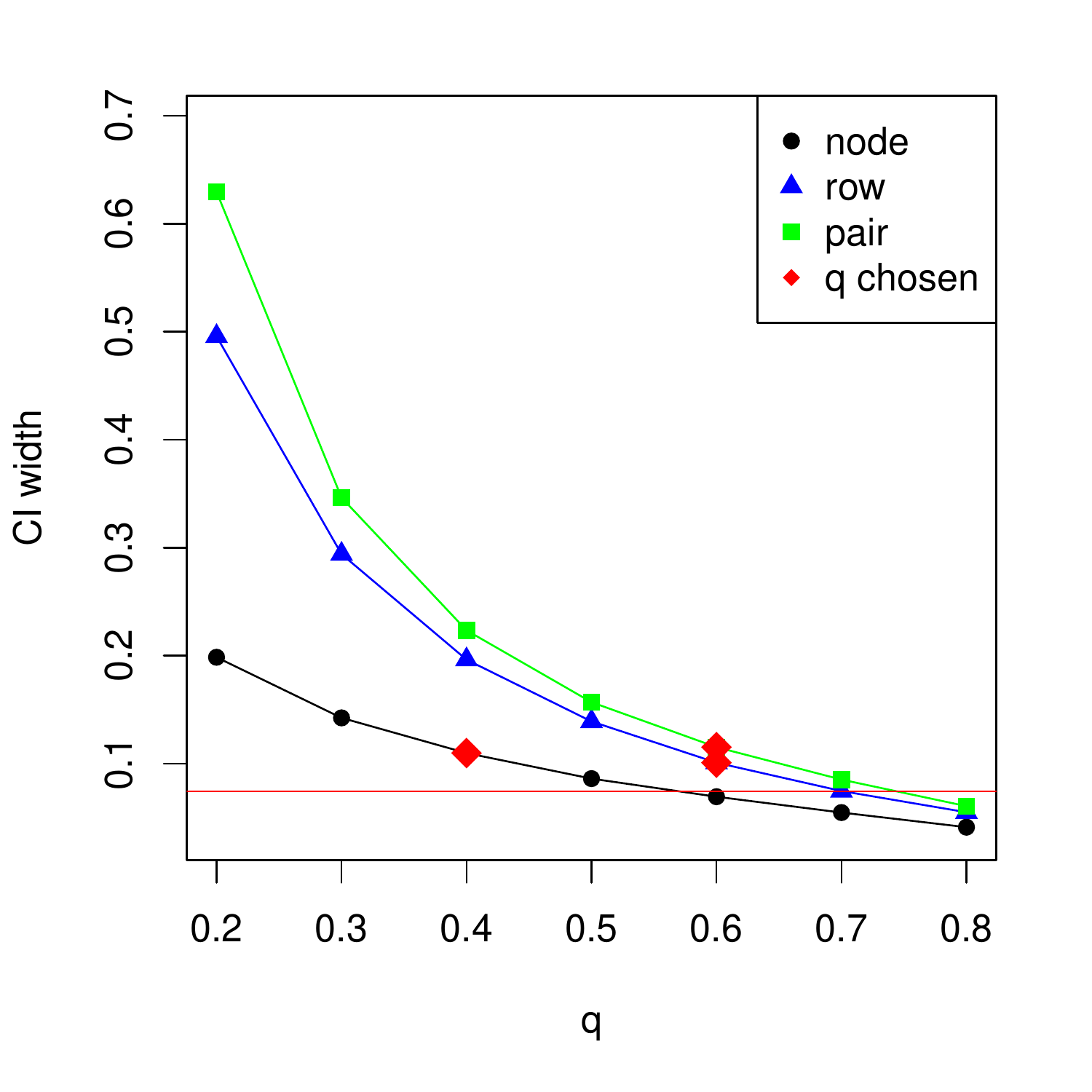} 
   \end{minipage}  
  \begin{minipage}{0.4\linewidth}  
   \centering
   \includegraphics[scale=0.35]{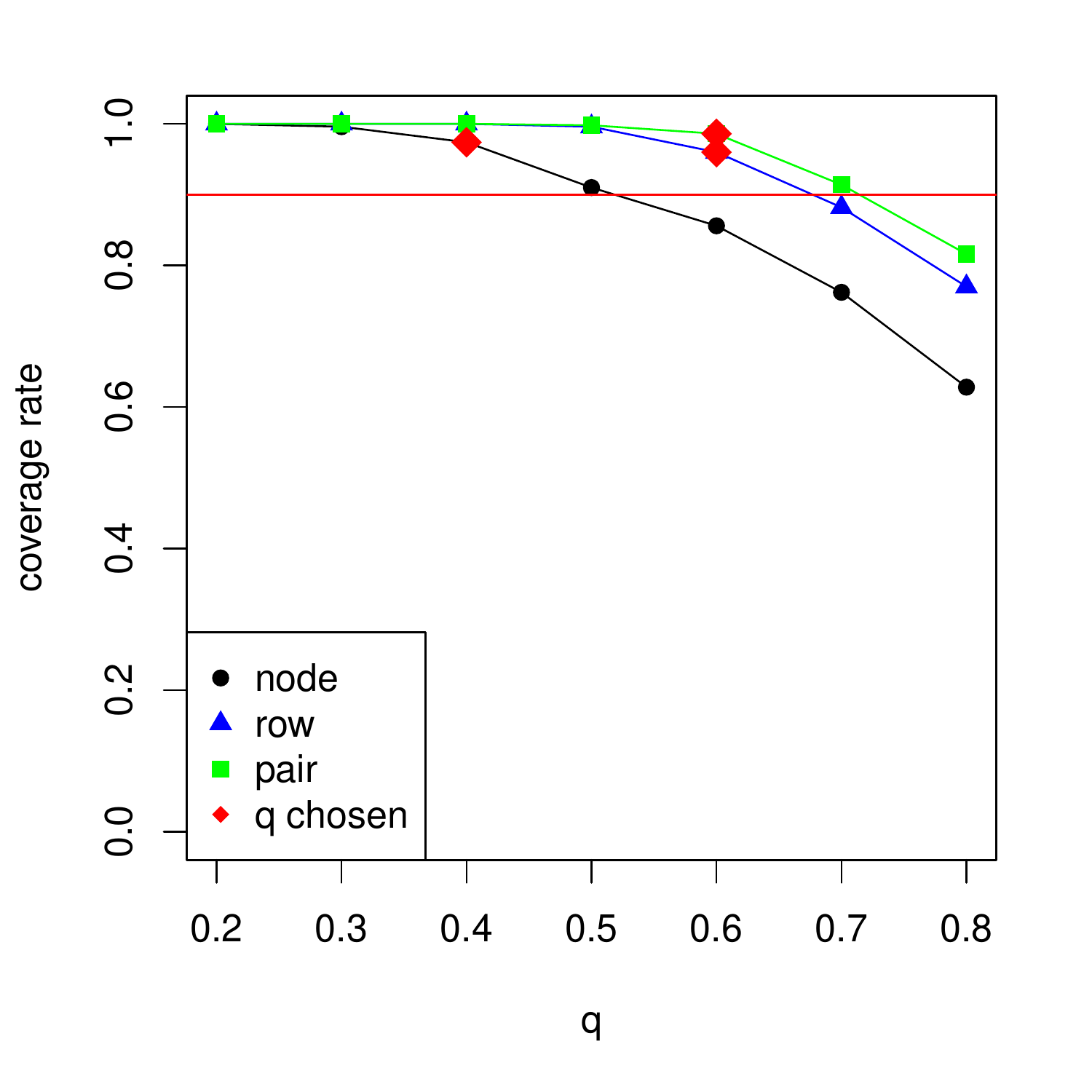}  
   \end{minipage} 	
	    \caption{Confidence intervals width  (left) and coverage (right) for the three different subsampling schemes on SBM graphs with $K=3$ equal-sized communities, and various values of $n$ and edge density $\rho$.   The value of $q$ chosen by our algorithm is shown as a red dot.}
	     \label{Triden2}
    \end{figure}

For a fair comparison, we match the proportion of adjacency matrix observed, $q$, in all subsampling schemes, and vary $q$ from 0.2 to 0.8. Figures \ref{Triden1} and \ref{Triden2} show the widths and coverage rates for the confidence intervals constructed by the three subsampling procedures over the full range of $q$, as well as the value of $q$  chosen by our proposed double bootstrap procedure.    

There are several general conclusions we can draw from the results in Figures~\ref{Triden1} and \ref{Triden2}.

(1) The confidence intervals generally get shorter as $q$ increases, since the subsampled graphs overlap more, and this dependence reduces variance.    The coverage rate correspondingly goes down as $q$ goes up.  Still, all resampling procedures achieve nominal or higher coverage for $q \le 0.5$.    The one exception to this is node pair sampling at the lowest value of $q$ with  the lowest edge density, where most subsampled graphs are too sparse to have any triangles.  In this case the procedure fails, producing estimated zero triangle density and confidence intervals with length and coverage both equal to zero.

(2)   For a given value of $q$, a larger effective sample size, either from higher density $\rho$ or from a larger network size $n$, leads to shorter intervals, as one would expect.

(3) Node sampling produces shorter intervals than the other two schemes, while maintaining similar coverage rates for smaller values of $q$.  The coverage rate falls faster for node sampling for larger values of $q$ outside the optimal range.

(4) The double bootstrap procedure selects a value of $q$ that achieves a good balance of coverage rate and confidence interval width, maintaining close to nominal coverage without being overly conservative.  It does get misled by the pathological sparse case with no triangles, highlighting the need of not deploying resampling algorithms without checking that there is enough data to preserve the feature of interest (triangles in this case)  in the subsampled graphs.

\begin{figure}[ht!]
\centering
\subfloat[ER, $n=300$]{\includegraphics[width=0.4\textwidth]{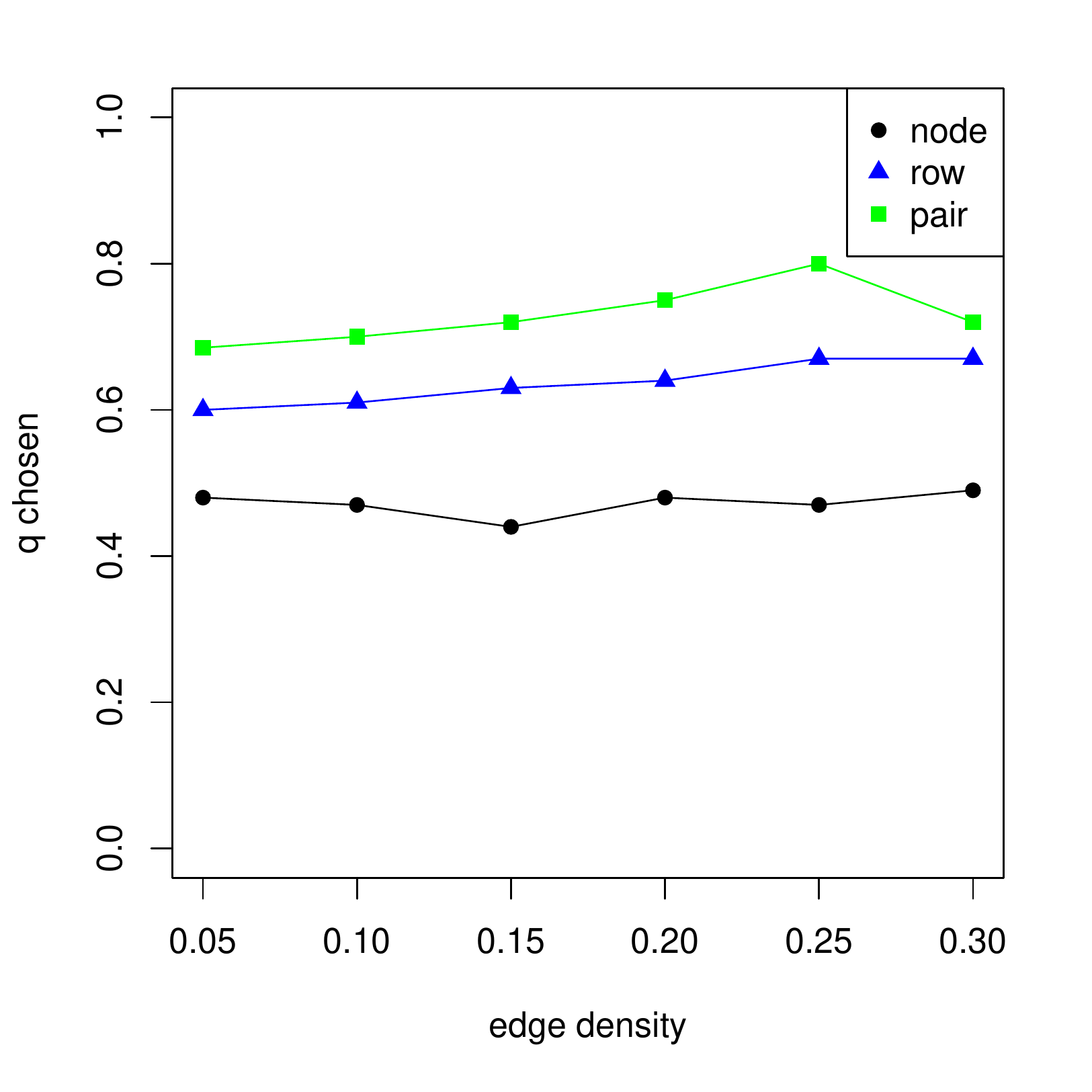}}
\subfloat[ER, $\rho = 0.05$] {\includegraphics[width=0.4\textwidth]{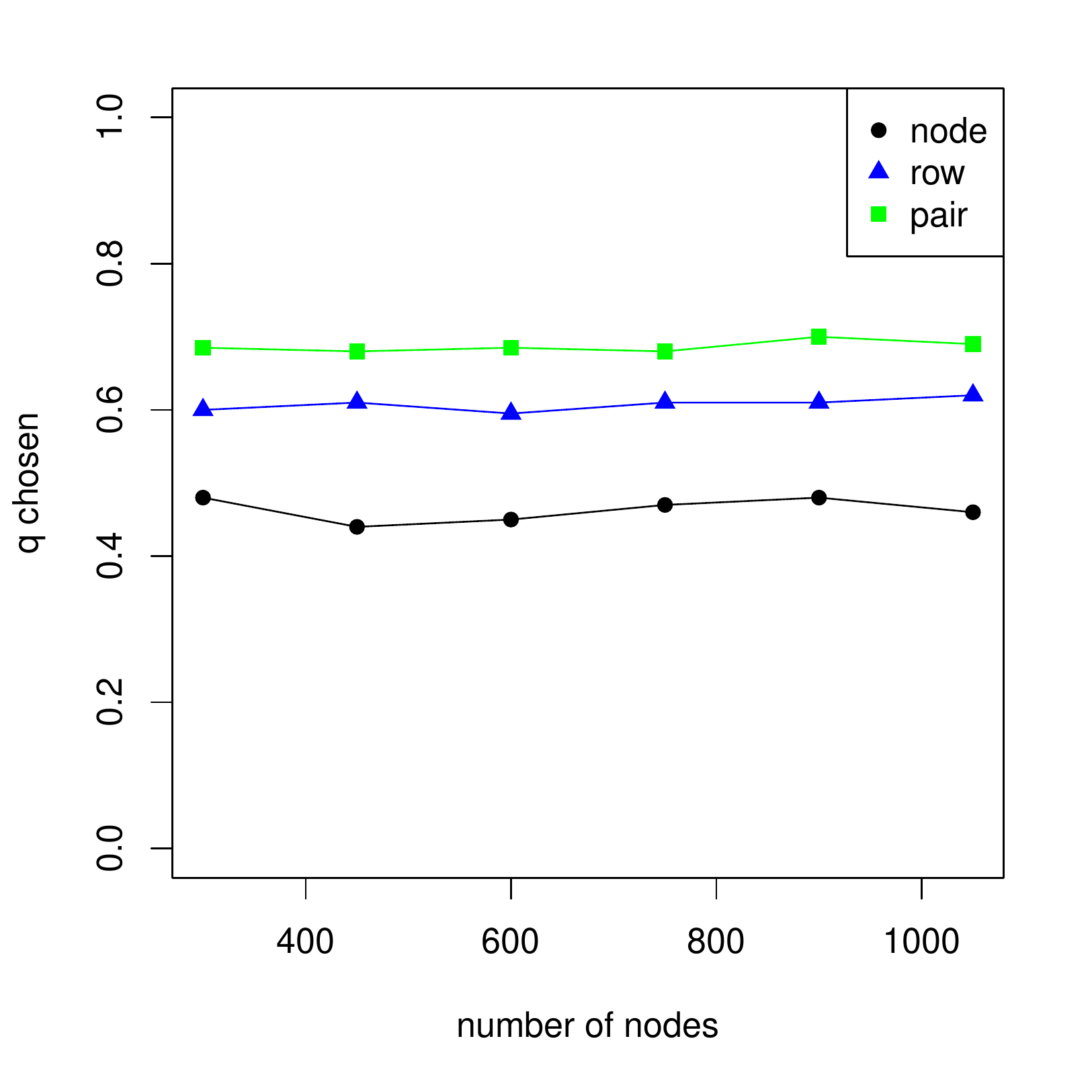}}\\
\subfloat[SBM, $n=300$]{\includegraphics[width=0.4\textwidth]{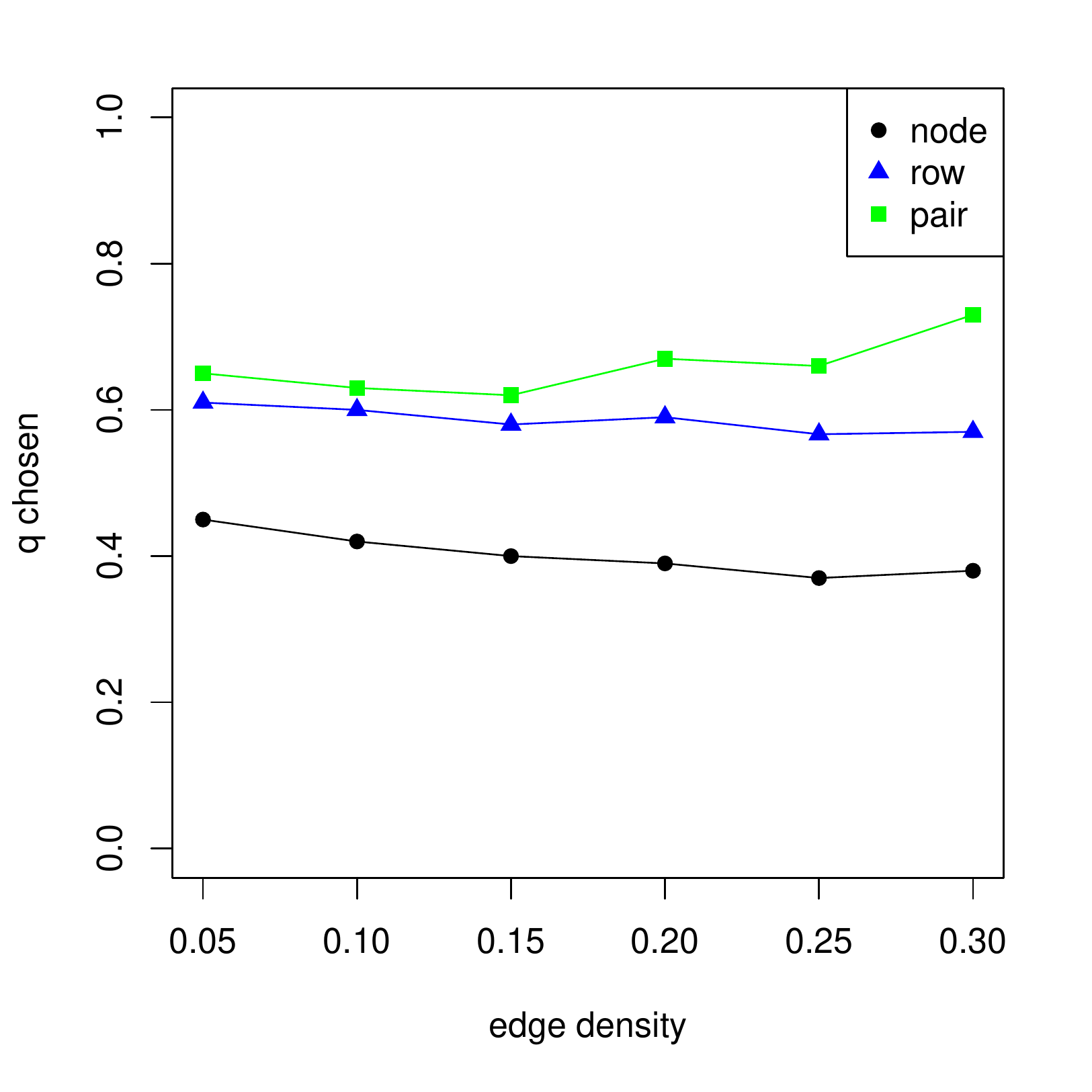}}
\subfloat[SBM, $\rho = 0.05$] {\includegraphics[width=0.4\textwidth]{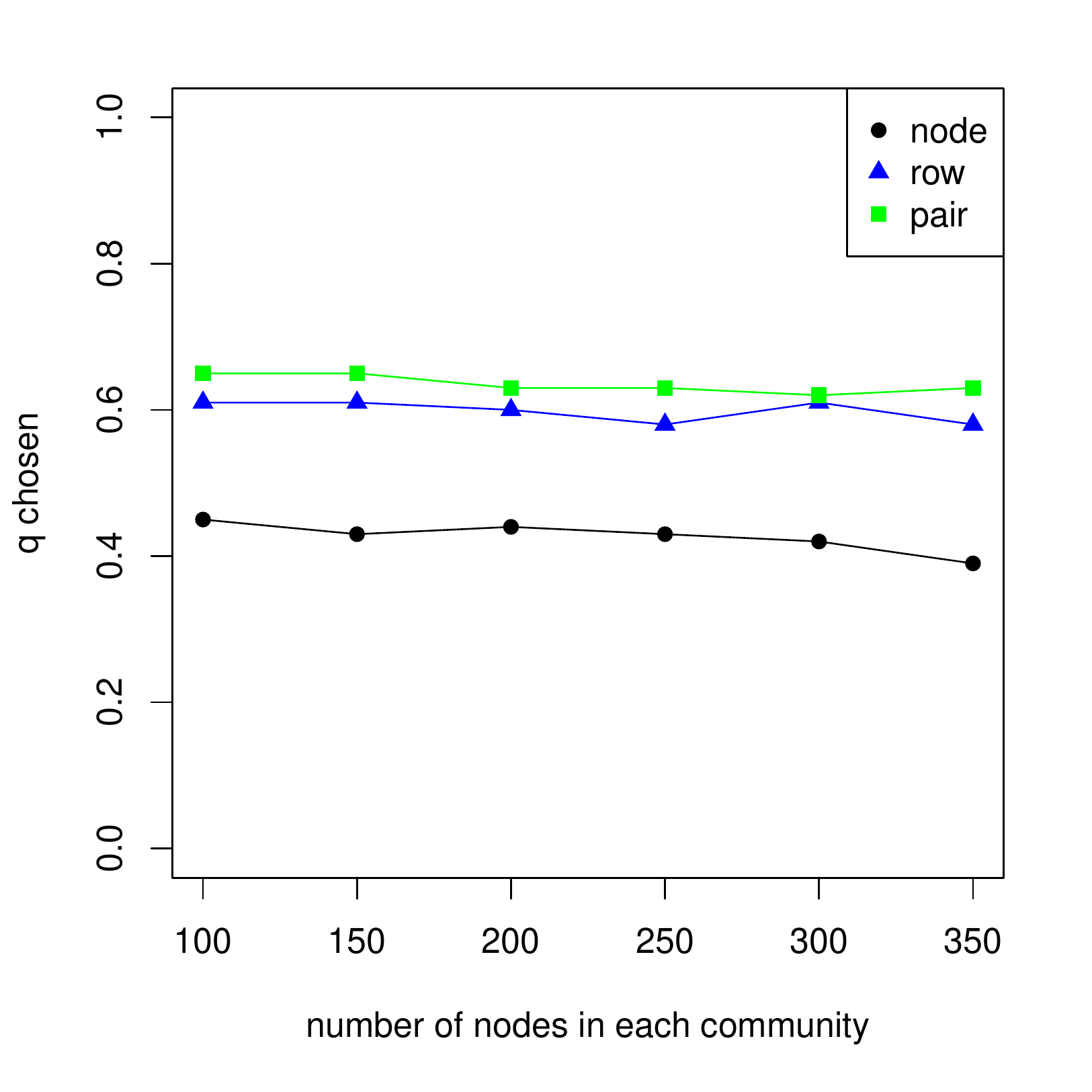}}
\caption{The optimal value $q$ chosen by double bootstrap, as a function of edge density $\rho$ (left) and number of nodes $n$ (right).  }
	 \label{Triden3}
\end{figure}

  Figure \ref{Triden3} shows the value of $q$ chosen by the double bootstrap procedure as a function of graph size and edge density;  both appear to not have much effect on the value of $q$ within the range considered. 
  This suggests that as long as the structure of a network is preserved, a reasonable value of $q$ could be chosen on a smaller subsampled network, resulting in computational savings with little change in accuracy.    And we can see that the $q$ chosen for node sampling is always smaller than that for the other two subsampling methods, so node sampling can be more computationally efficient, giving smaller subsampled graphs.

\FloatBarrier

	\subsection{Number of Communities}
Finding communities in a network often helps understand and interpret the network patterns, and the number of communities $K$ is required as input for most community detection algorithms \cite{bhattacharyya_2014_community, fortunato_2010_community}.   A number of methods have been proposed to estimate $K$:   as a function of the observed network's spectrum \cite{le_estimating_2015}, through sequential hypothesis testing \cite{bickel_hypothesis_2016}, or subsampling \cite{li_network_2020,chen_network_2018}.   Subsampling has been shown to be one of the most reliable methods to estimate $K$, but little attention has been paid to choosing the subsampling fraction and understanding its effects.   Here we investigate the performance of the three subsampling schemes on this task and the choice of the subsampling fraction on the task of understanding uncertainty in $K$, beyond the earlier focus on point estimation.  Again, we match the proportion of adjacency matrix observed, $q$, across all subsampling schemes.
	
For this task, we generate networks from an SBM with $K=3$ equal-sized communities. We vary   the number of nodes $n$, the expected edge density $\rho$, and the ratio of within and between communities edge probabilities $t$, which together control the problem difficulty.   

For subgraphs generated by node sampling and row sampling, we estimate the number of communities using the spectral method based on the Bethe-Hessian matrix \cite{le_estimating_2015, saade_matrix_2016}.   For a discrete variable like $K$, confidence intervals make little sense, and thus we look at their bootstrap distributions instead.    These distributions are not symmetric, and rarely overestimate the number of communities.   
	For subgraphs generated from node pair sampling, there is no complete adjacency matrix to compute the Bethe-Hessian for, and computing it with missing values replaced by zeros is clearly introducing additional noise. We instead estimate the number of communities from node-pair samples using the network cross-validation approach \cite{li_network_2020}, first applying low-rank matrix completion with values of rank $K$ ranging from 1 to 6, and choosing the best $K$ by maximizing the AUC score for reconstructing the missing edges.

\begin{figure}[ht!]
\centering
\vspace{-8pt} 
\subfloat[$\rho = 0.05$, $t=2$]{\includegraphics[width=0.3\textwidth]{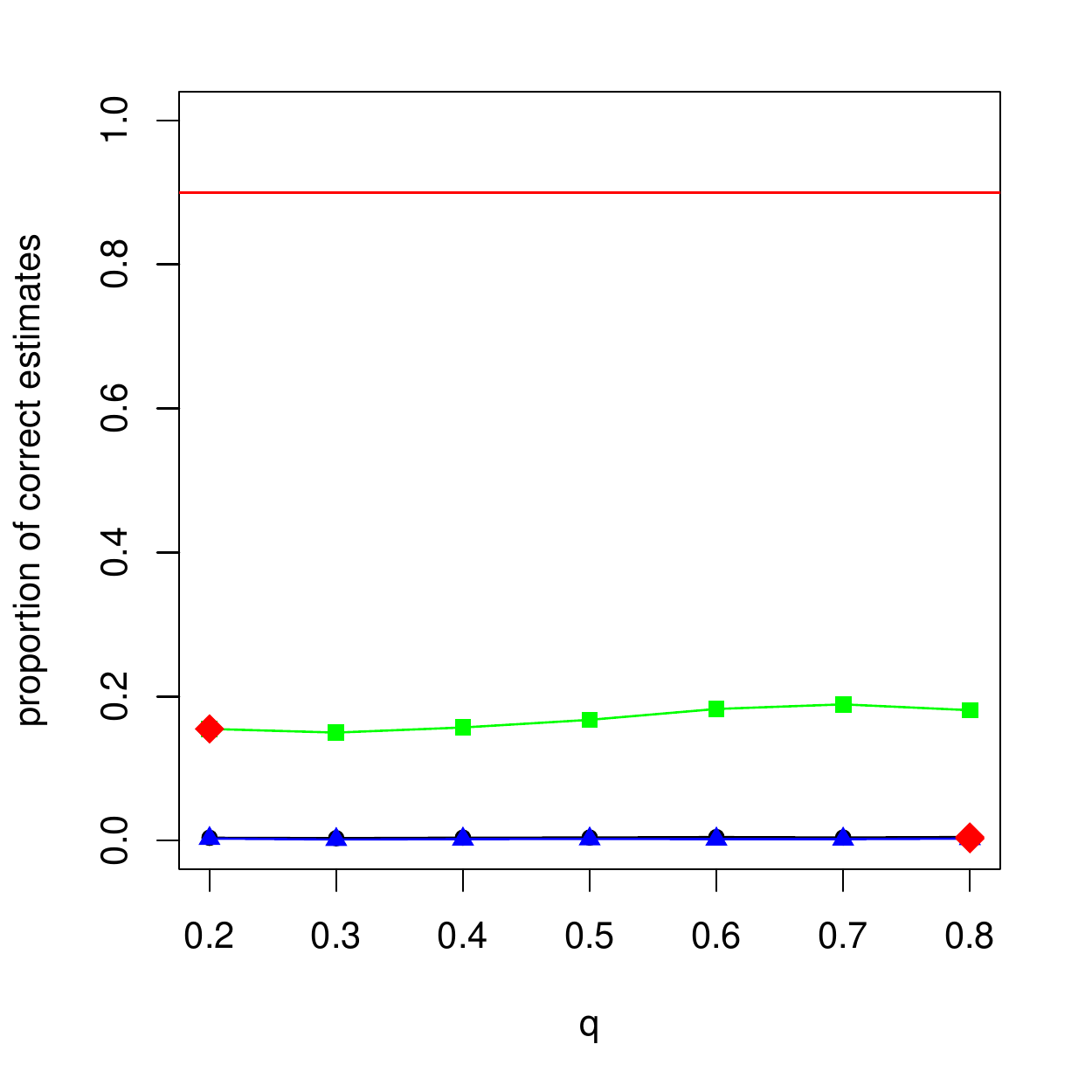}}
\subfloat[$\rho = 0.1$, $t=2$] {\includegraphics[width=0.3\textwidth]{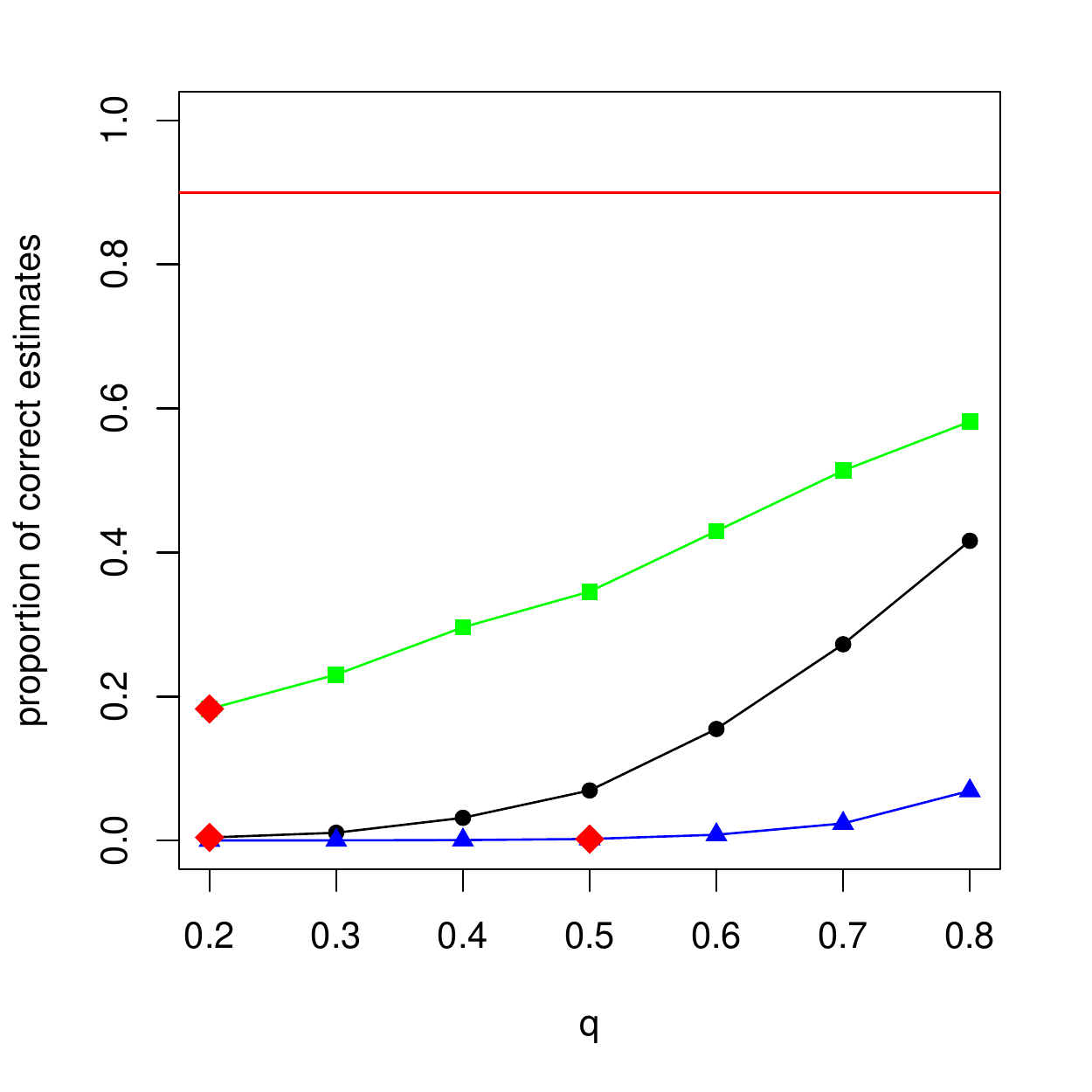}}\vspace{-5pt} \\
\vspace{-8pt} 
\subfloat[$\rho = 0.05$, $t=3$]{\includegraphics[width=0.3\textwidth]{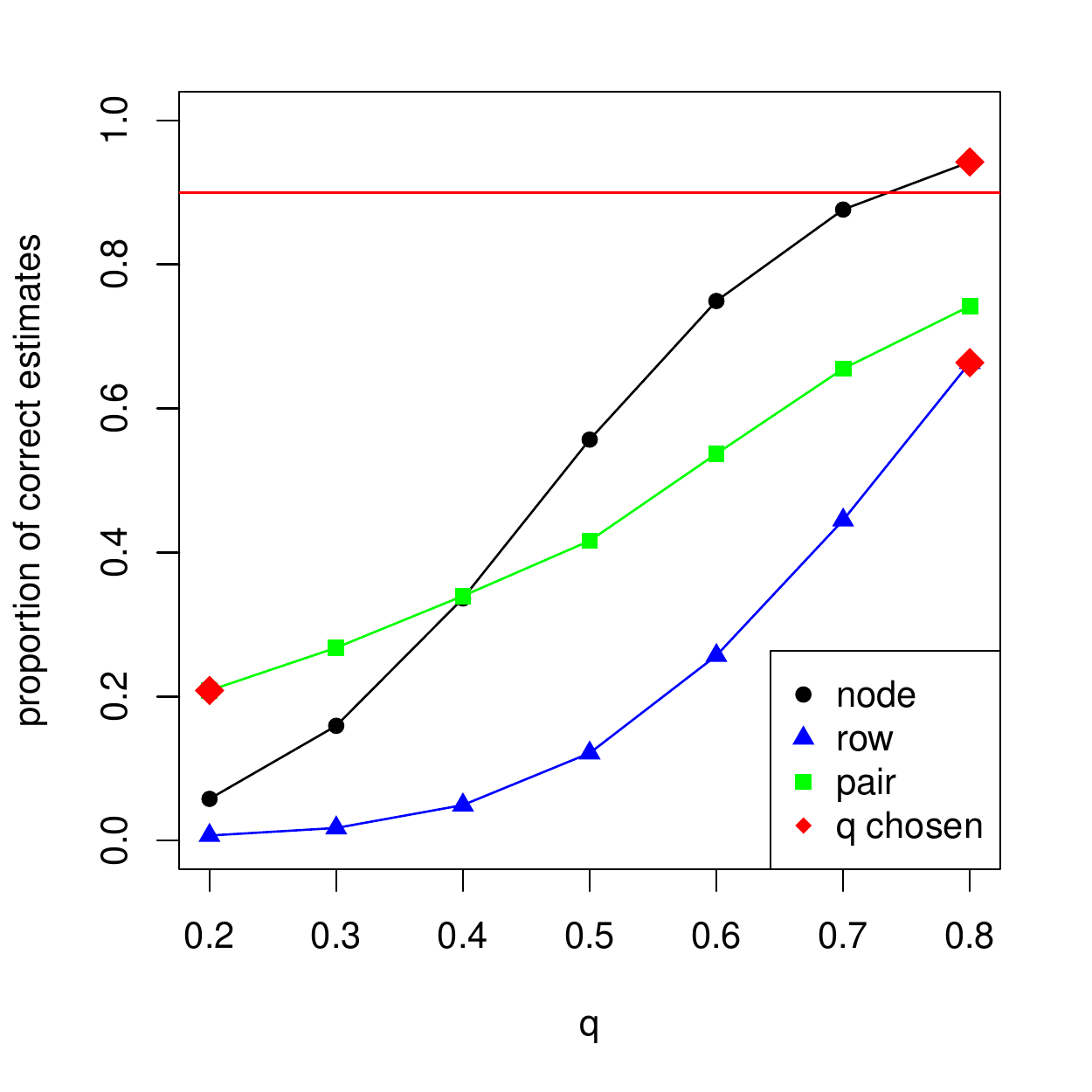}}
\subfloat[$\rho = 0.1$, $t=3$] {\includegraphics[width=0.3\textwidth]{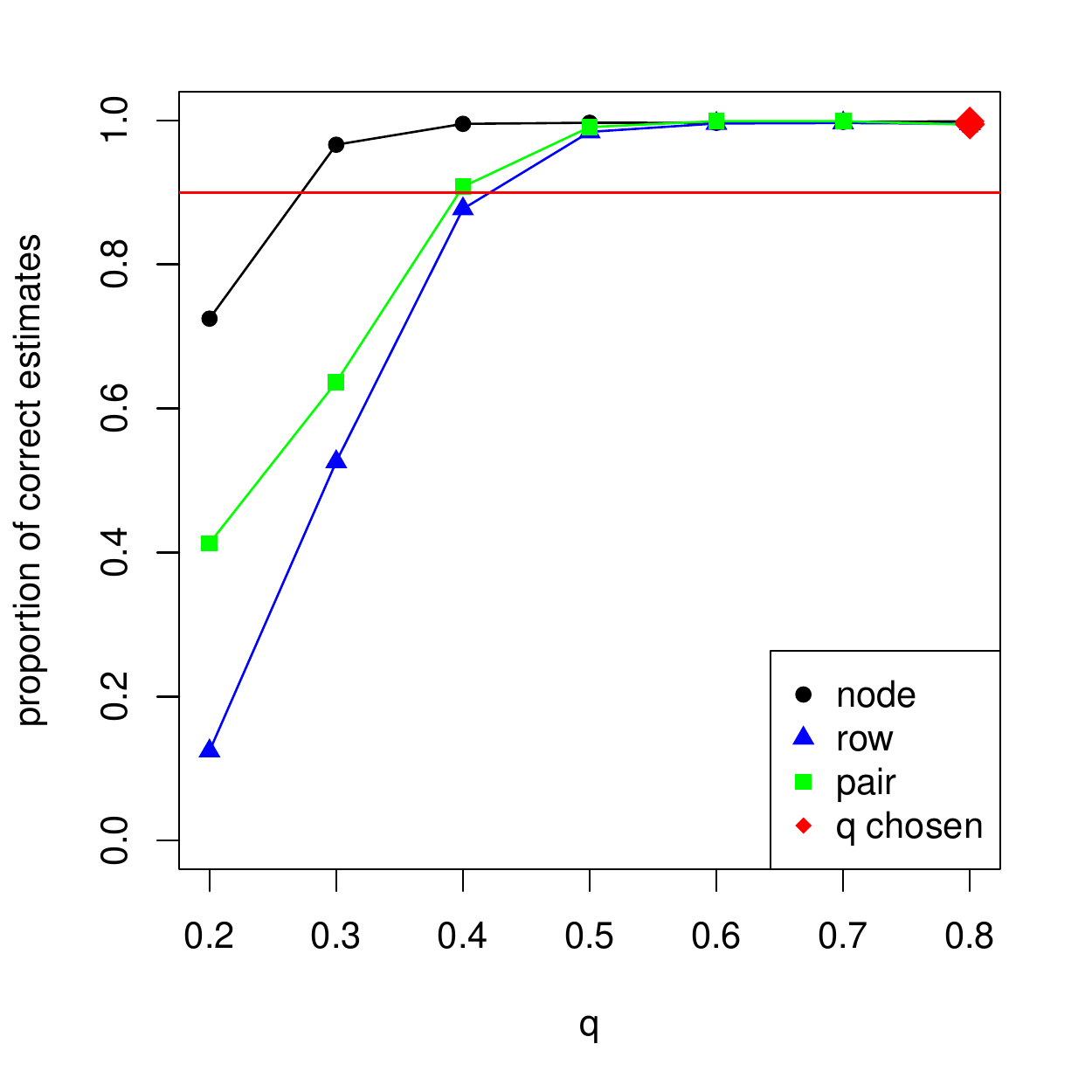}}\vspace{-5pt} \\
\vspace{-8pt}
\subfloat[$\rho = 0.05$, $t=4$]{\includegraphics[width=0.3\textwidth]{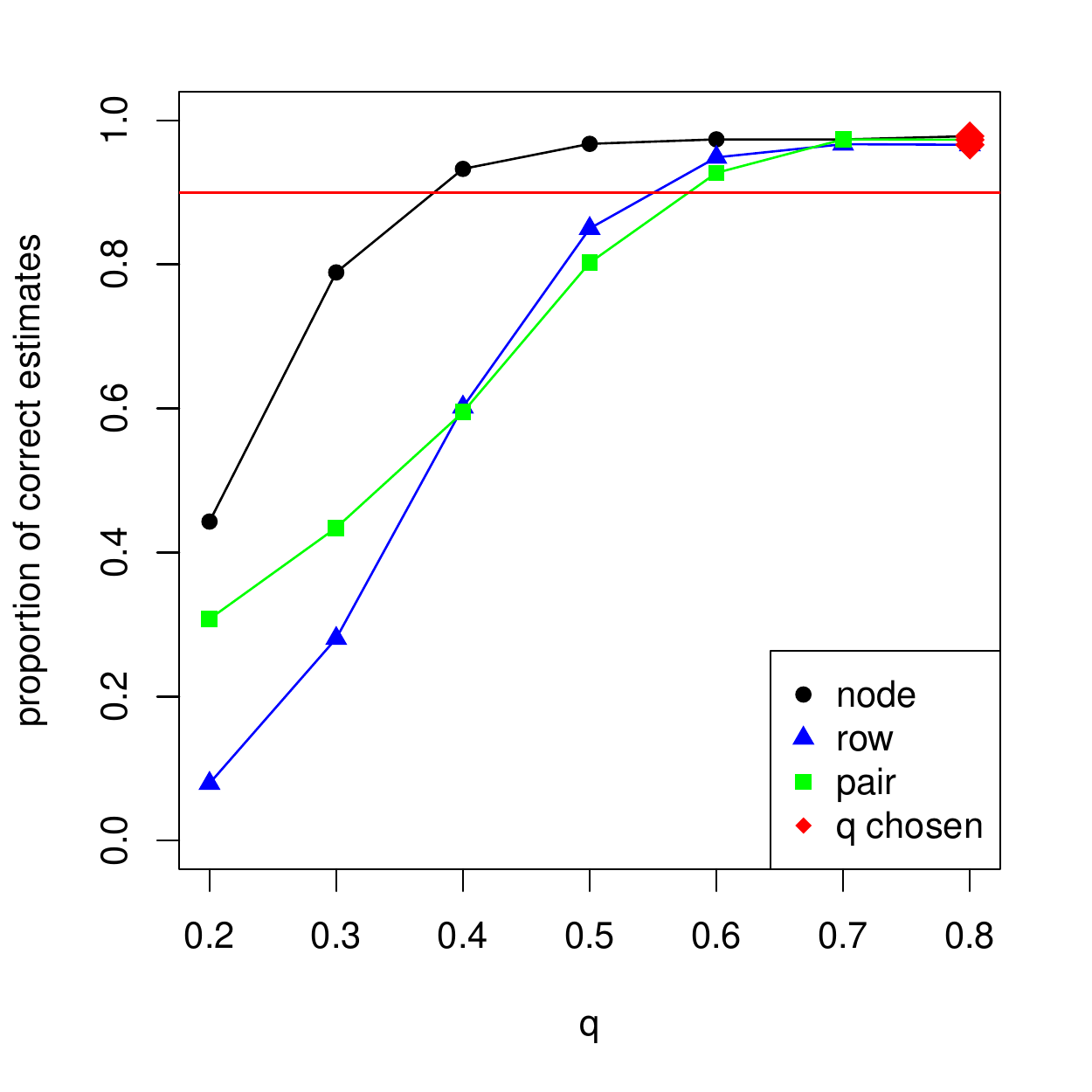}}
\subfloat[$\rho = 0.1$, $t=4$] {\includegraphics[width=0.3\textwidth]{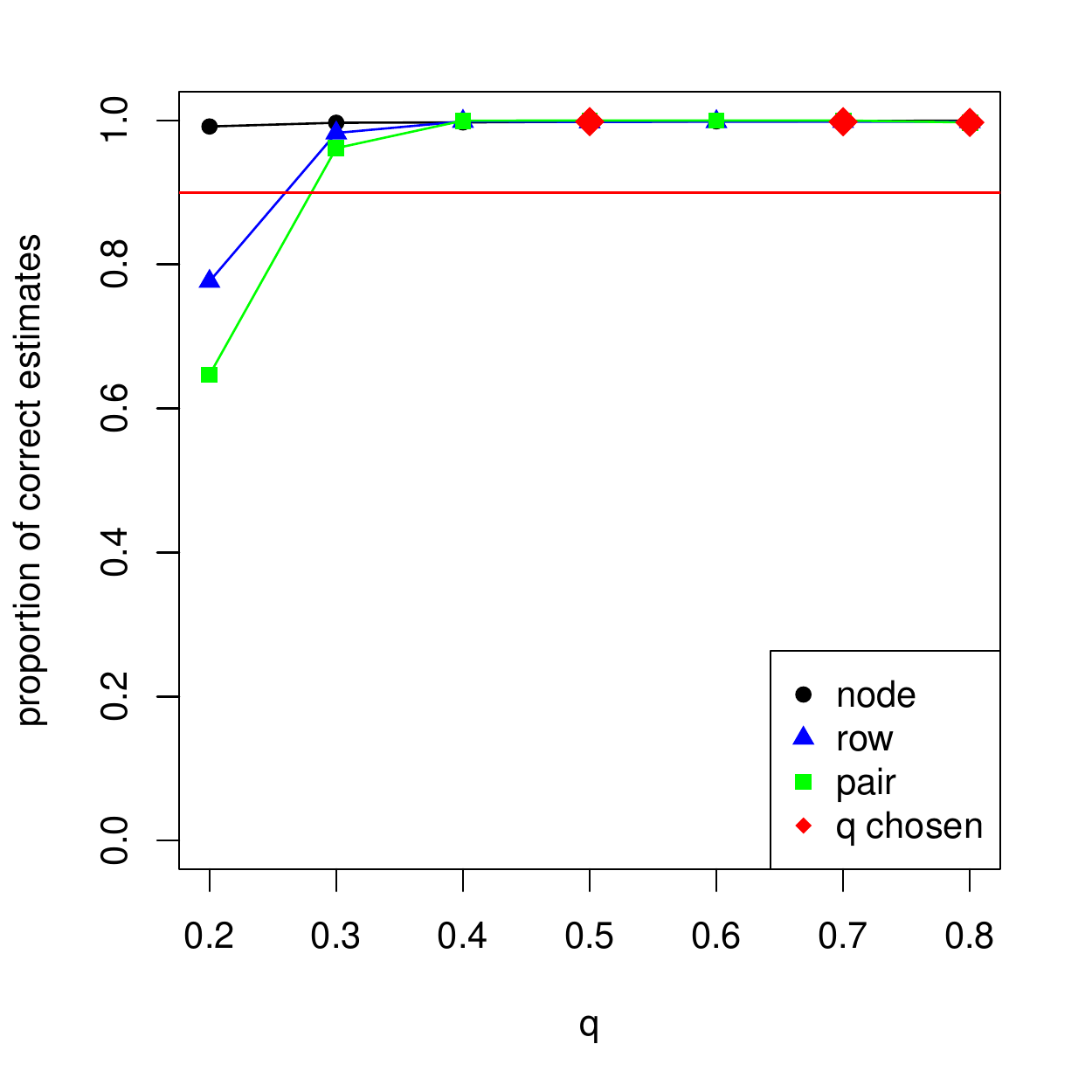}}\vspace{-5pt} \\

\subfloat[$\rho = 0.05$, $t=5$]{\includegraphics[width=0.3\textwidth]{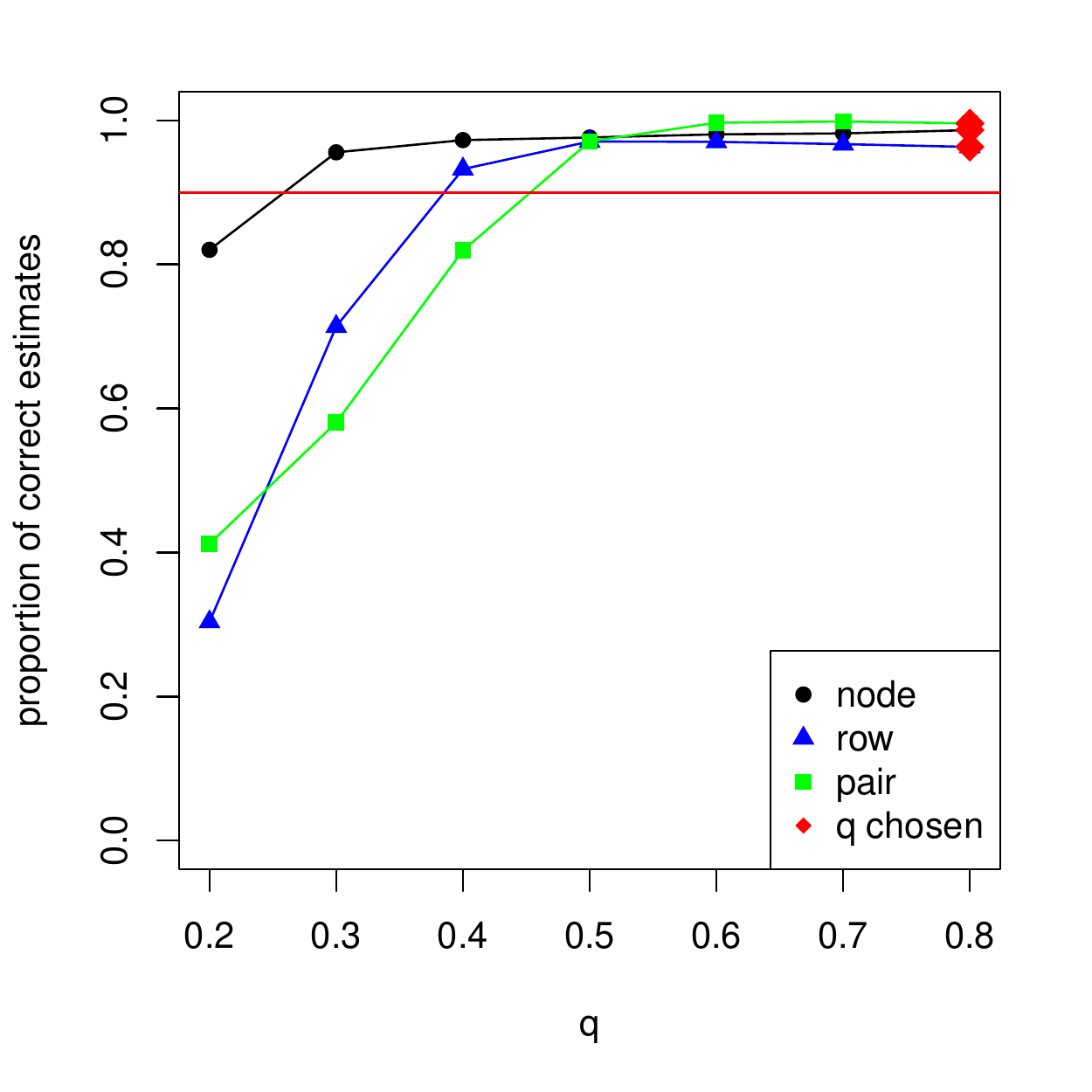}}
\subfloat[$\rho = 0.1$, $t=5$] {\includegraphics[width=0.3\textwidth]{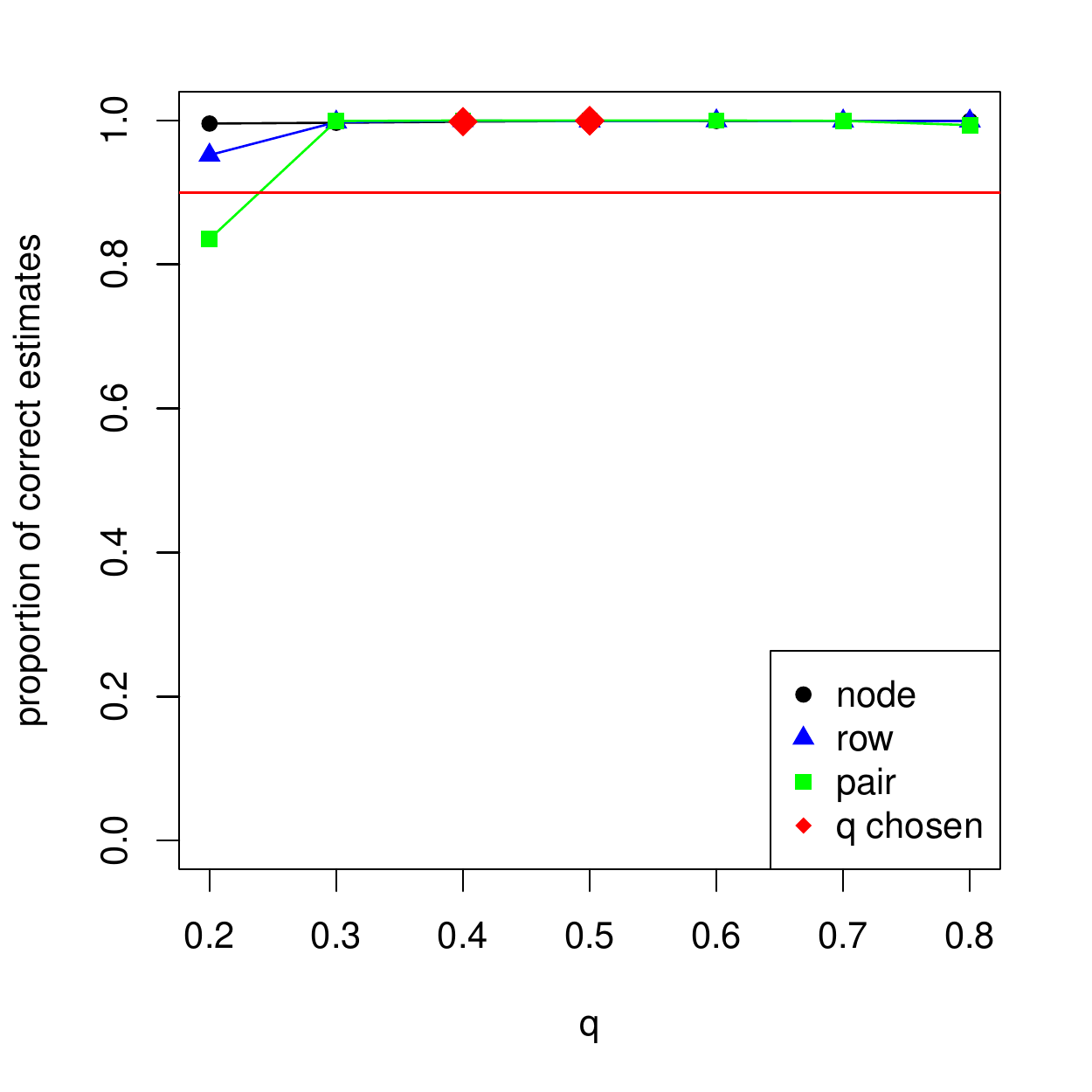}}

\caption{Simulation results for SBMs with $K = 3$ and $n = 300$.}
\label{Community1}
\end{figure}

\begin{figure}[ht!]
\centering
\vspace{-8pt} 
\subfloat[$\rho = 0.05$, $t=2$]{\includegraphics[width=0.3\textwidth]{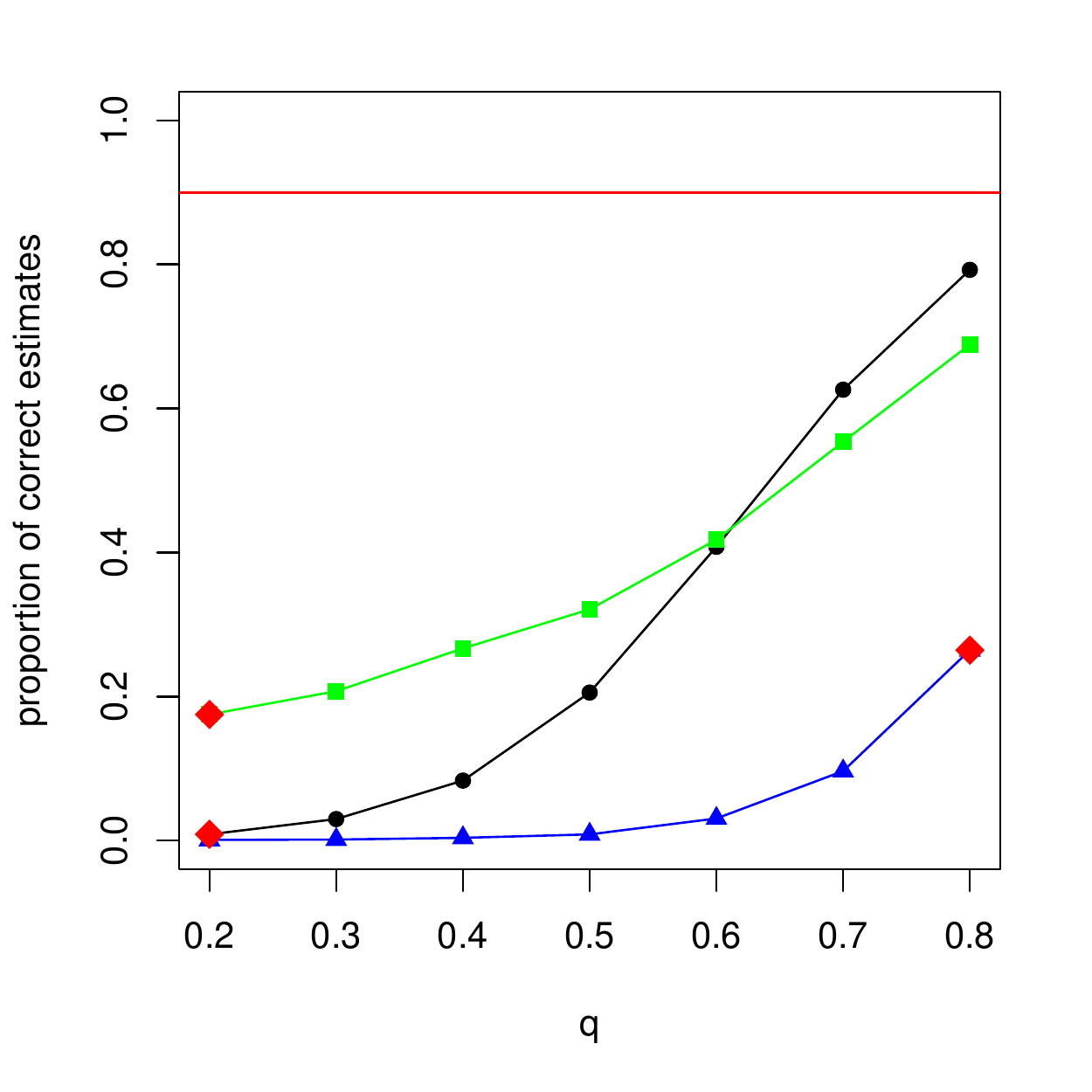}}
\subfloat[$\rho = 0.1$, $t=2$] {\includegraphics[width=0.3\textwidth]{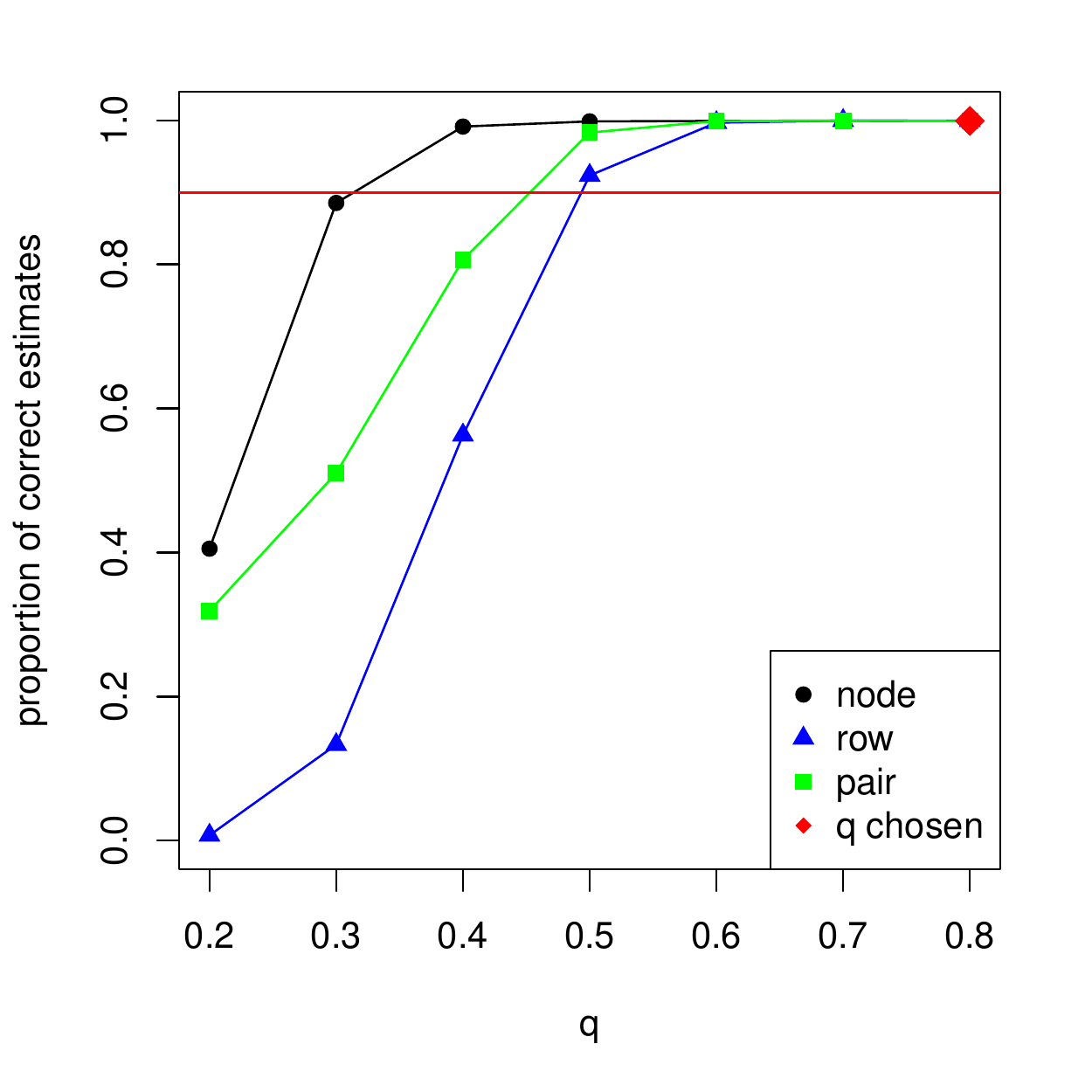}}\vspace{-5pt} \\
\vspace{-8pt} 
\subfloat[$\rho = 0.05$, $t=3$]{\includegraphics[width=0.3\textwidth]{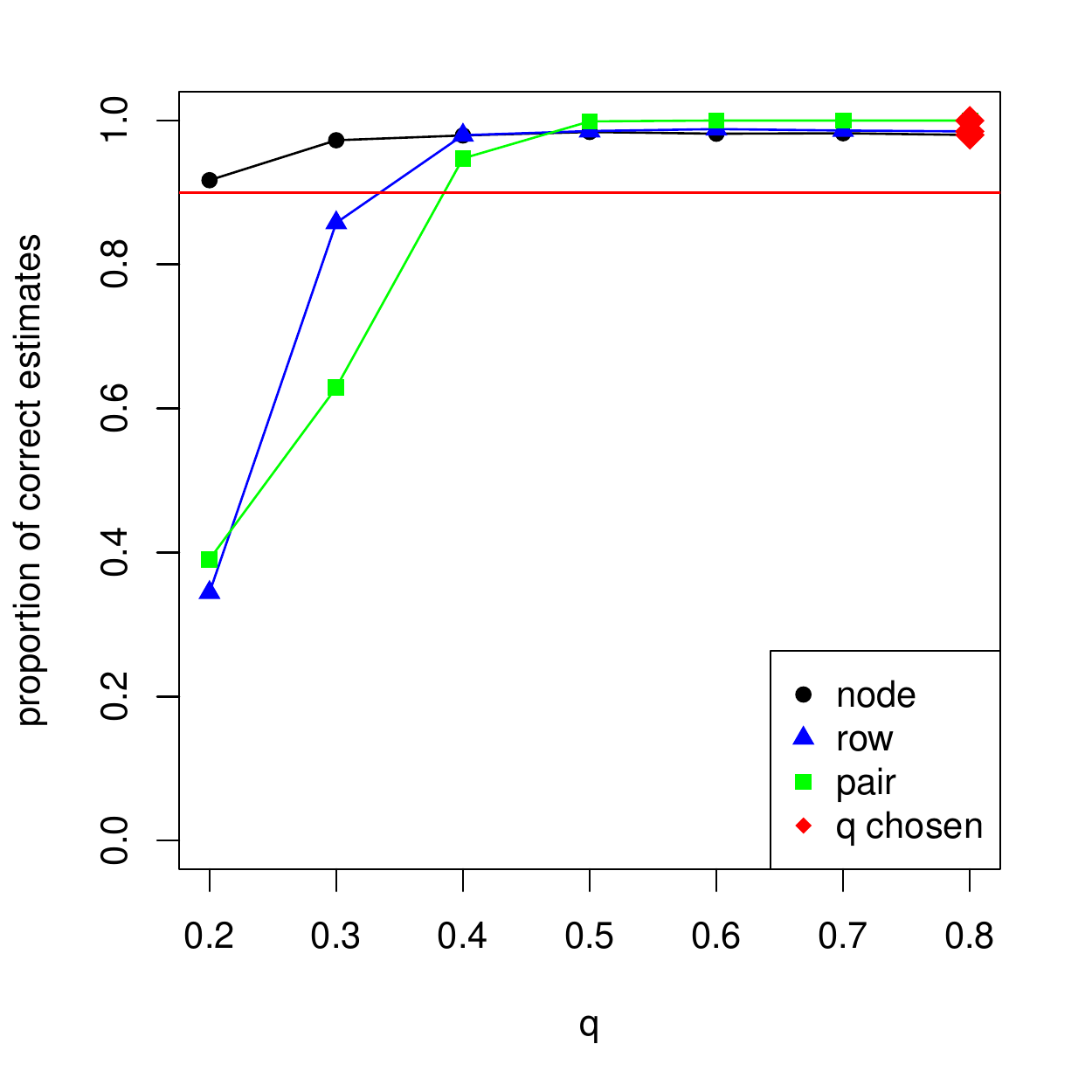}}
\subfloat[$\rho = 0.1$, $t=3$] {\includegraphics[width=0.3\textwidth]{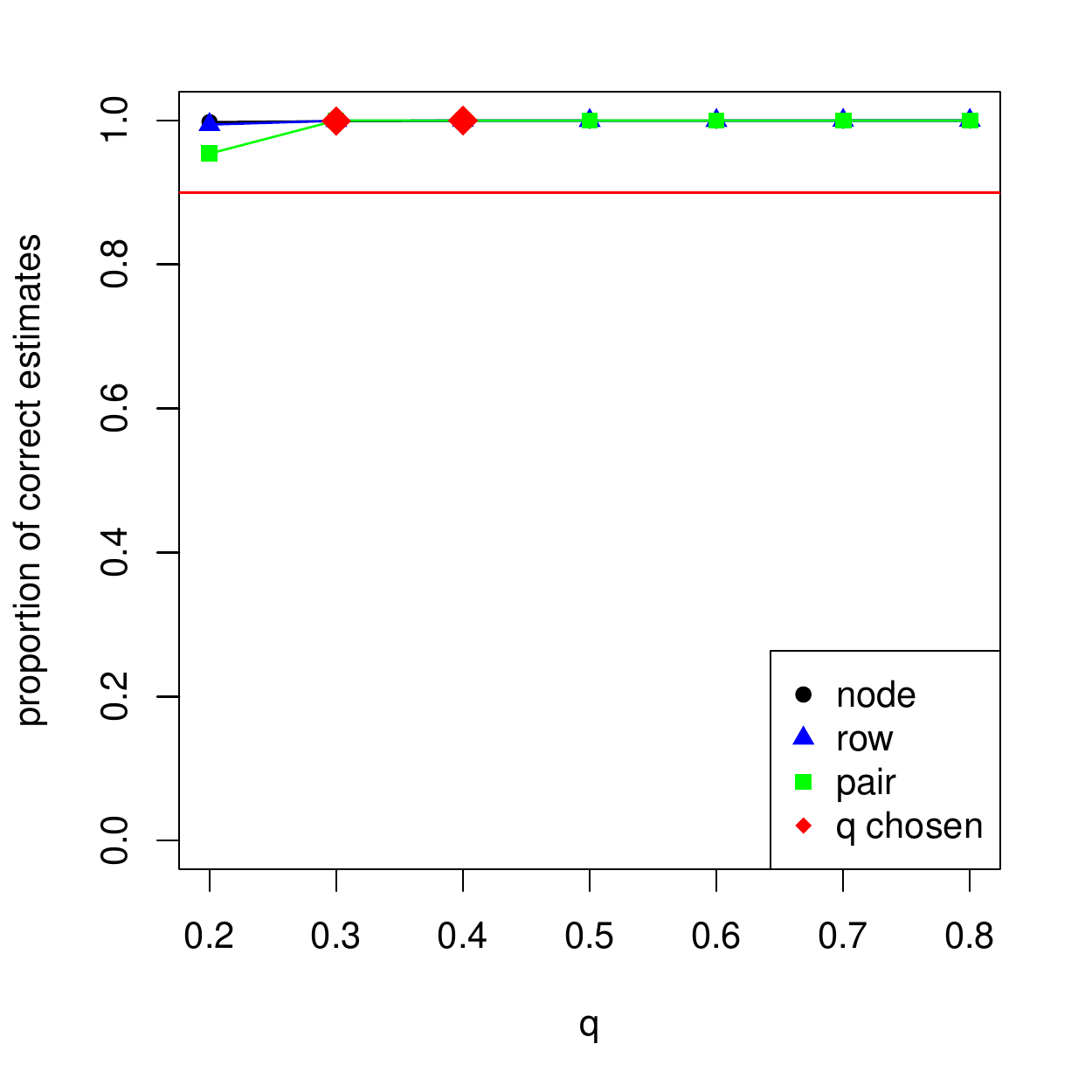}}\vspace{-5pt} \\
\vspace{-8pt}
\subfloat[$\rho = 0.05$, $t=4$]{\includegraphics[width=0.3\textwidth]{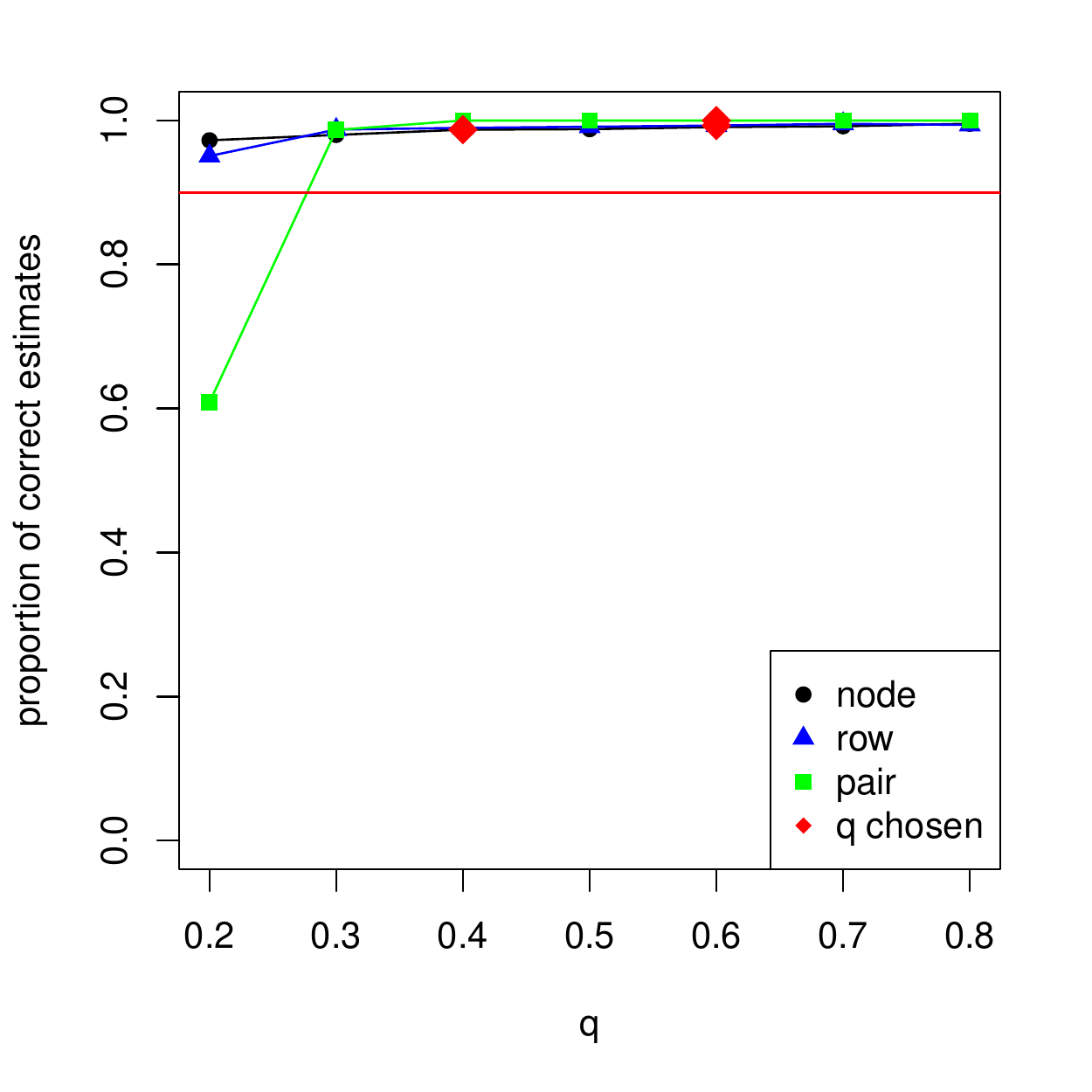}}
\subfloat[$\rho = 0.1$, $t=4$] {\includegraphics[width=0.3\textwidth]{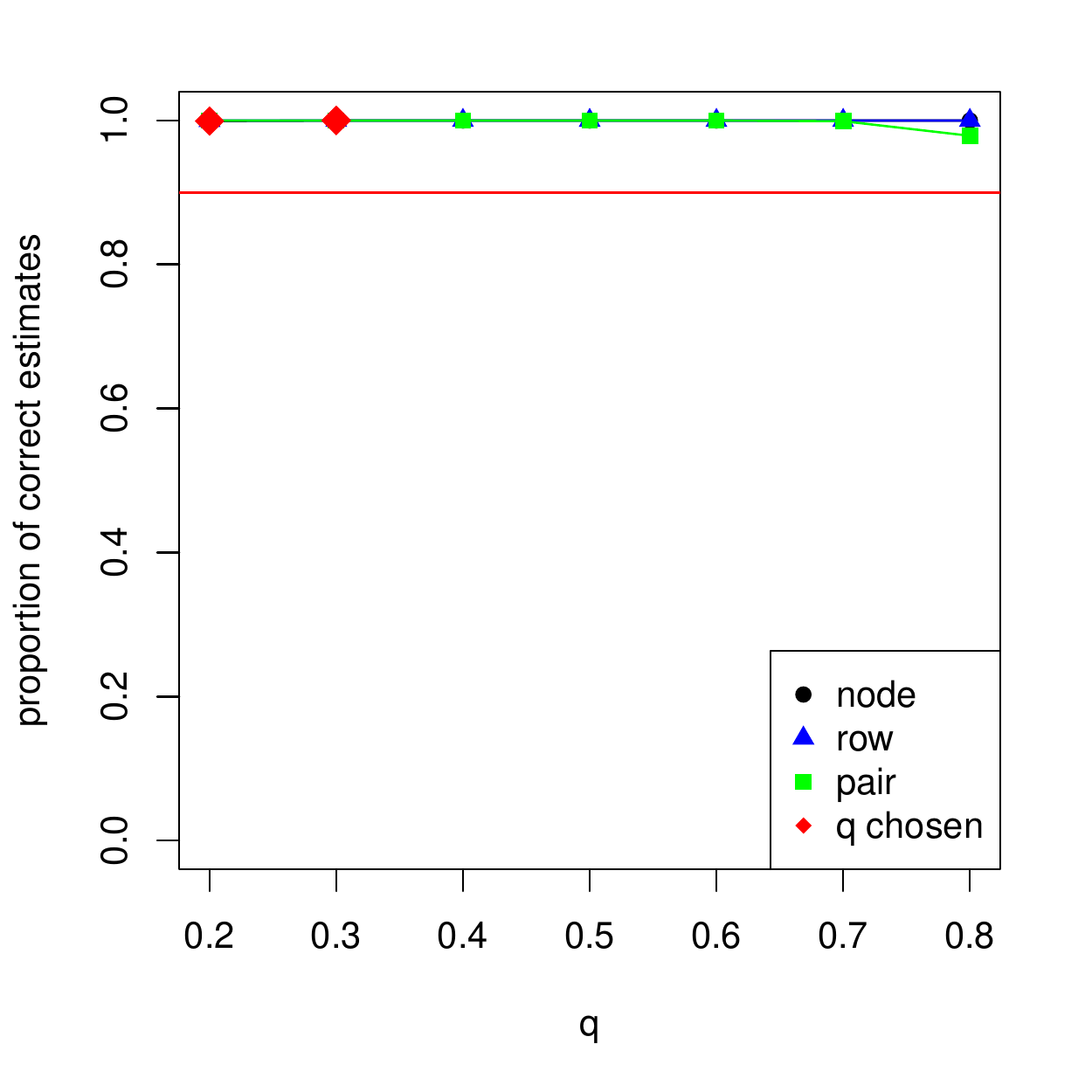}}\vspace{-5pt} \\

\subfloat[$\rho = 0.05$, $t=5$]{\includegraphics[width=0.3\textwidth]{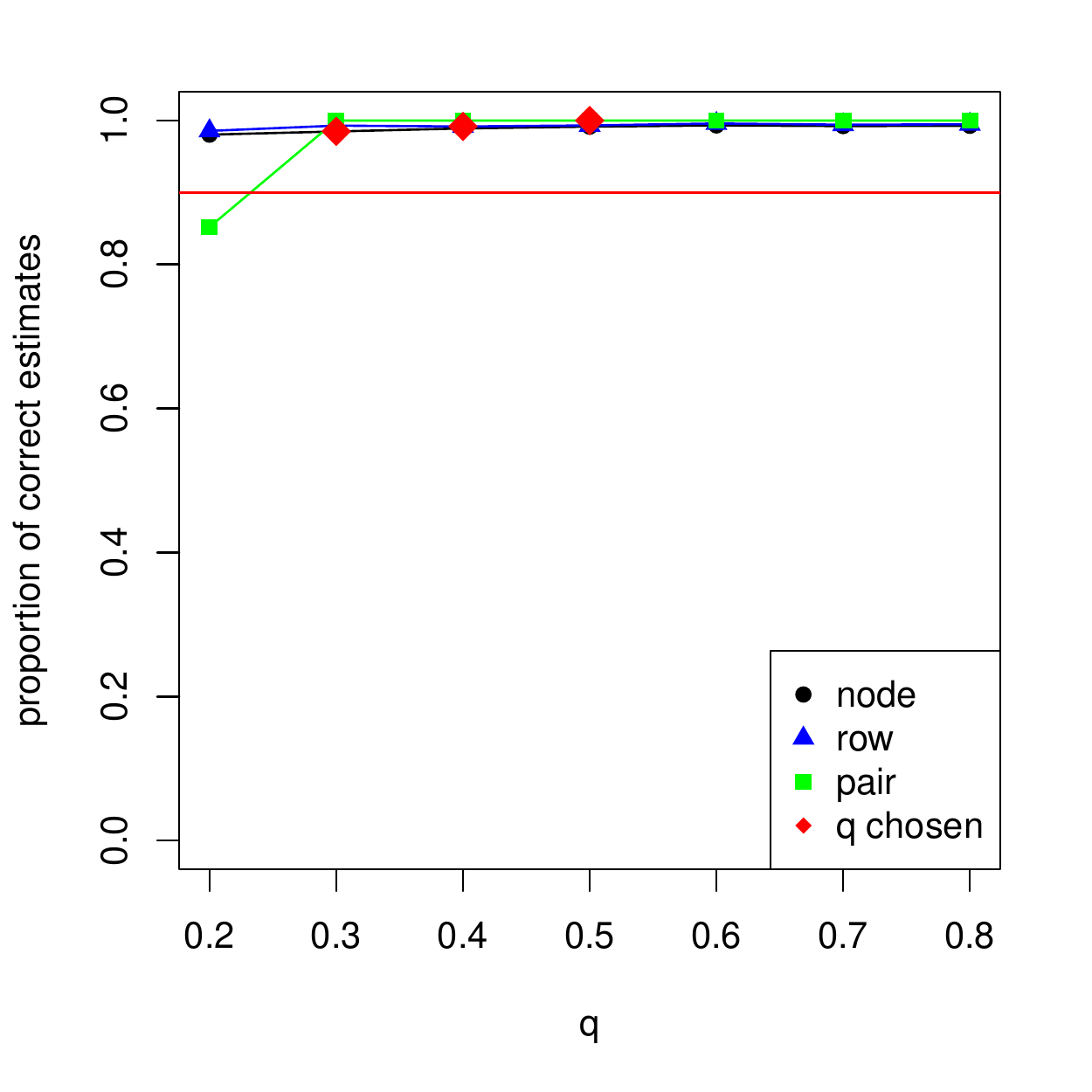}}
\subfloat[$\rho = 0.1$, $t=5$] {\includegraphics[width=0.3\textwidth]{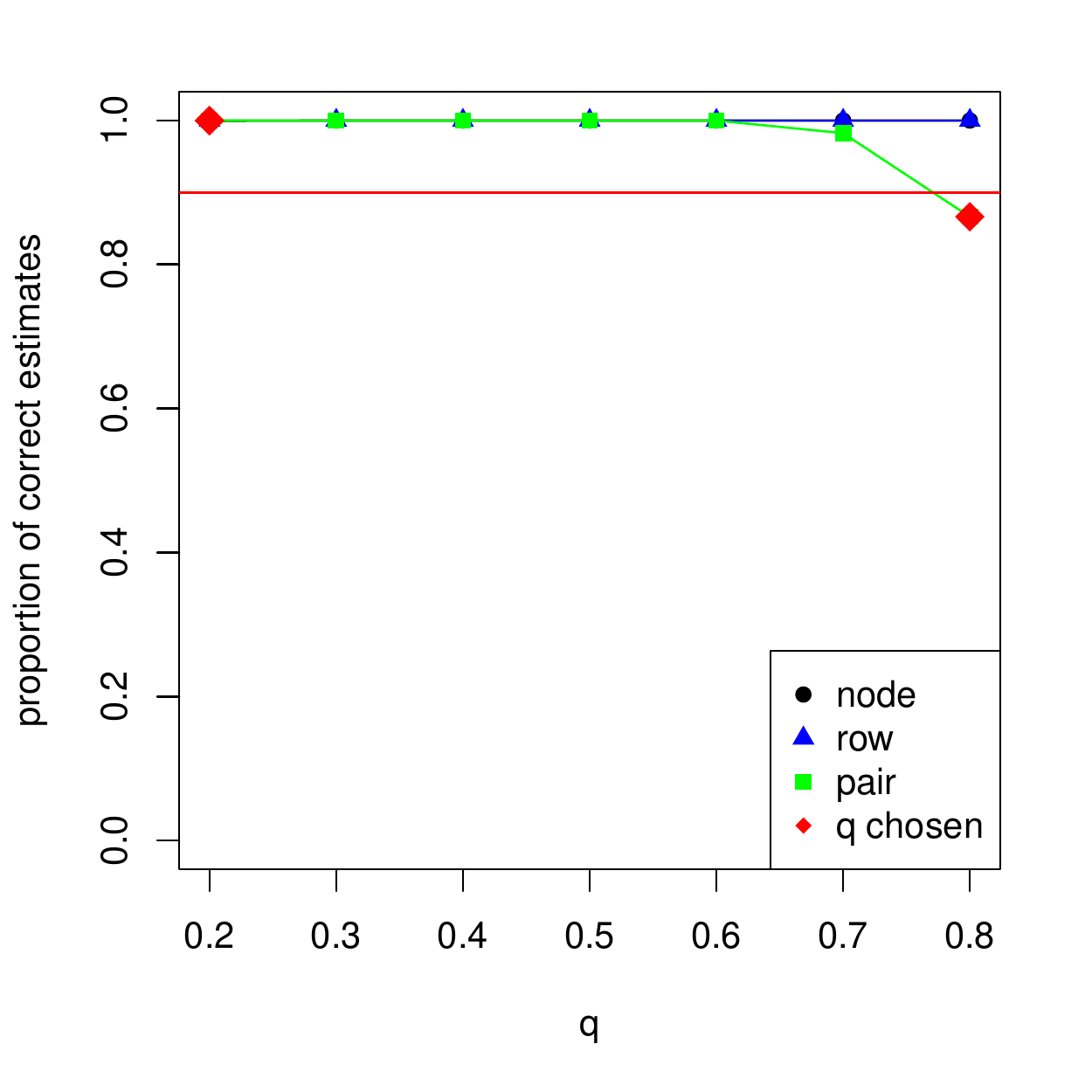}}

\caption{Simulation results for SBMs with $K = 3$ and $n = 600$.}
\label{Community2}

\end{figure}

	Figure \ref{Community1} and \ref{Community2} shows the results under different subsampling schemes.  Since $K$ is discrete, instead of considering coverage rate, we compare methods by the proportion of correctly estimated $K$.   
	It is clear that whenever the community estimation problem itself is hard (the lowest values of $\rho$,  $t$, and $n$), the number of communities is also hard to estimate.   In these hardest cases, $K$ tends to be severely underestimated by node and row sampling, and varies highly for node pair sampling, and there is no good choice of $q$.  In all other scenarios, our proposed algorithm does choose a good value of $q$ balancing the bias and variability of $\hat{K}$ by maintaining the empirical coverage rate above 90\% and provide a set of candidate rank estimates.  

%
%
	
	\subsection{Regression with node cohesion}
	In many network applications, we often observe covariates associated with nodes. For example, in a school friendship network survey, one may also have the students' demographic information, grades, family background, and so on \cite{addhealth_2018}.   In such settings, the network itself is often used as an aid in answering the main question of interest, rather than the primary object of analysis.  For instance,  a network cohesion penalty was proposed as a tool to improve linear regression when observations are connected by a network \cite{li_prediction_2019}.  They consider the predictive model
        $$Y=\mathbf{\alpha}+ X\beta+\epsilon,$$ where  $X\in \mathbb{R}^{n\times p}$, $Y\in \mathbb{R}^n$, and $\beta \in \mathbb{R}^p$ are the usual matrix of predictors, response vector, and regression coefficients, respectively, and $\mathbf{\alpha}\in \mathbb{R}^n$ is the individual effect for each node associated with the network.  They proposed fitting this model by minimizing  the objective function
	\begin{equation}
	    \|Y-\alpha-X\beta\|_2+\lambda_1\alpha^T L\alpha,
	    \label{eqn:rncreg}
	\end{equation}
        where $\lambda_1$ is a tuning parameter, and $L = D-A$ is the Laplacian of the network, with $D = \text{diag}(d_1,\dots,d_n)$ the degree matrix.  The Laplacian-based network cohesion penalty has the effect of shrinking individual effects of connected nodes towards each other, ensuring more similar behavior for neighbors, and improving prediction  \cite{li_prediction_2019}.

  This method provides a point estimate of the coefficients $\beta$, but no measure of uncertainty, which makes interpretation difficult.   Our subsampling methods can add a measure of uncertainty to point estimates $\hat{\beta}$. Since different subsampling methods will lead to different numbers of remaining nodes and thus different sample sizes for regression, we only replace the penalty term in \eqref{eqn:rncreg} with $\lambda_1\alpha^T \tilde{L}\alpha$, where $\tilde{L}$ is the corresponding Laplacian of the subsampled graph, and retain the full size $n$ sample of $Y$ and $X$ for regression, to isolate the uncertainly associated with the network penalty.   Alternatively, one could resample $(X,Y)$ pairs together with the network to assess uncertainty overall.  If we view the observed graph $A$ as a realization from a distribution $F$, then the subsampled graphs under different subsampling methods can also be considered as a smaller sample from the same distribution $F$ and the estimation based on these smaller samples should still be consistent. 
    
 In addition to the three subsampling methods, we also include a naive bootstrap method.   Let $A$ be the original adjacency matrix with $n$ nodes.   Sample $n$ nodes  with replacement, obtaining $n_1, n_2, \dots, n_n$, and create a new adjacency matrix $\tilde{A}$ by setting $\tilde{A}_{ij} = 1$ if $A_{n_i n_j}=1$ or $n_i = n_j$, otherwise, $\tilde{A}_{ij}=0$. Note that with this method we cannot isolate the network as a source of uncertainty, and thus the uncertainty assessment will include the usual bootstrap sample variability of regression coefficients.  
    
\begin{figure}[ht!]
\centering
\subfloat[$\rho = 0.2$, $t = 10$, $\sigma_{\alpha} = 0$]{
\includegraphics[width=0.3\textwidth]{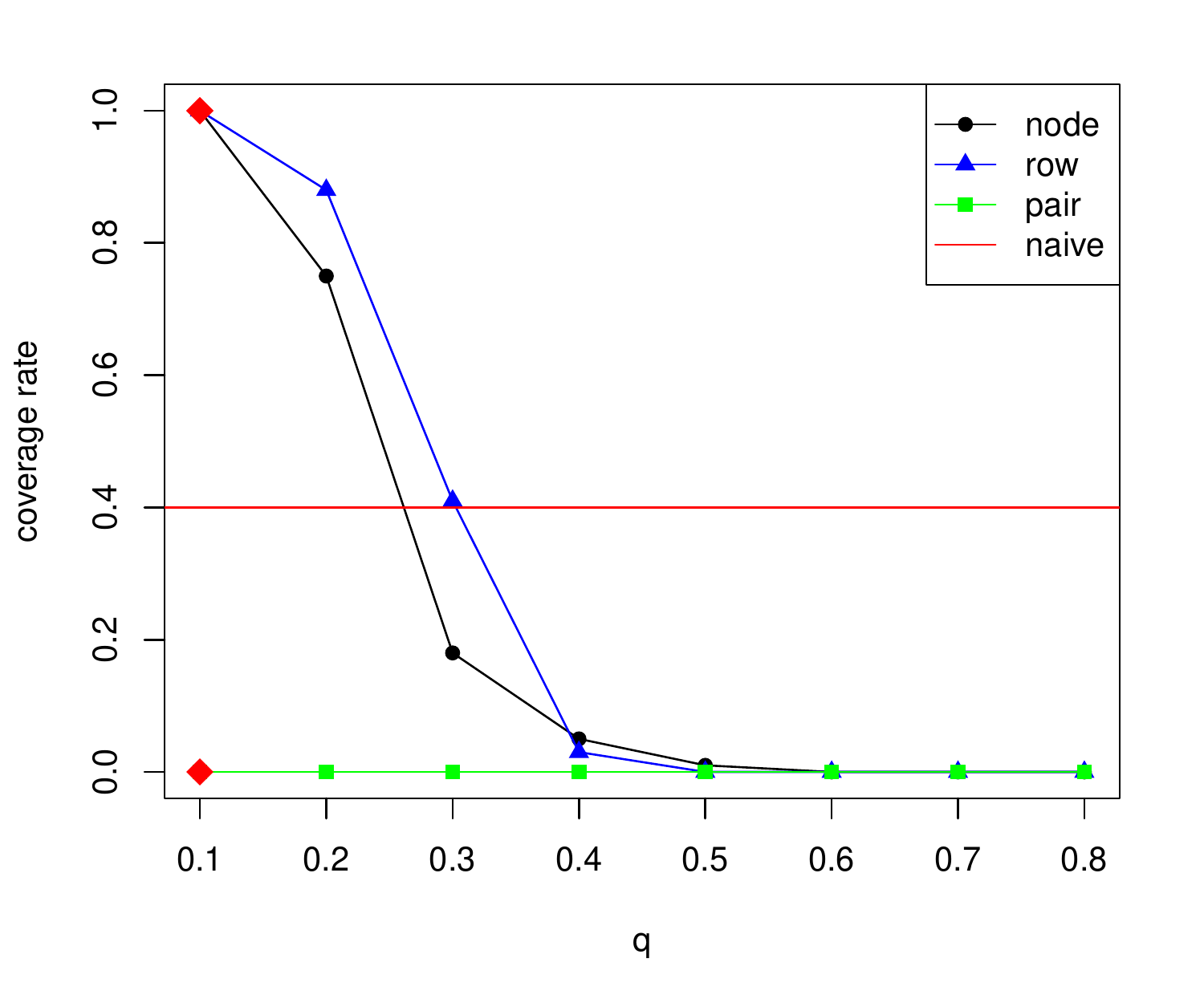}
\includegraphics[width=0.3\textwidth]{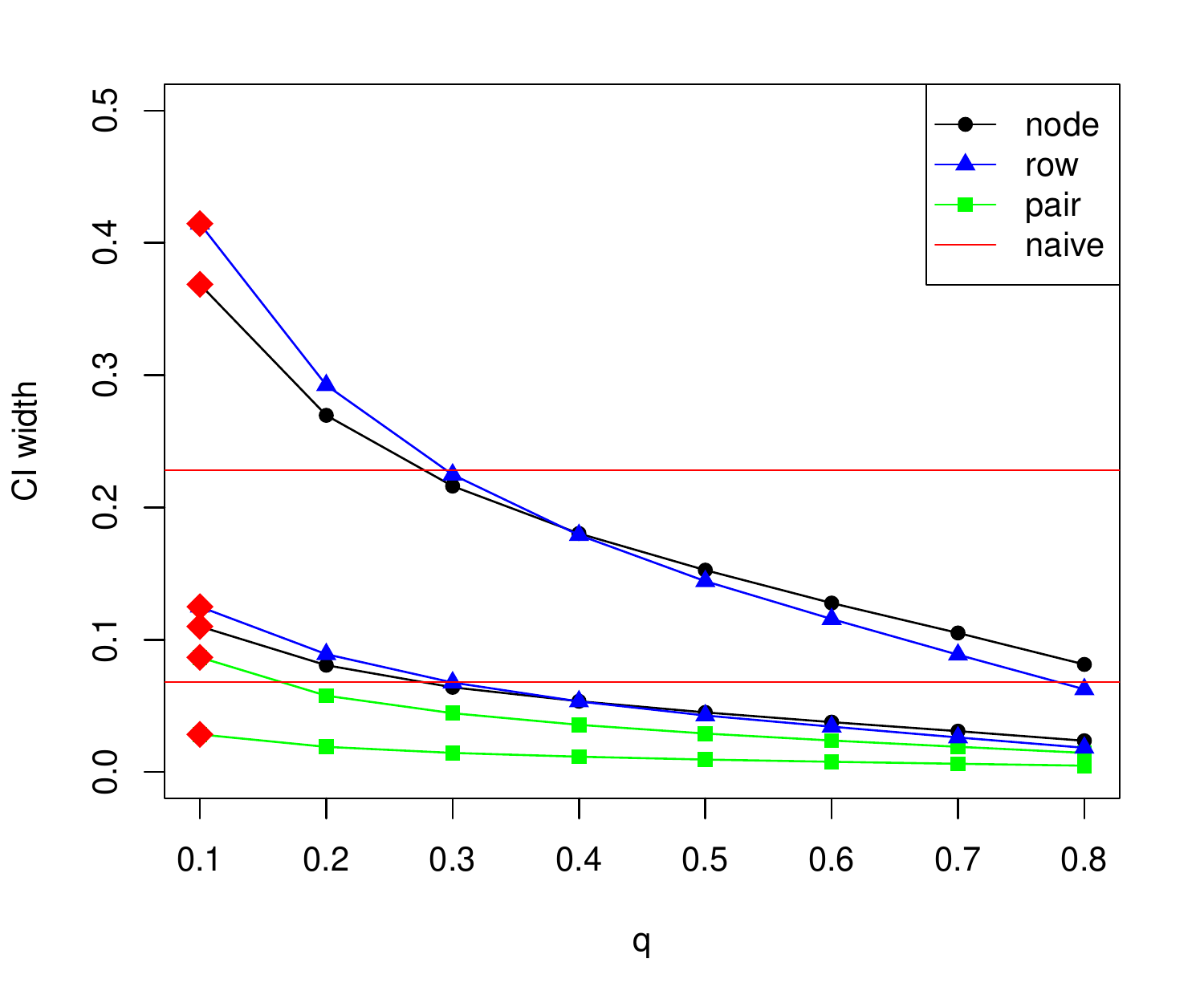}
\includegraphics[width=0.3\textwidth]{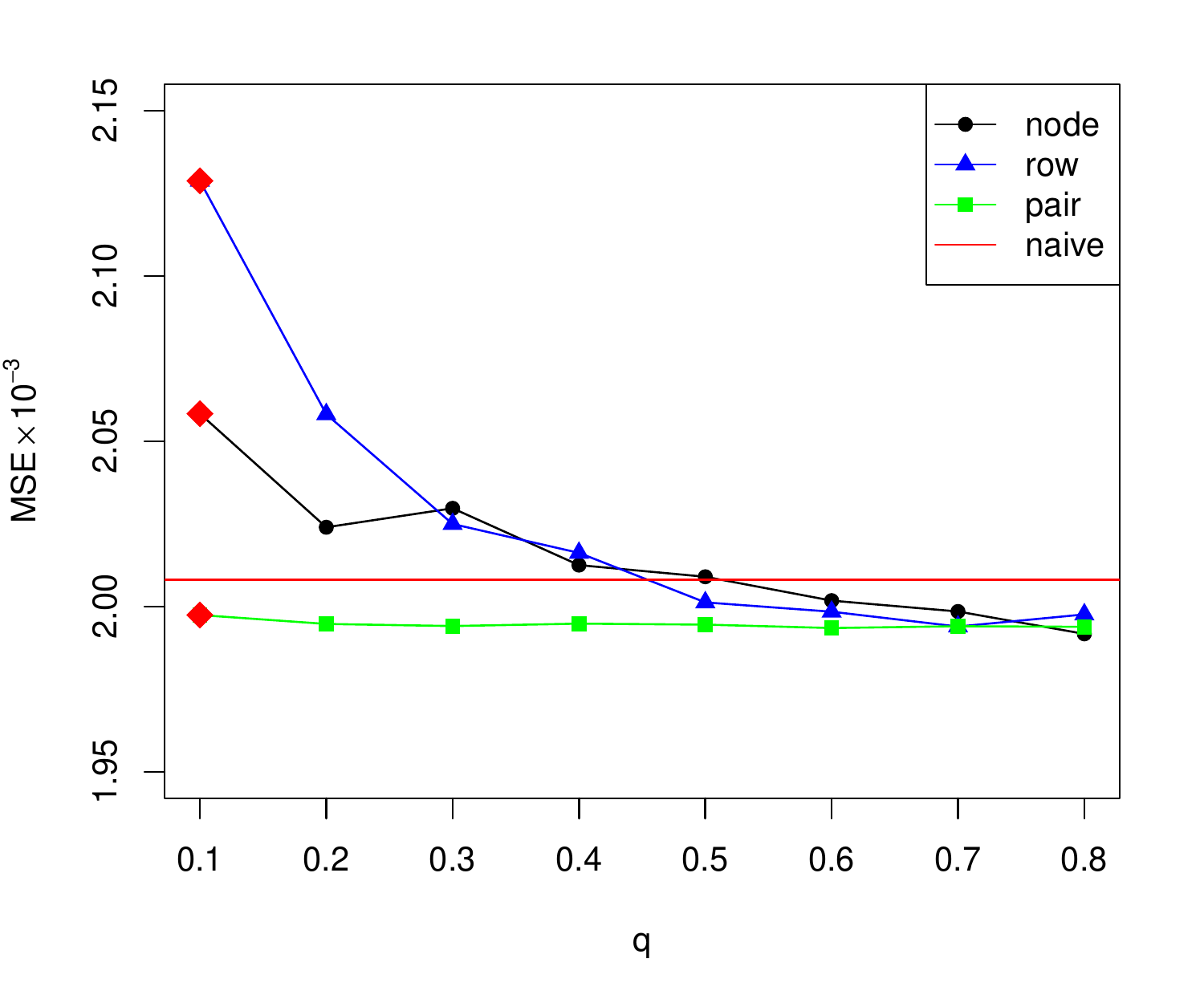}
}\\
\subfloat[$\rho = 0.2$, $t = 10$, $\sigma_{\alpha} = 0.1$]{
\includegraphics[width=0.3\textwidth]{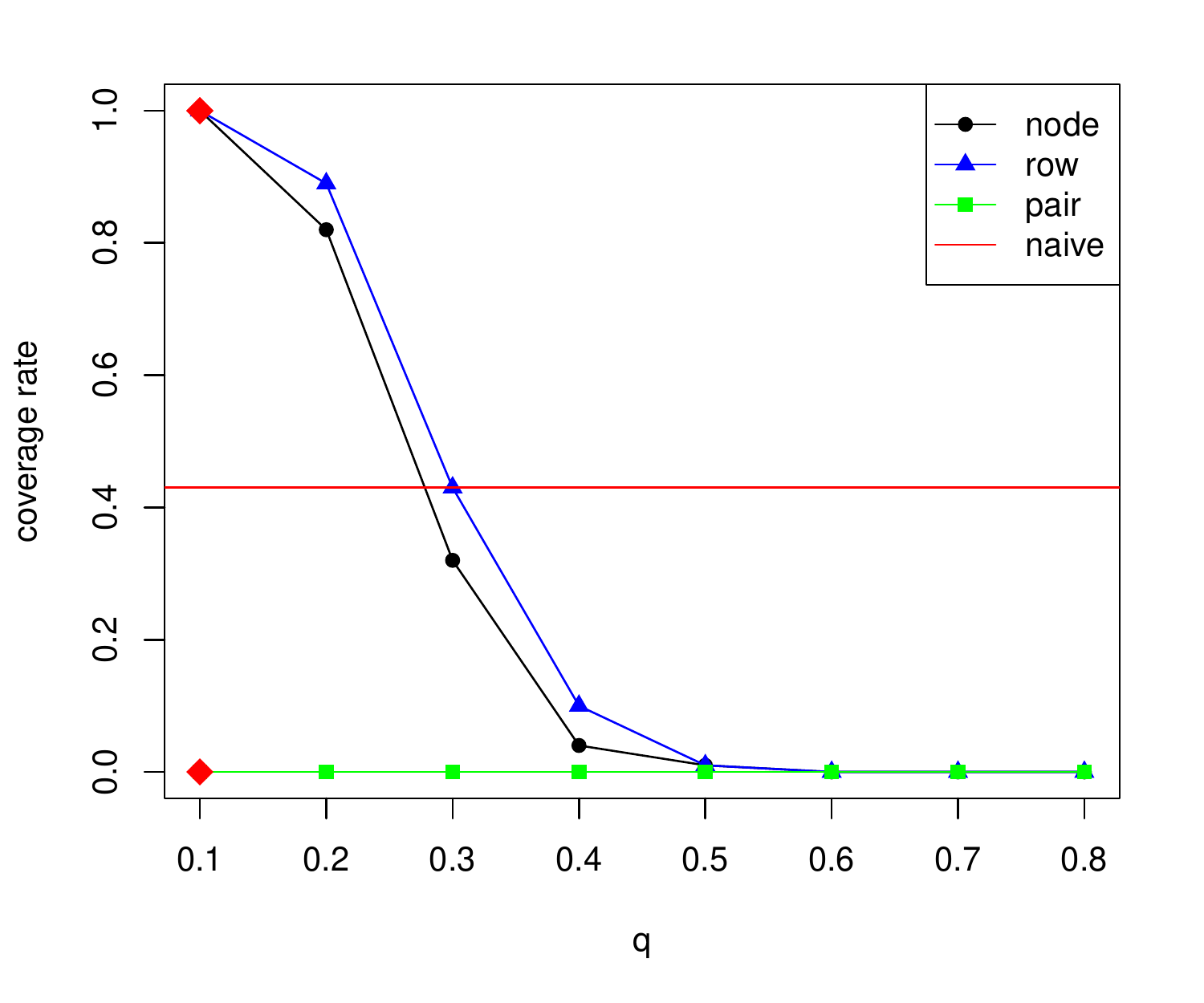}
\includegraphics[width=0.3\textwidth]{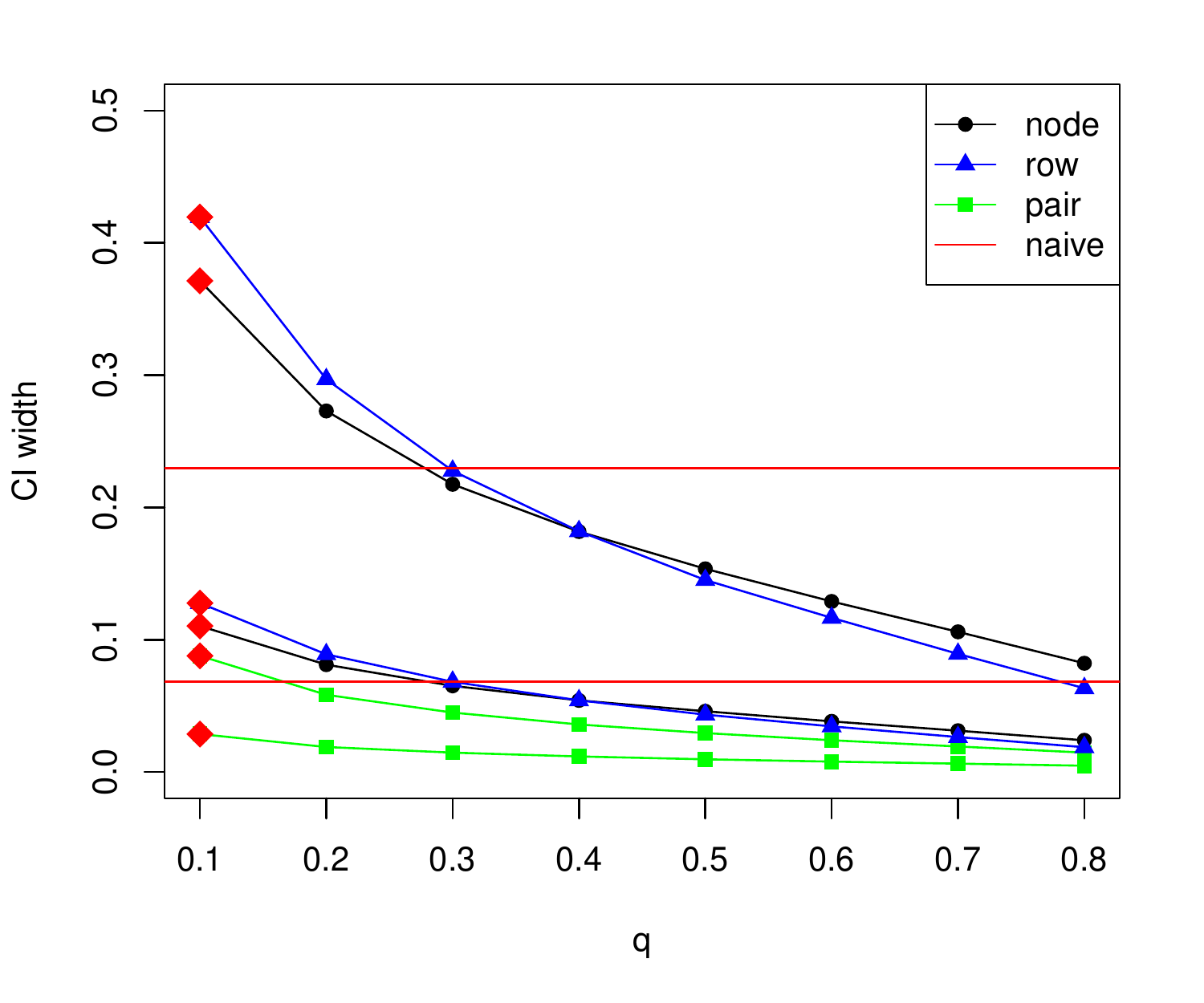}
\includegraphics[width=0.3\textwidth]{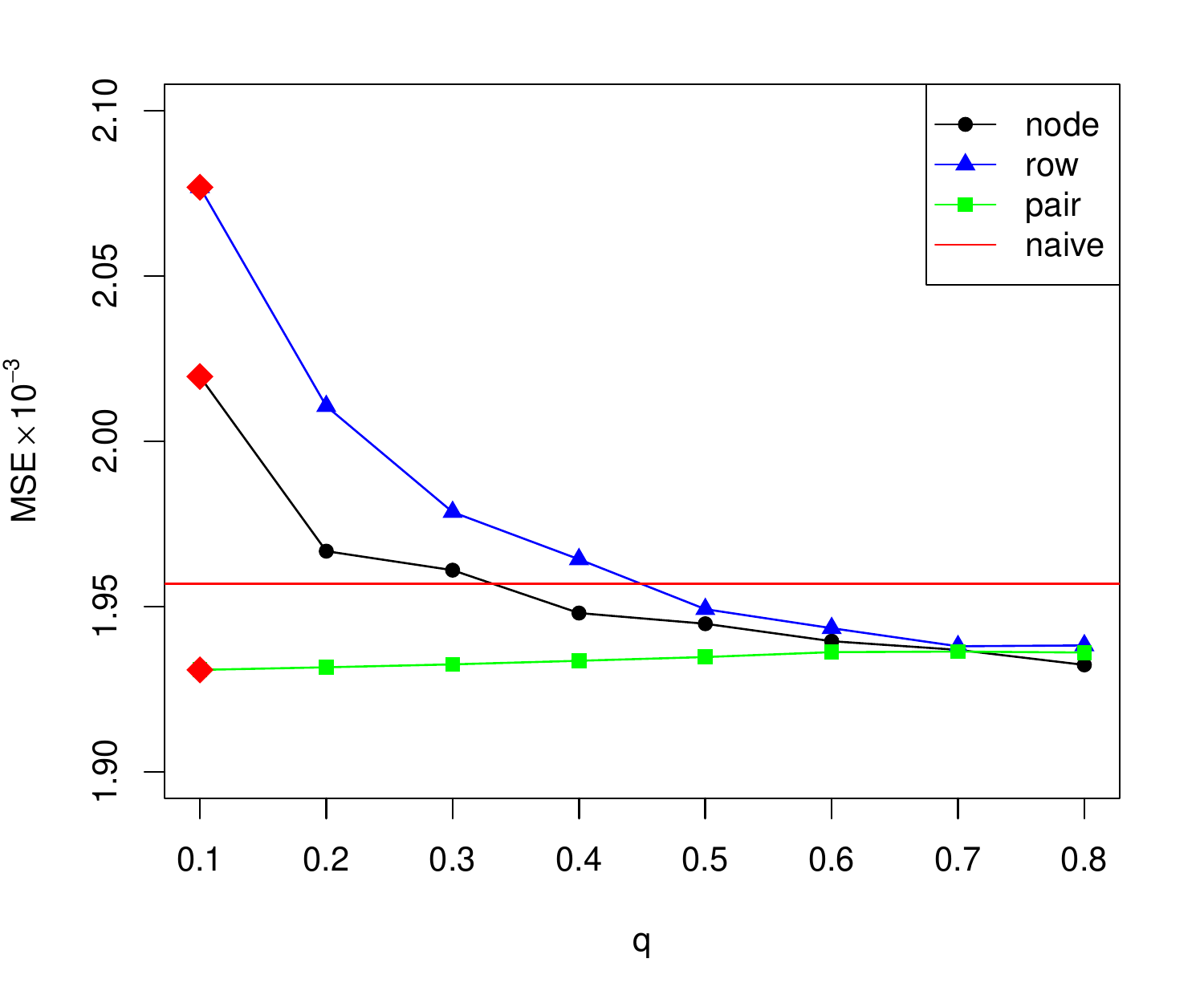}
}\\
\subfloat[$\rho = 0.2$, $t = 10$, $\sigma_{\alpha} = 0.5$]{
\includegraphics[width=0.3\textwidth]{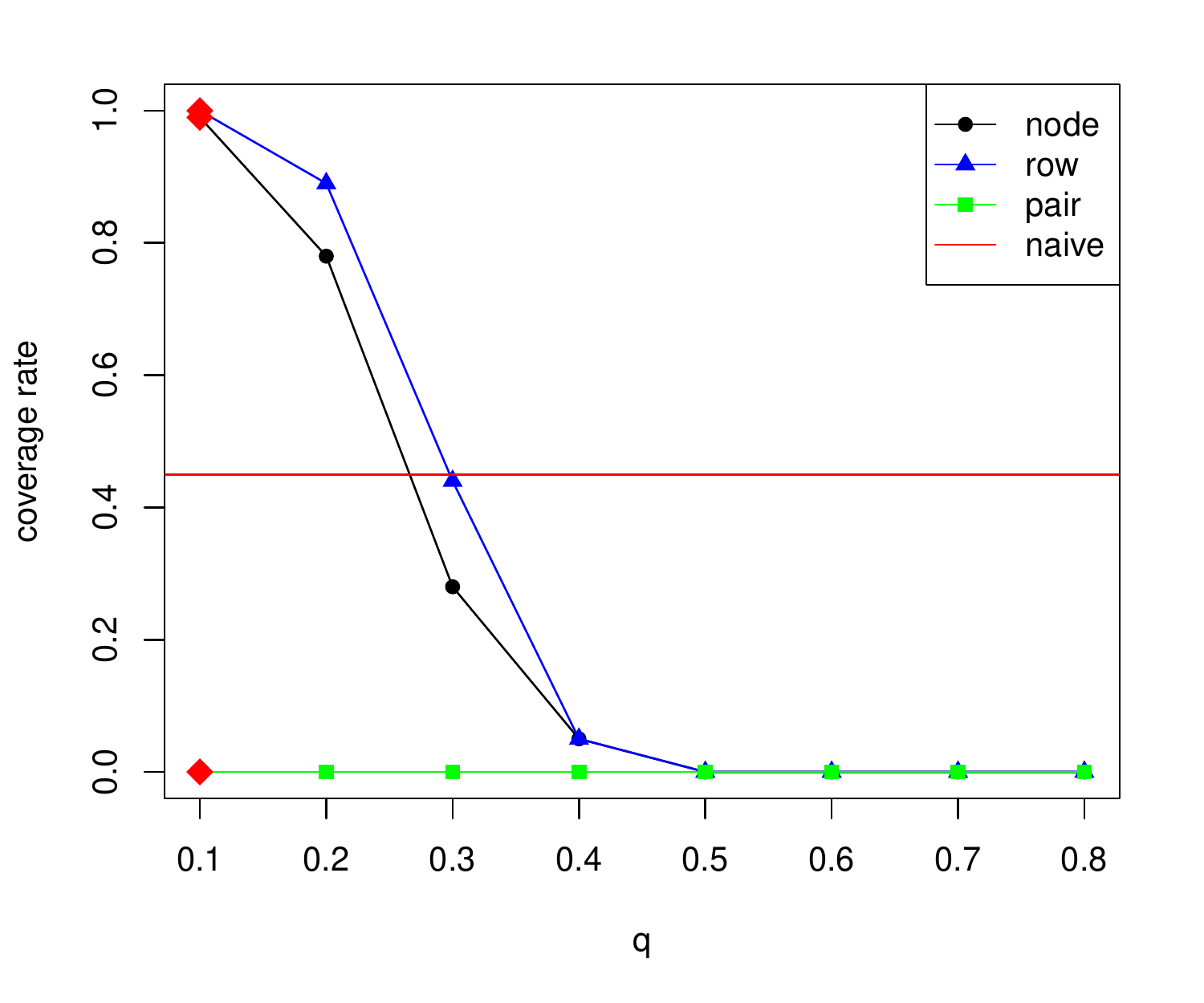}
\includegraphics[width=0.3\textwidth]{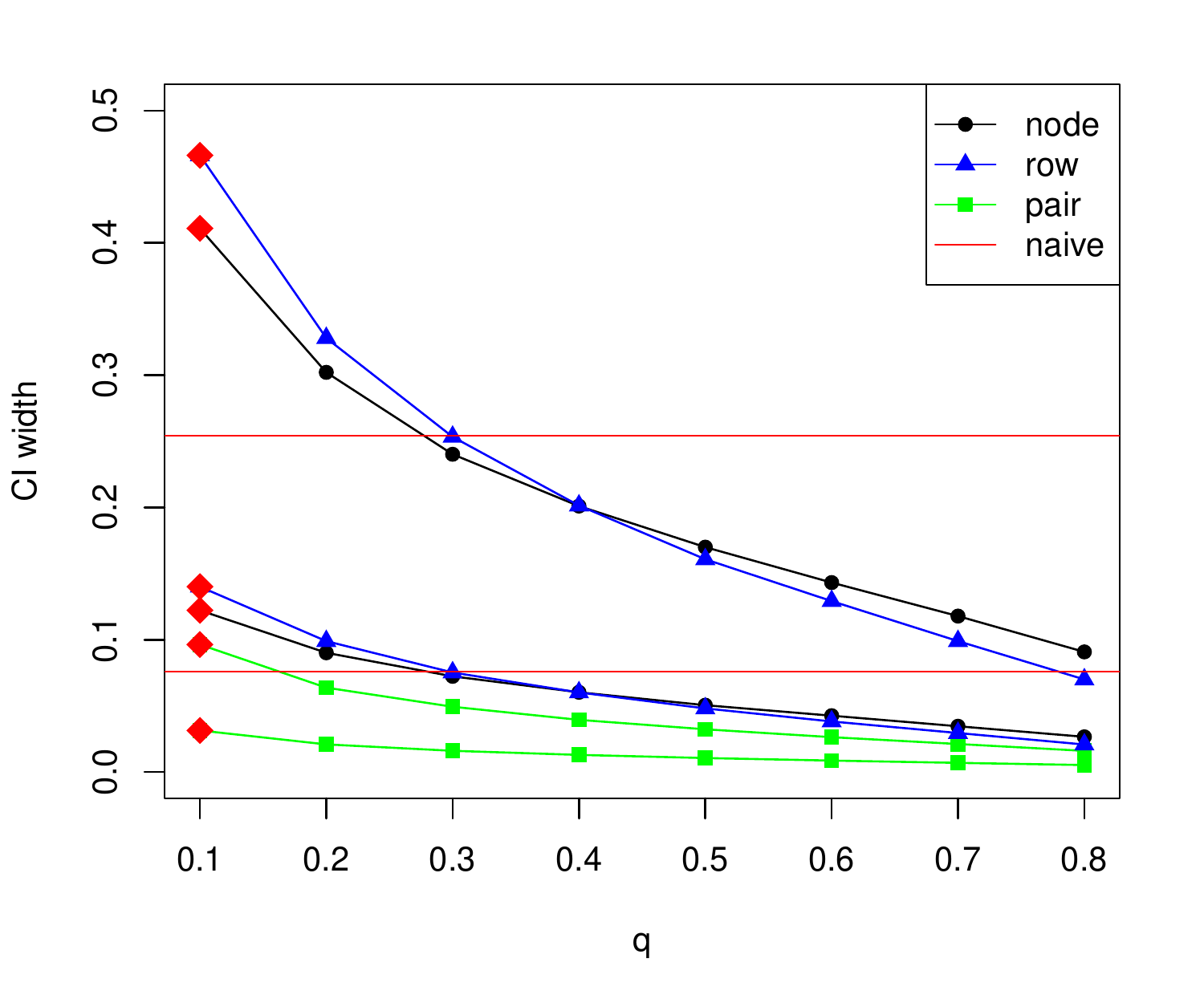}
\includegraphics[width=0.3\textwidth]{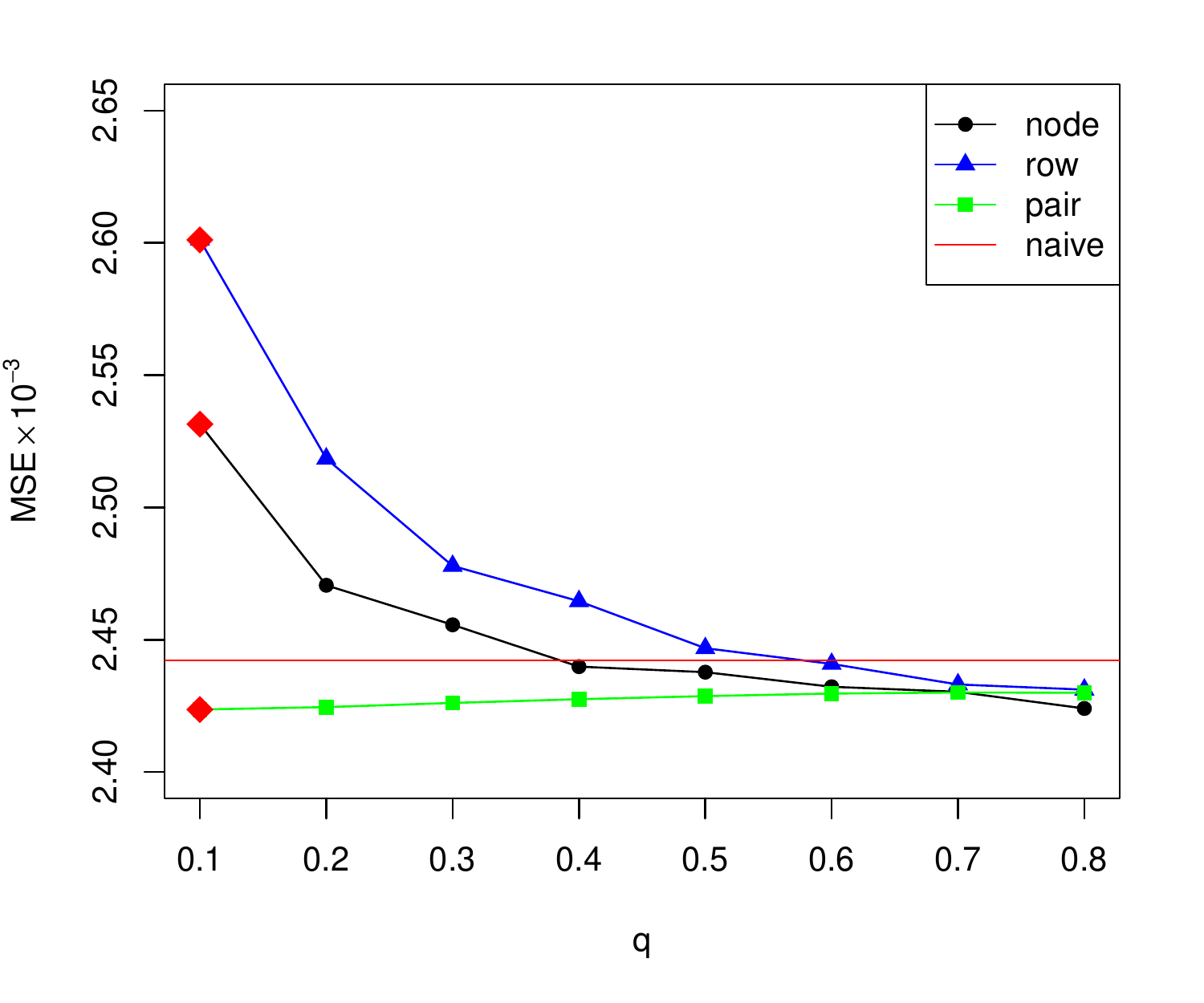}
}

\caption{Resampling performance as a function of $\sigma_\alpha$.   Coverage rate (left), widths of confidence intervals along the longest and shortest directions (middle), mean squared error of the mean of resampling estimates of $\beta$ (right). }
\label{RNC1}
\end{figure}

\begin{figure}[ht!]
\centering
\subfloat[$\rho = 0.2$, $t = 2$, $\sigma_{\alpha} = 0.1$]{
\includegraphics[width=0.3\textwidth]{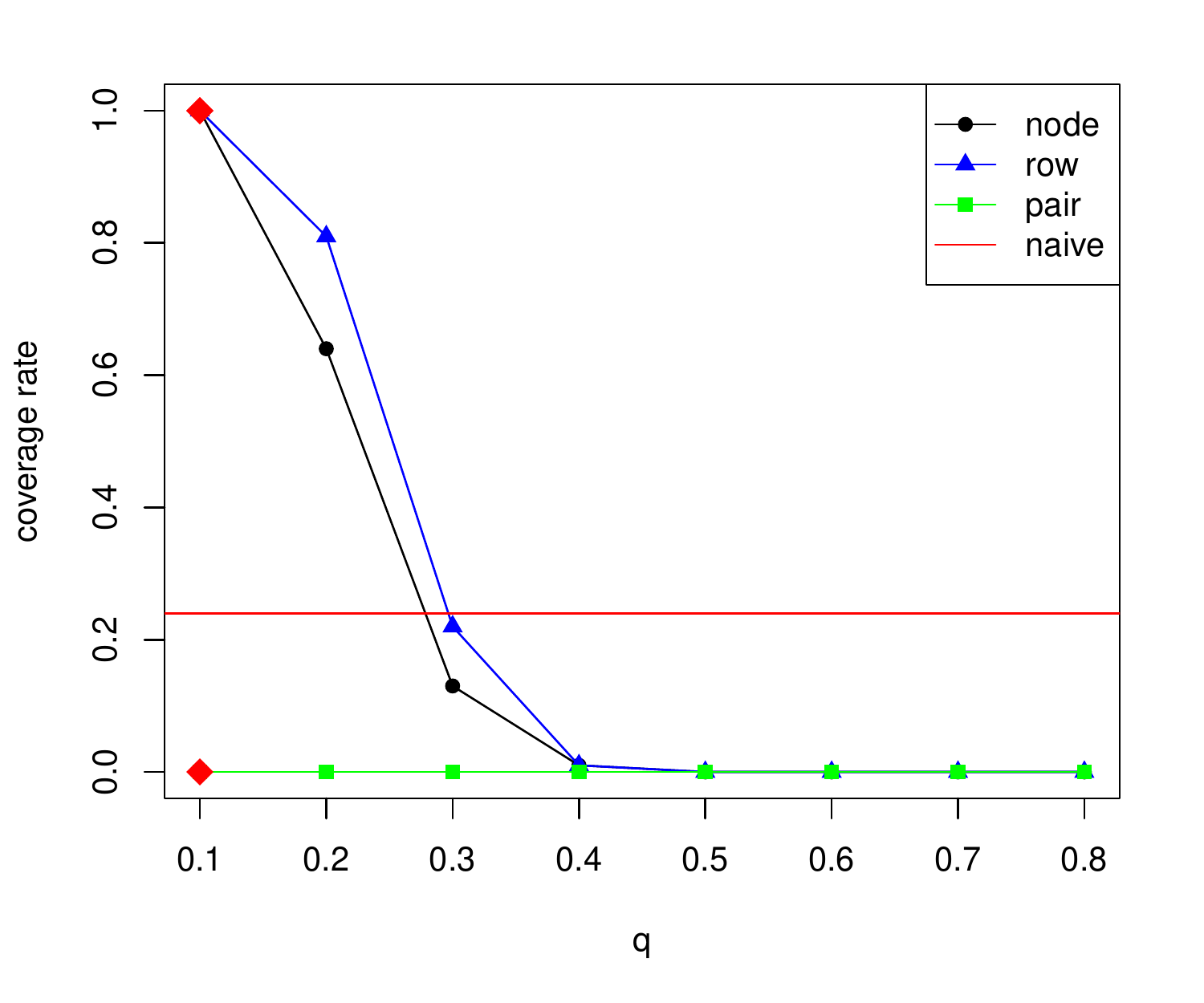}
\includegraphics[width=0.3\textwidth]{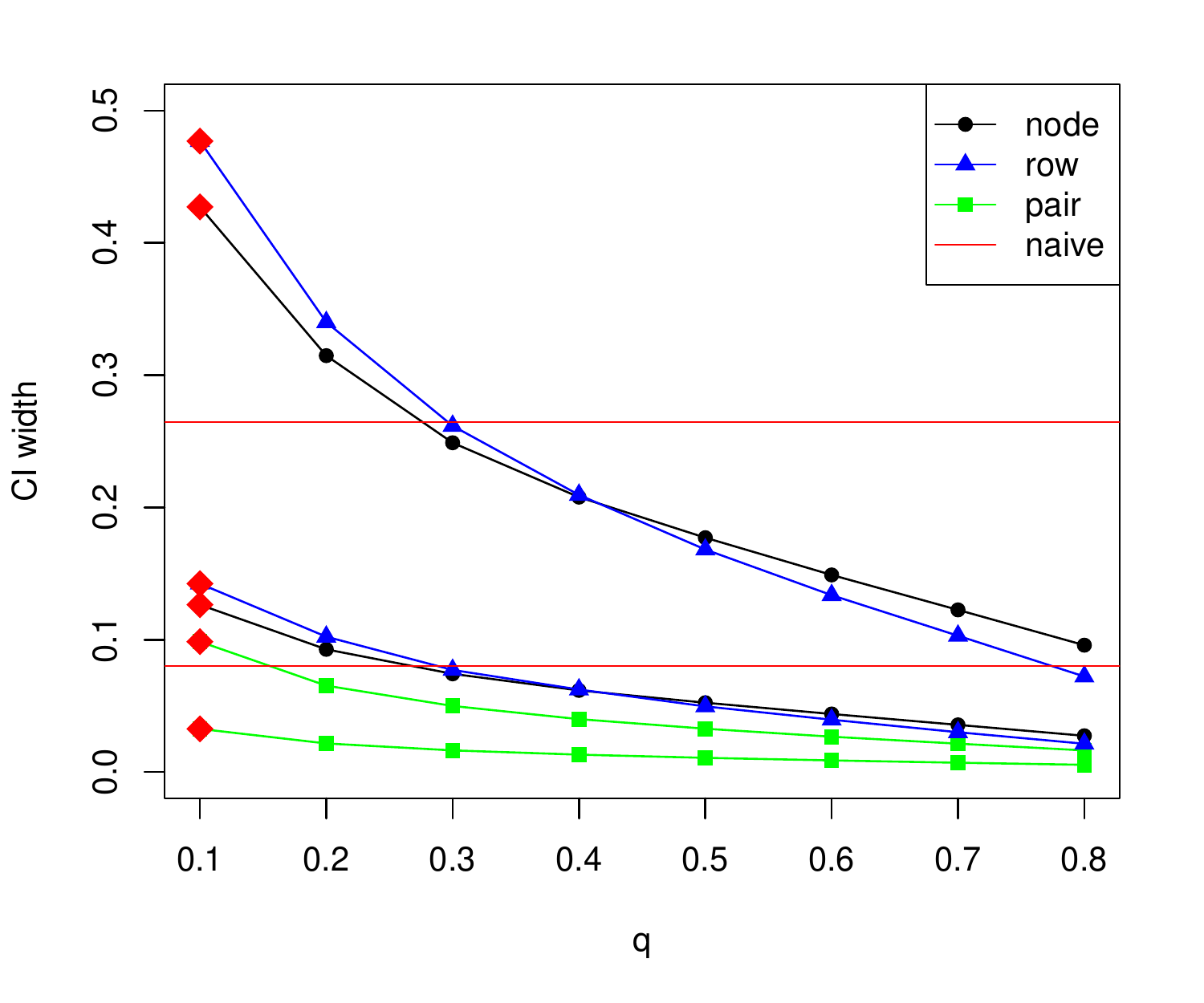}
\includegraphics[width=0.3\textwidth]{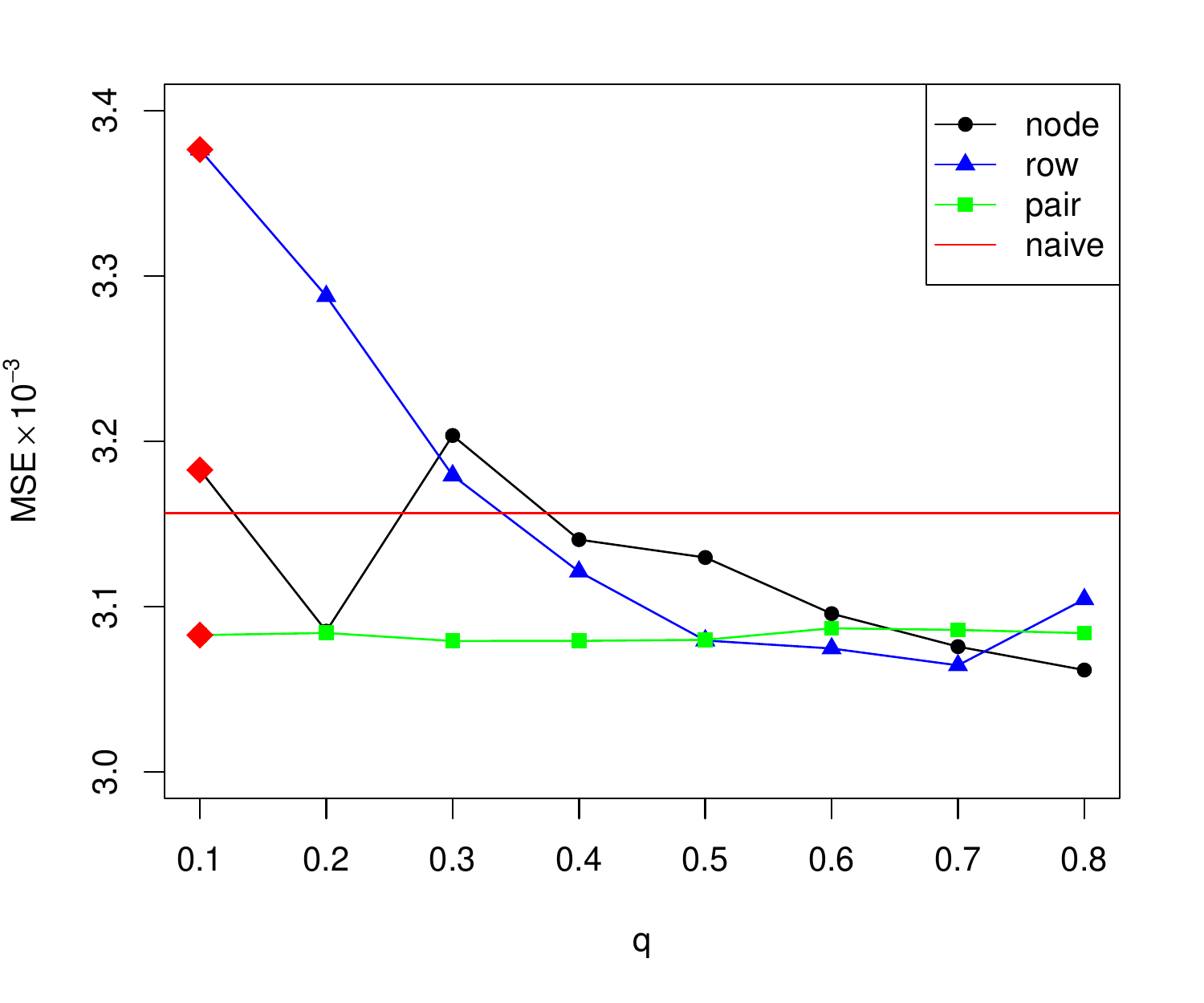}
}\\
\subfloat[$\rho = 0.2$, $t = 5$, $\sigma_{\alpha} = 0.1$]{
\includegraphics[width=0.3\textwidth]{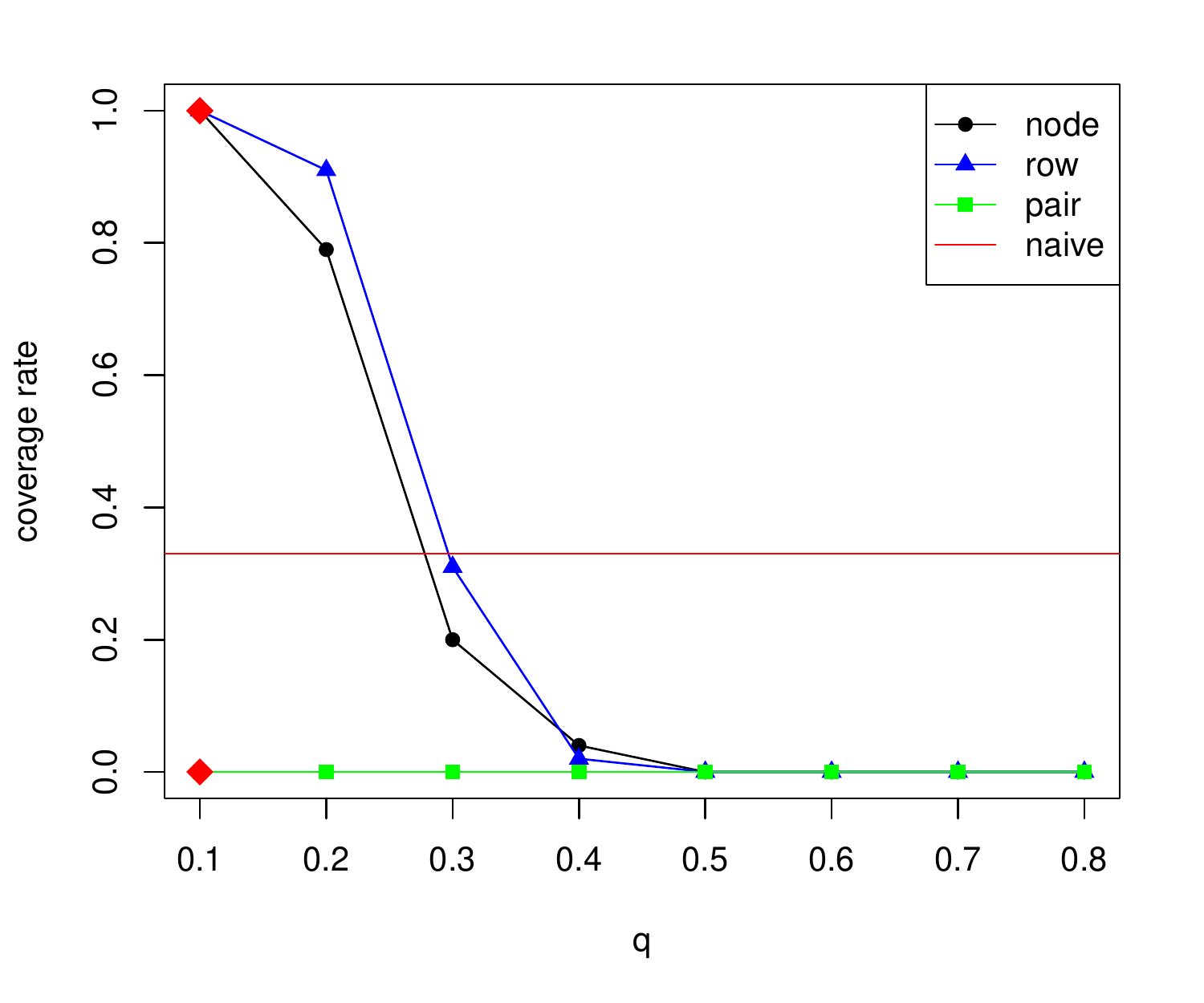}
\includegraphics[width=0.3\textwidth]{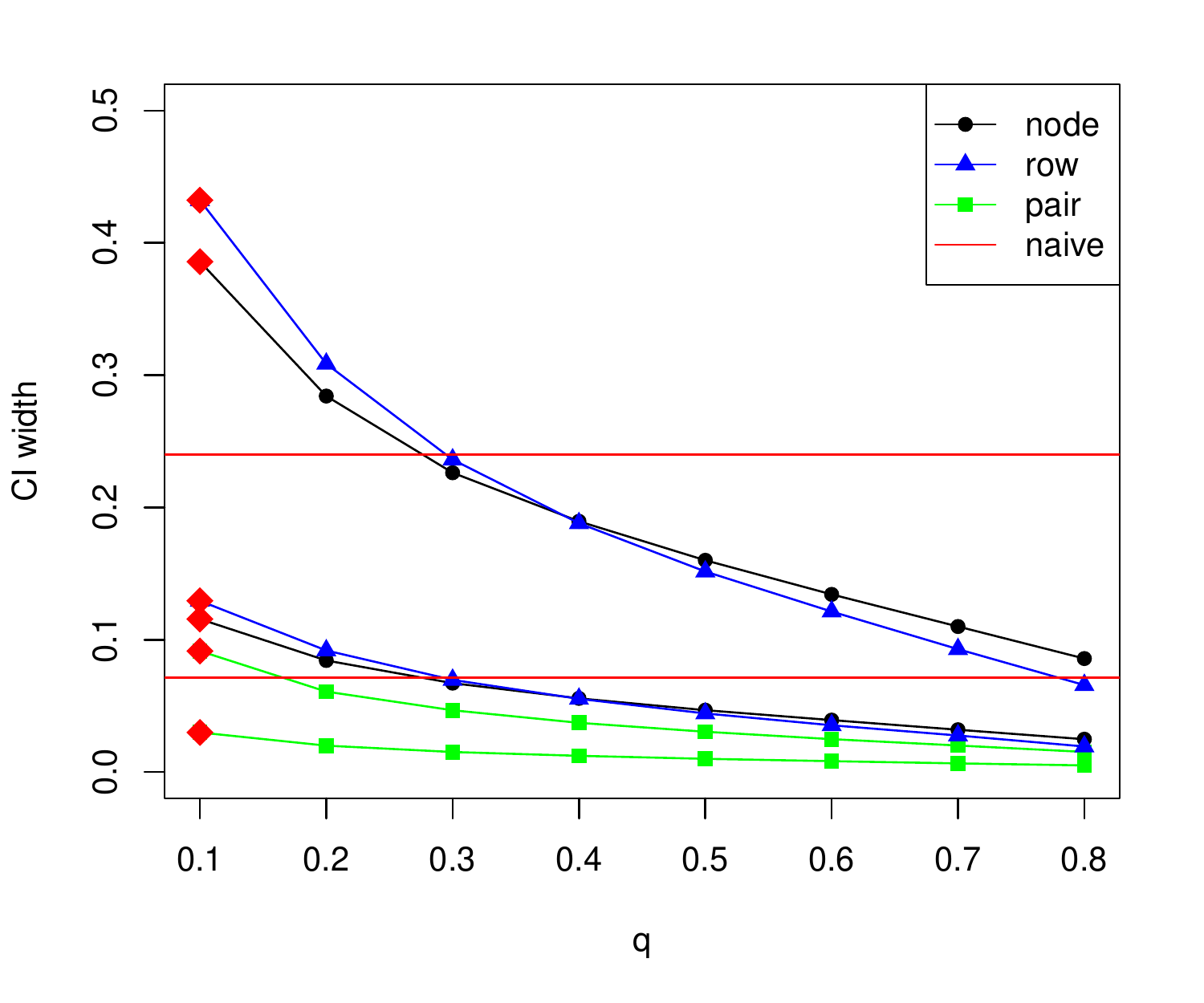}
\includegraphics[width=0.3\textwidth]{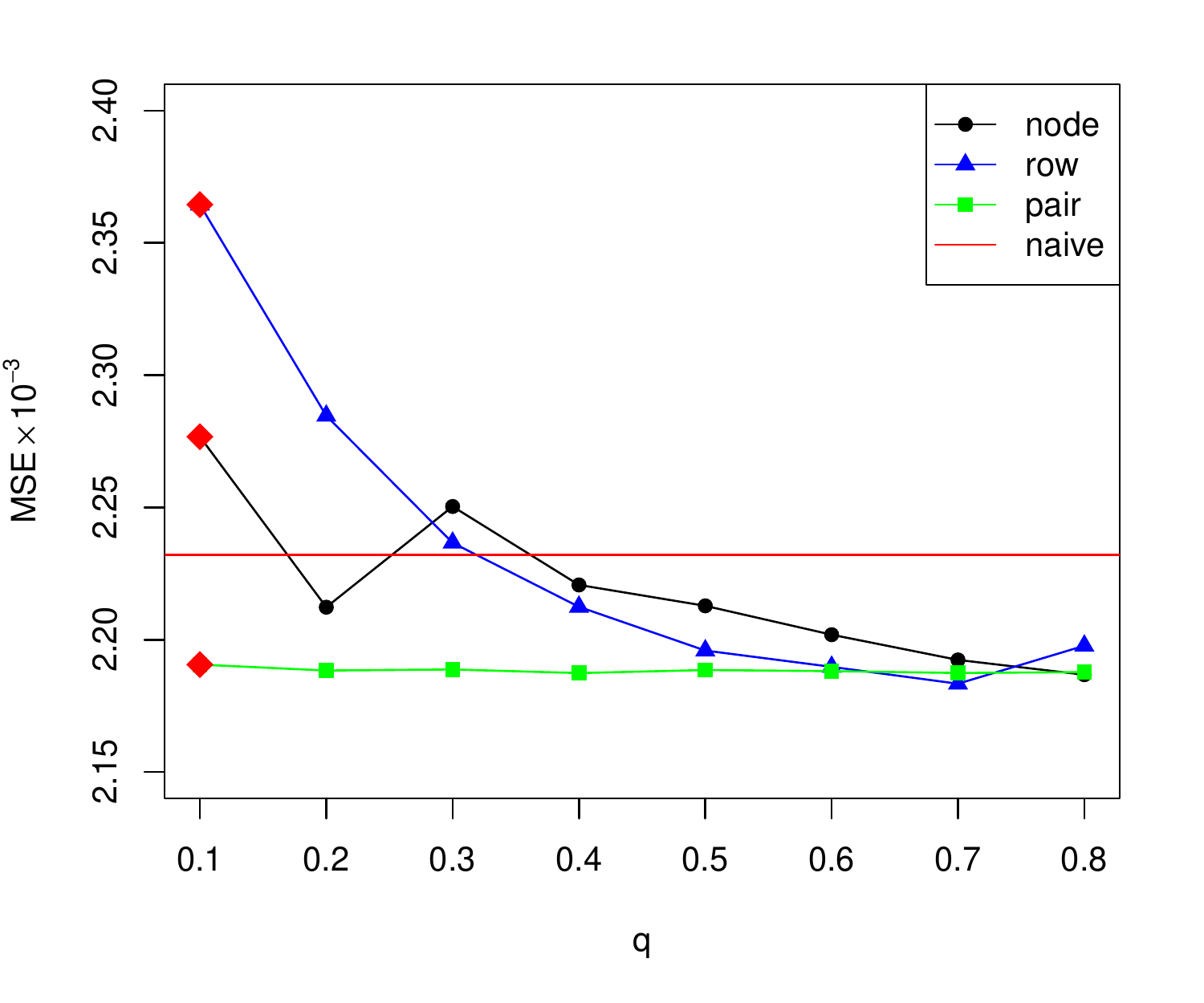}
}\\
\subfloat[$\rho = 0.2$, $t = 10$, $\sigma_{\alpha} = 0.1$]{
\includegraphics[width=0.3\textwidth]{plot/rnc/rho20/ratio10/sd1/coverage.pdf}
\includegraphics[width=0.3\textwidth]{plot/rnc/rho20/ratio10/sd1/CIwidth.pdf}
\includegraphics[width=0.3\textwidth]{plot/rnc/rho20/ratio10/sd1/mse.pdf}
}

 \caption{Resampling performance as a function of $t$. From left to right: coverage rate, widths of the confidence interval along the longest and shortest direction, mean squared error of the mean of resampling estimates of $\beta$.}
	\label{RNC2}
\end{figure}

\begin{figure}[ht!]
\centering
\subfloat[$\rho = 0.05$, $t = 10$, $\sigma_{\alpha} = 0.1$]{
\includegraphics[width=0.3\textwidth]{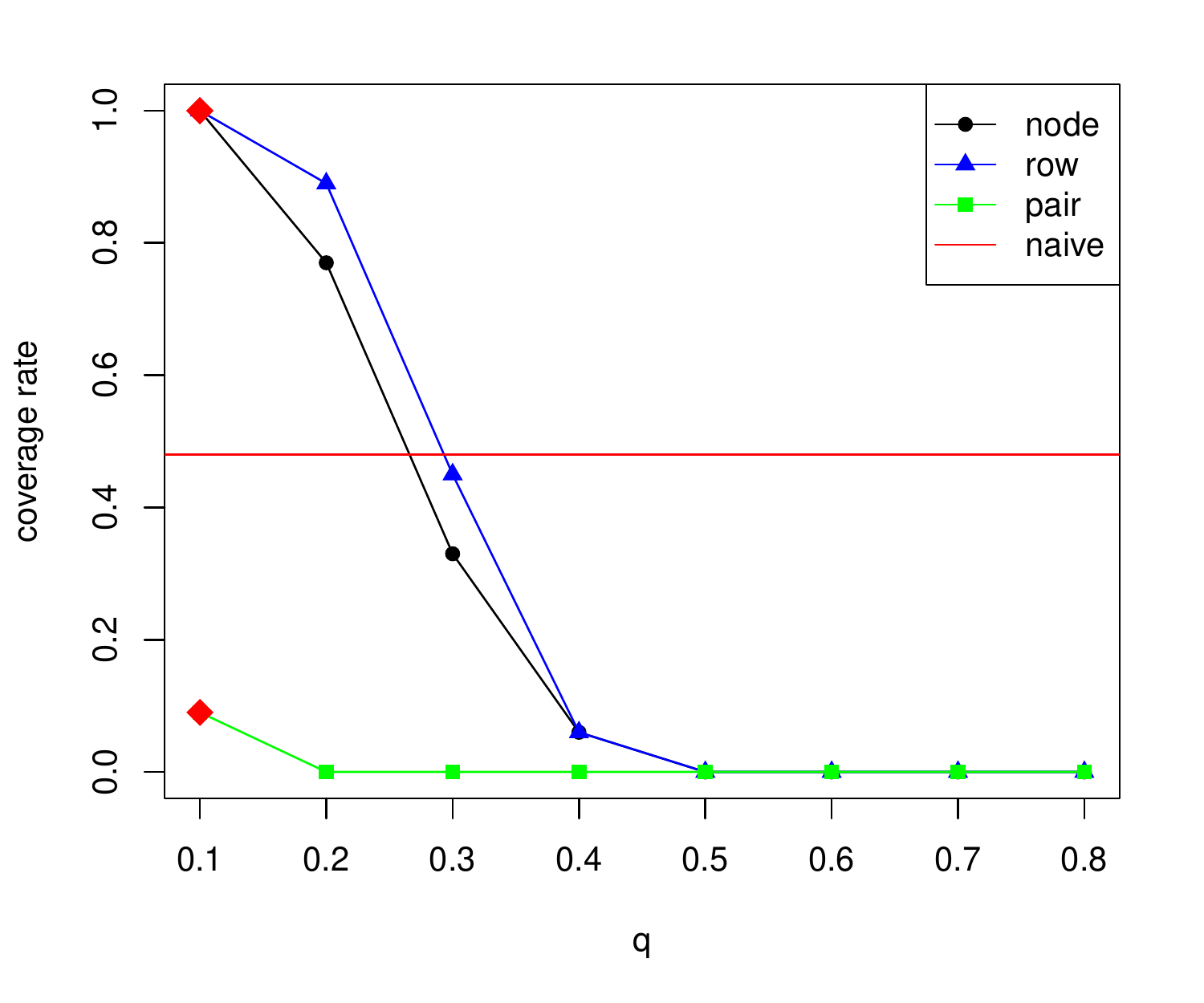}
\includegraphics[width=0.3\textwidth]{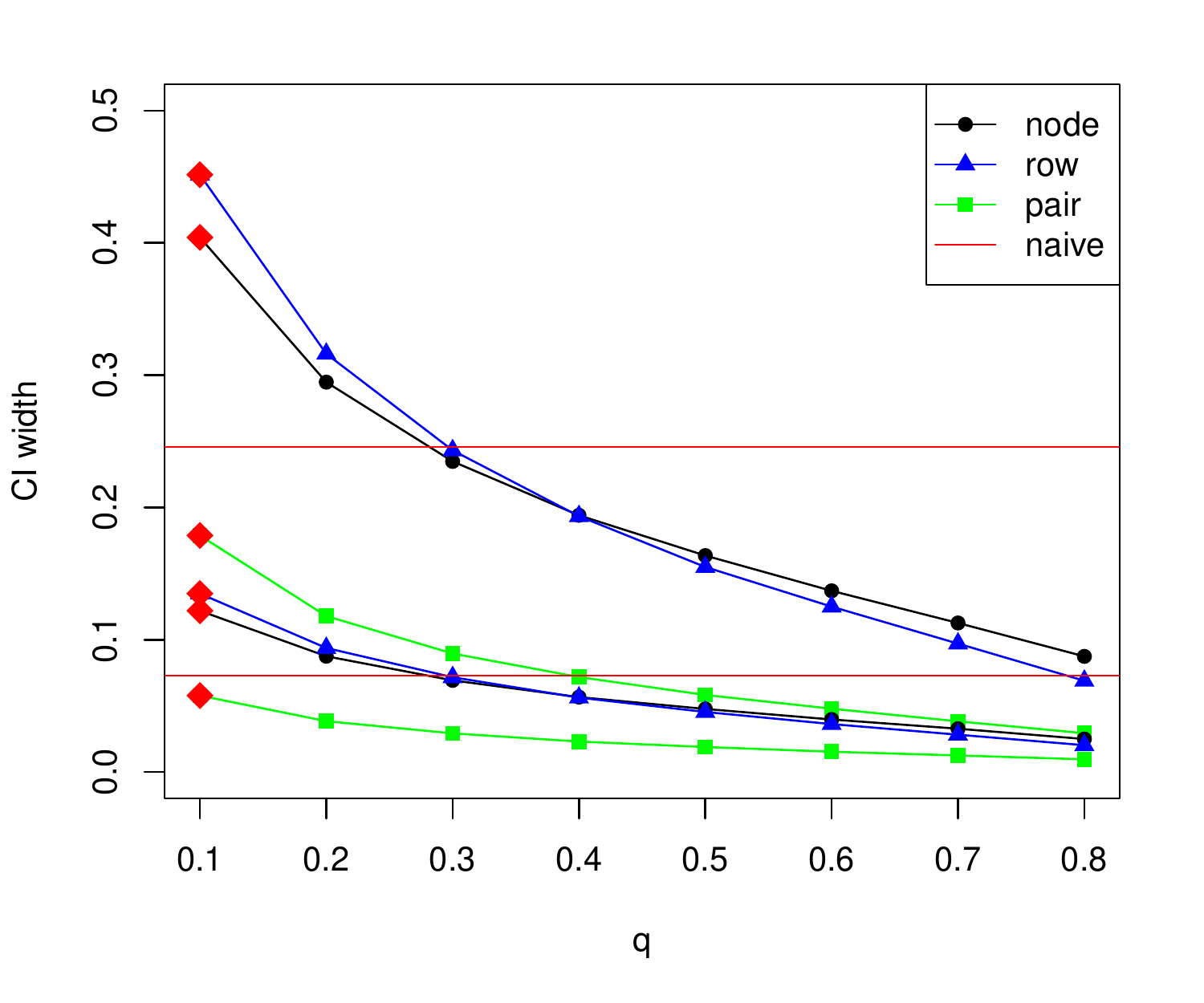}
\includegraphics[width=0.3\textwidth]{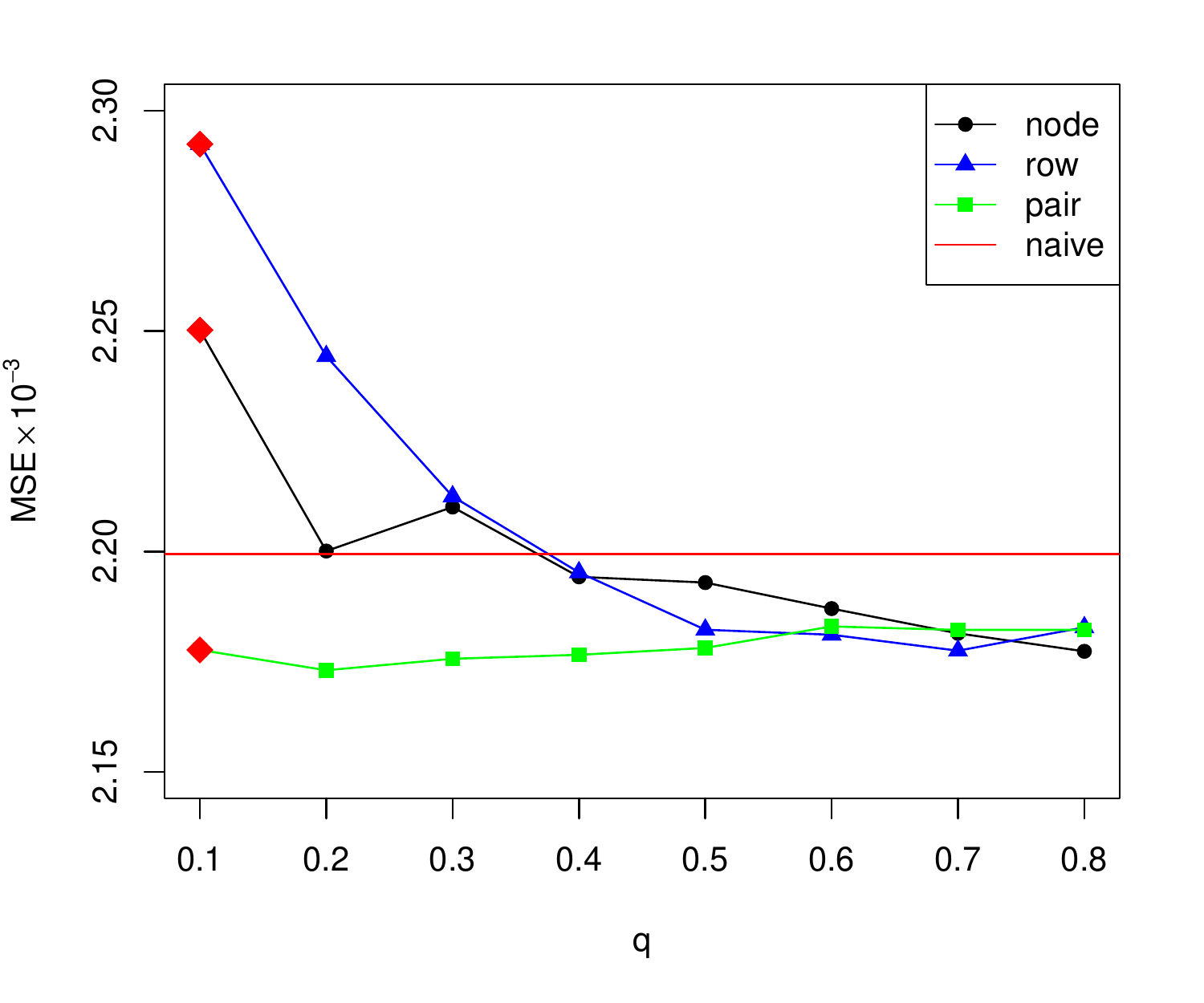}
}\\
\subfloat[$\rho = 0.1$, $t = 10$, $\sigma_{\alpha} = 0.1$]{
\includegraphics[width=0.3\textwidth]{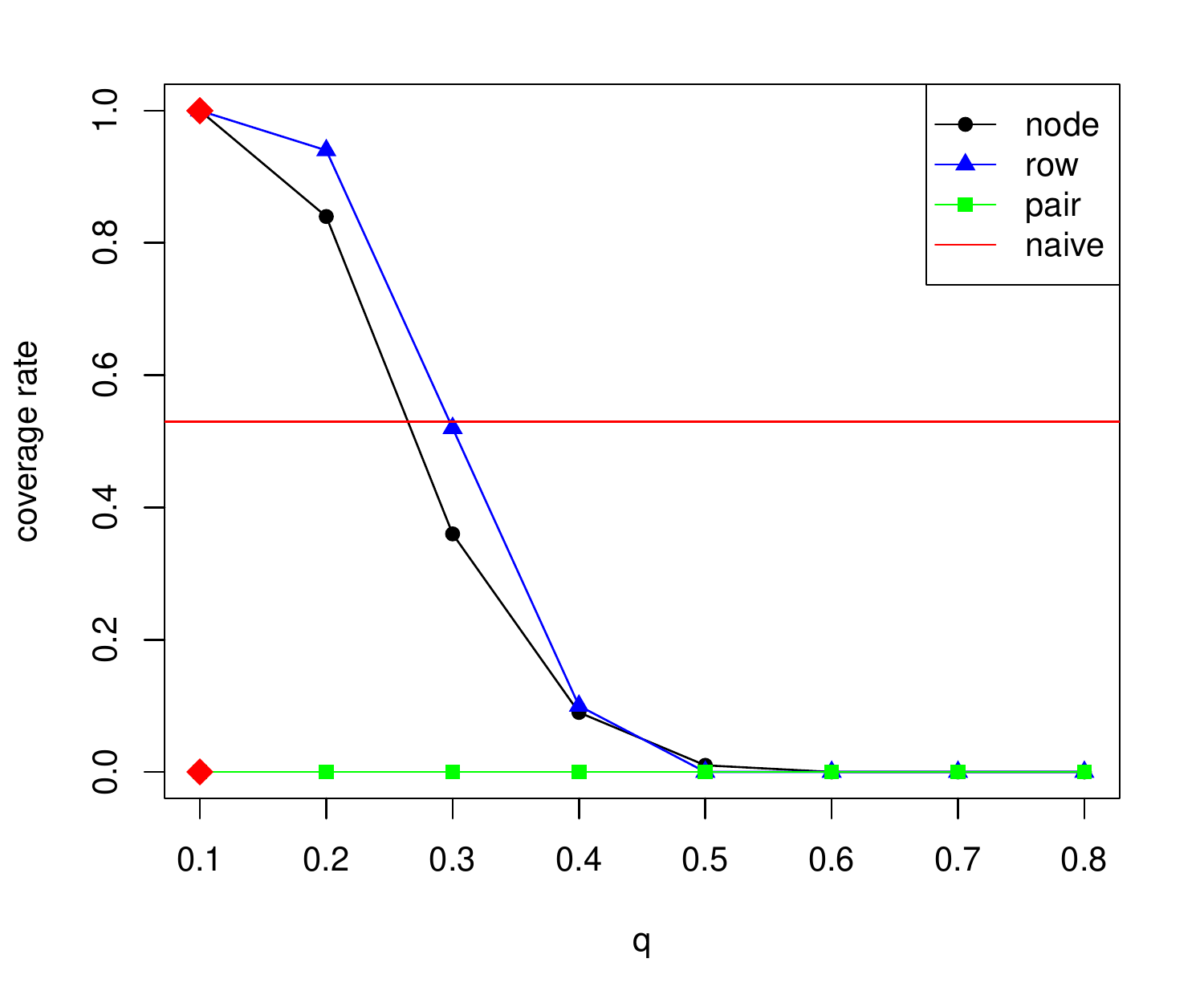}
\includegraphics[width=0.3\textwidth]{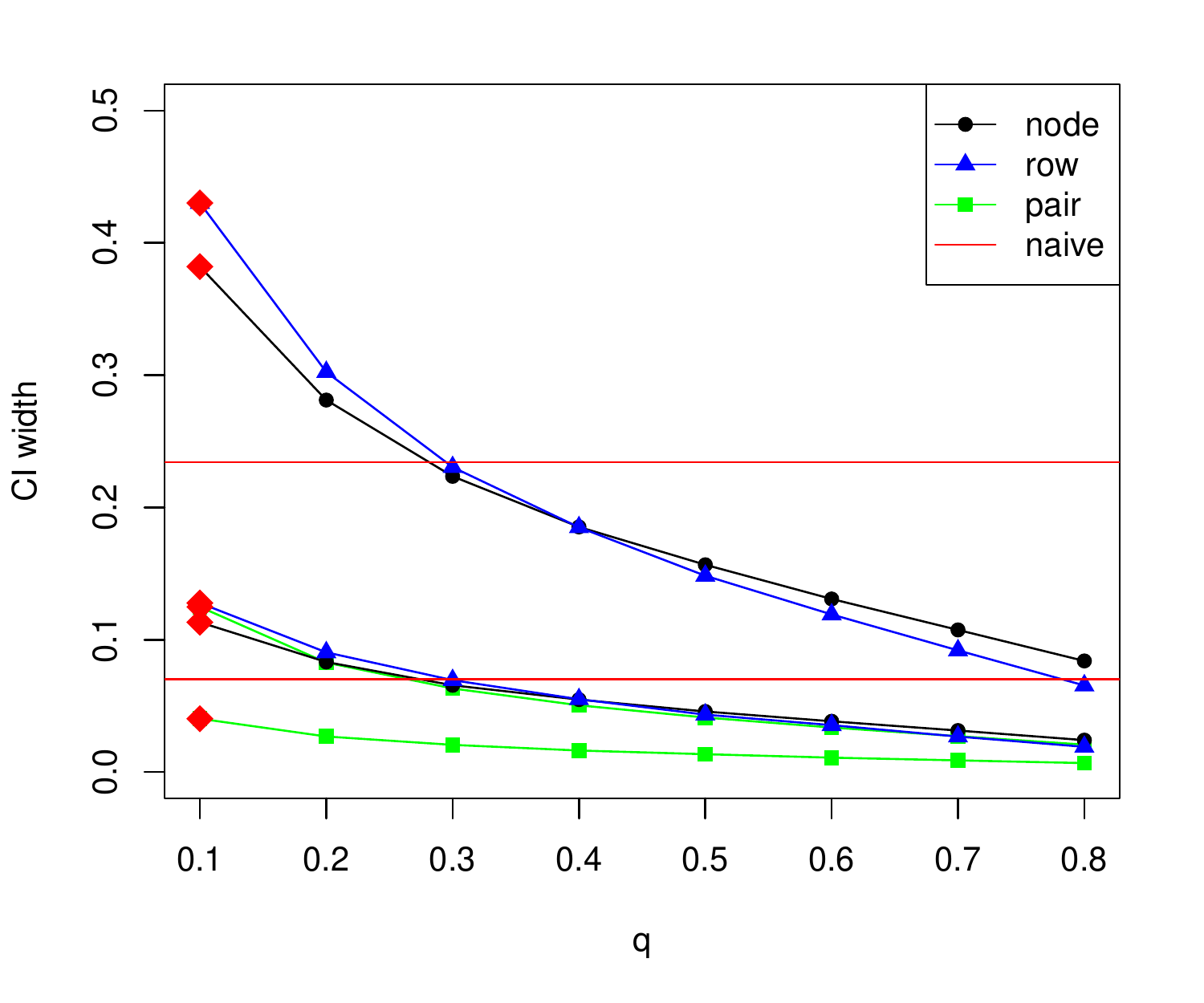}
\includegraphics[width=0.3\textwidth]{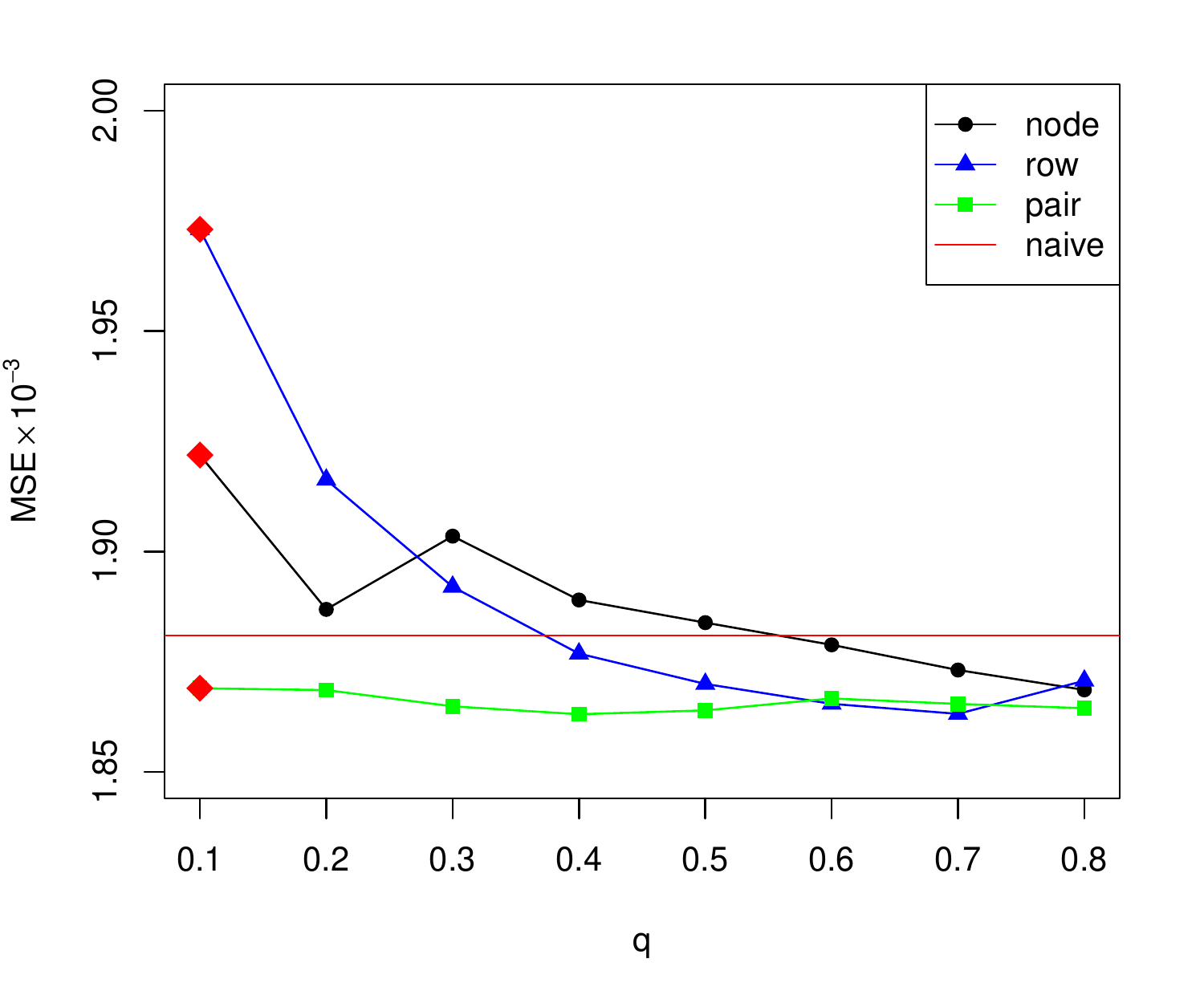}
}\\
\subfloat[$\rho = 0.2$, $t = 10$, $\sigma_{\alpha} = 0.1$]{
\includegraphics[width=0.3\textwidth]{plot/rnc/rho20/ratio10/sd1/coverage.pdf}
\includegraphics[width=0.3\textwidth]{plot/rnc/rho20/ratio10/sd1/CIwidth.pdf}
\includegraphics[width=0.3\textwidth]{plot/rnc/rho20/ratio10/sd1/mse.pdf}
}

 \caption{Resampling performance as a function of $\rho$.  From left to right: coverage rate, widths of the confidence interval along the longest and shortest direction, mean squared error of the mean of resampling estimates of $\beta$.}
	\label{RNC3}
\end{figure}

 We generate the adjacency matrix $A$ as an SBM with with $K=3$ communities, with 200 nodes each. The individual effect $\alpha$ is generated from the normal distribution $N(c_k, \sigma_{\alpha}^2)$, where $c_k$ are $-1$, $0$, and $1$ for nodes from communities $1$, $2$, and $3$, respectively. We vary the ratio of within and between edge probabilities $t$ and the standard deviation $\sigma_\alpha$, to obtain different signal-to-noise ratios. Regression coefficients $\beta_j$'s are drawn from $N(1,1)$ independently, and each response $y_i$ is drawn from $N(\alpha_i + \beta^T x_i,1)$. Figures \ref{RNC1}, \ref{RNC2} and \ref{RNC3} show the results from  this simulation.  
    
Once again, as $q$ increases, the confidence intervals get narrower as subgraphs become more similar to the full network. However, for node and row sampling the coverage is also lower for the smallest value of $q=0.1$, since the variance of bootstrap estimates of $\beta$ is larger for smaller $q$, 
and the center of the constructed confidence interval gets further from the true $\beta$. This may be caused by the fact that only a small fraction of nodes contribute to the penalty term.    Node pair sampling is different: it gives better performance in terms of mean squared error for nearly all choices of $q$, but since its confidence intervals are much narrower than those of node and row sampling, their coverage rate is very poor.  
    All subsampling methods are anti-conservative and can only exceed the nominal coverage rate with small $q$. One possible explanation is that in a regression setting, the impact of node covariates dominates any potential impact of the network.   In all cases, Algorithm 2 chooses $q=0.1$, which is either the only choice of $q$ that exceeds the nominal coverage rate or the choice of $q$ that gives a coverage rate close to the nominal coverage rate.

%

    \subsection{Regression with node cohesion and variable selection}
    
    We now consider the case when the dimension of node covariates is larger. The LASSO in \cite{tibshirani_regression_1996} is a popular technique used for high dimensional i.i.d. data. By minimizing
    $$\|Y-X\beta\|^2+\lambda_2\|\beta\|_1,$$
    the resulted estimator $\hat{\beta}$ will have some entries shrunk to zero, and the set $S=\{i:\hat{\beta}_i\neq 0\}$ is known as the activated set. However, we need to balance the goodness-of-fit and complexity of the model properly by tuning $\lambda_2$, and model selection using cross validation is not always stable. In \cite{meinshausen_stability_2010}, the authors proposed stability selection, which uses subsampling to provide variable selection that is less sensitive to the choice of $\lambda$. 
    
    Here we propose a similar procedure for network data. With a graph $G$ observed, we first generate $B$ copies of $g_b$ using one of the  resampling schemes. Then we minimize the objective function 
    $$\|Y-\alpha-X\beta\|_2+\lambda_1\alpha^T L\alpha + \lambda_2\|\beta\|_1,$$
    over $g_b$ to construct activated set $S_b$, and for each $1\leq j \leq p$, calculate $\frac{1}{B}\sum \mathbf{1}\{j\in S_b\}$, the selection probability of each predictor.

\begin{figure}[ht!]
\centering
\subfloat[$\sigma_{\alpha}=0$]{\includegraphics[width=0.3\textwidth]{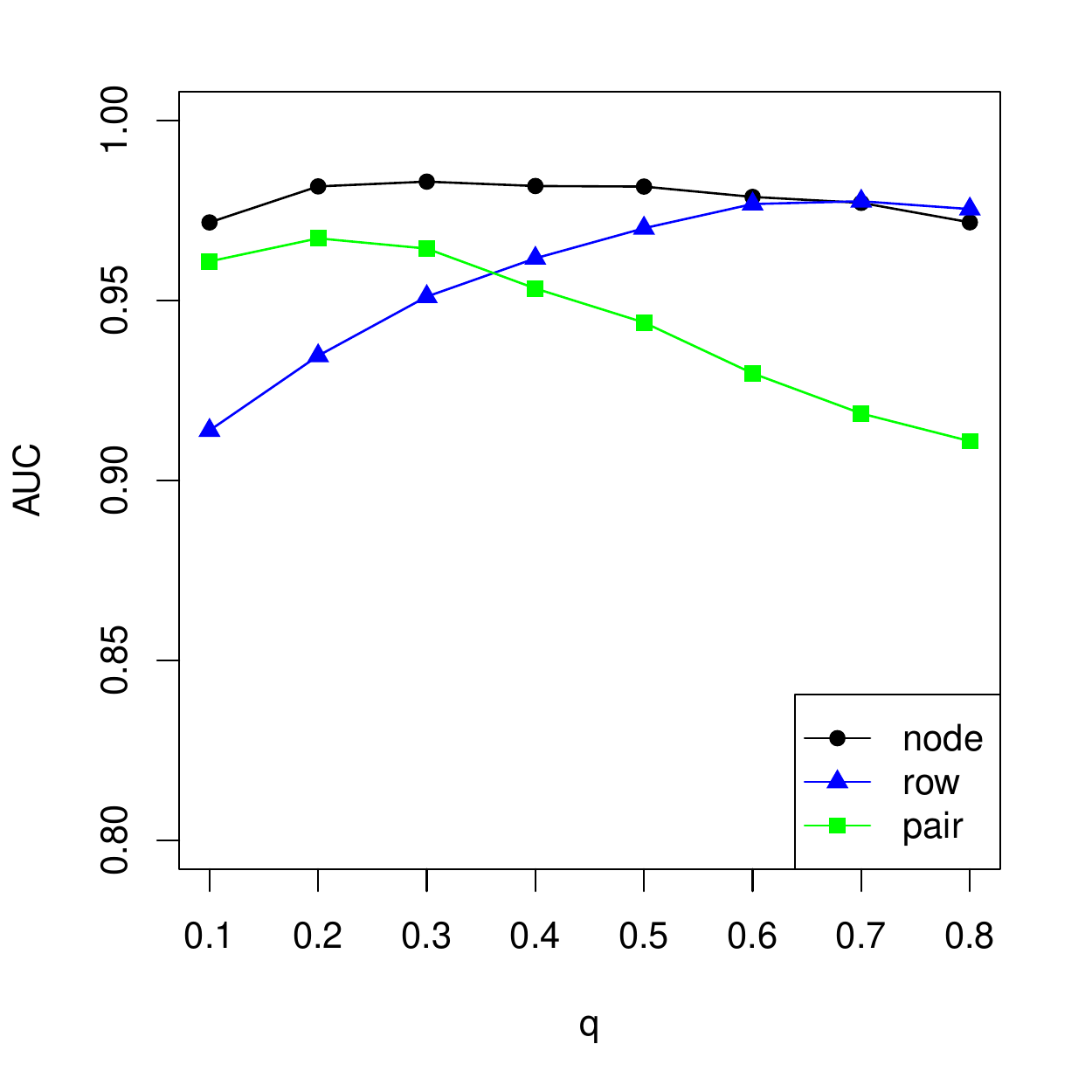}}
\subfloat[$\sigma_{\alpha}=0.1$]{\includegraphics[width=0.3\textwidth]{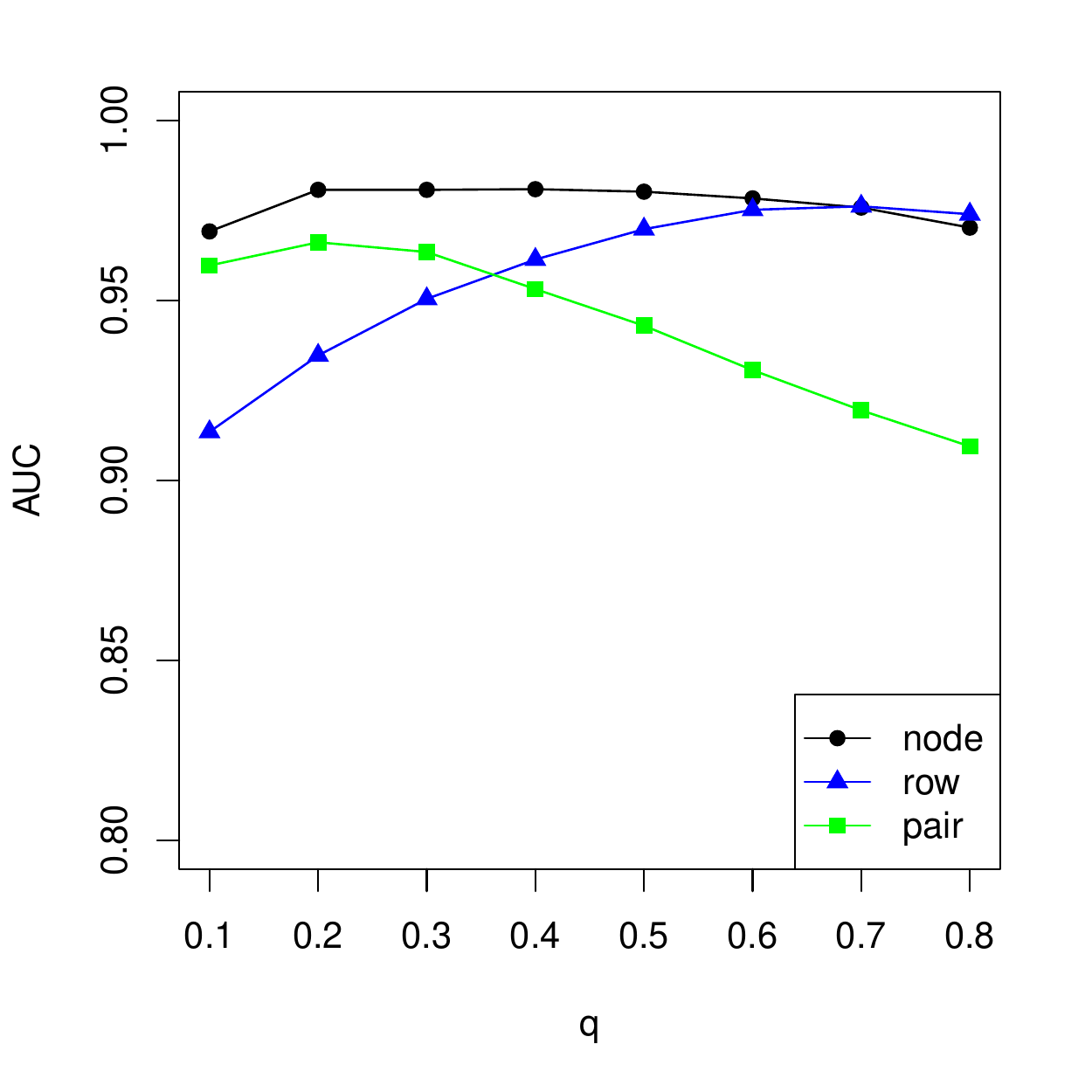}}
\subfloat[$\sigma_{\alpha}=0.5$]{\includegraphics[width=0.3\textwidth]{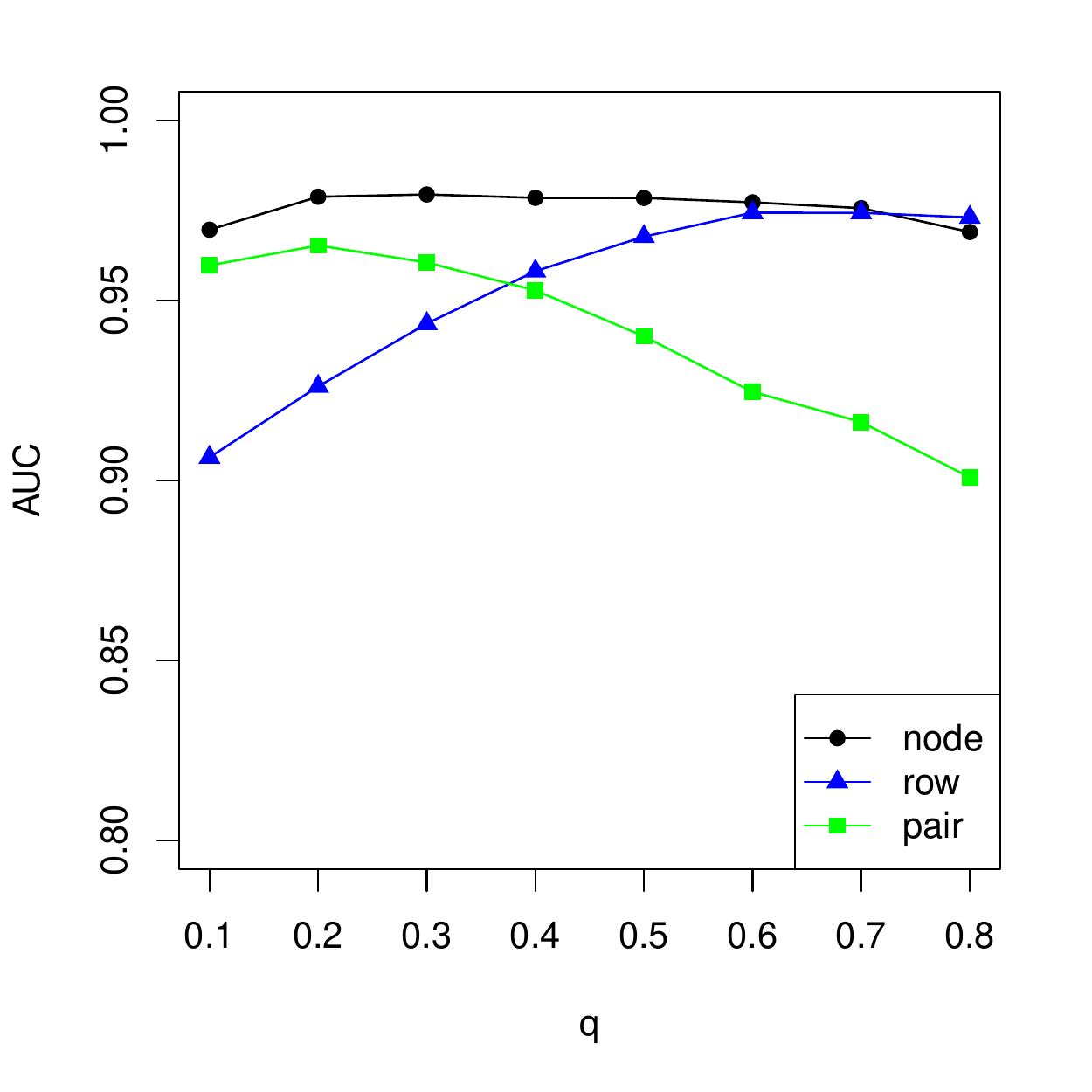}}

 \caption{Resampling performance as a function of $\sigma_\alpha$ with fixed $\rho = 0.2$ and $t = 10$.}
	\label{lasso1}
\end{figure}

\begin{figure}[ht!]
\centering
\subfloat[$t=2$]{\includegraphics[width=0.3\textwidth]{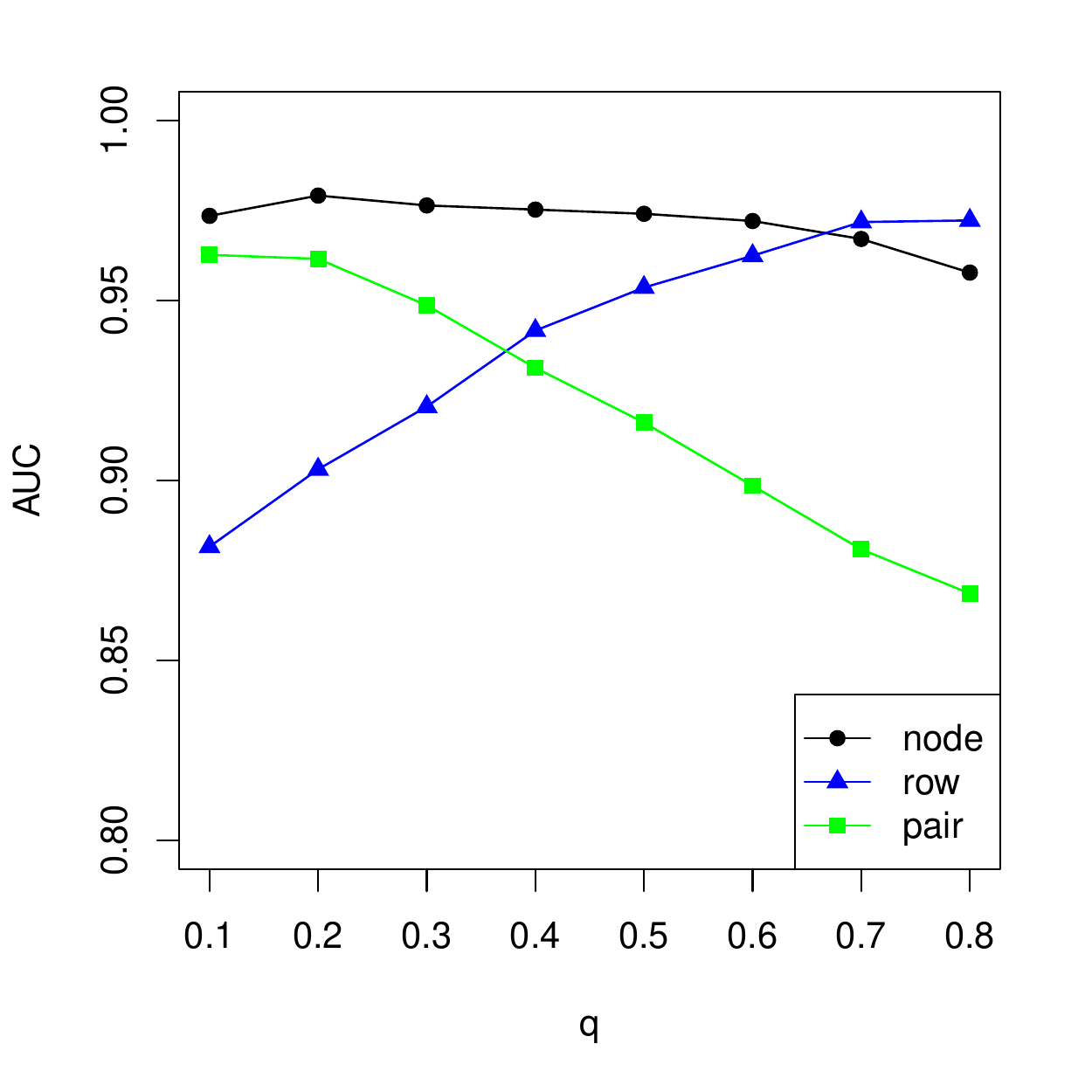}}
\subfloat[$t=5$]{\includegraphics[width=0.3\textwidth]{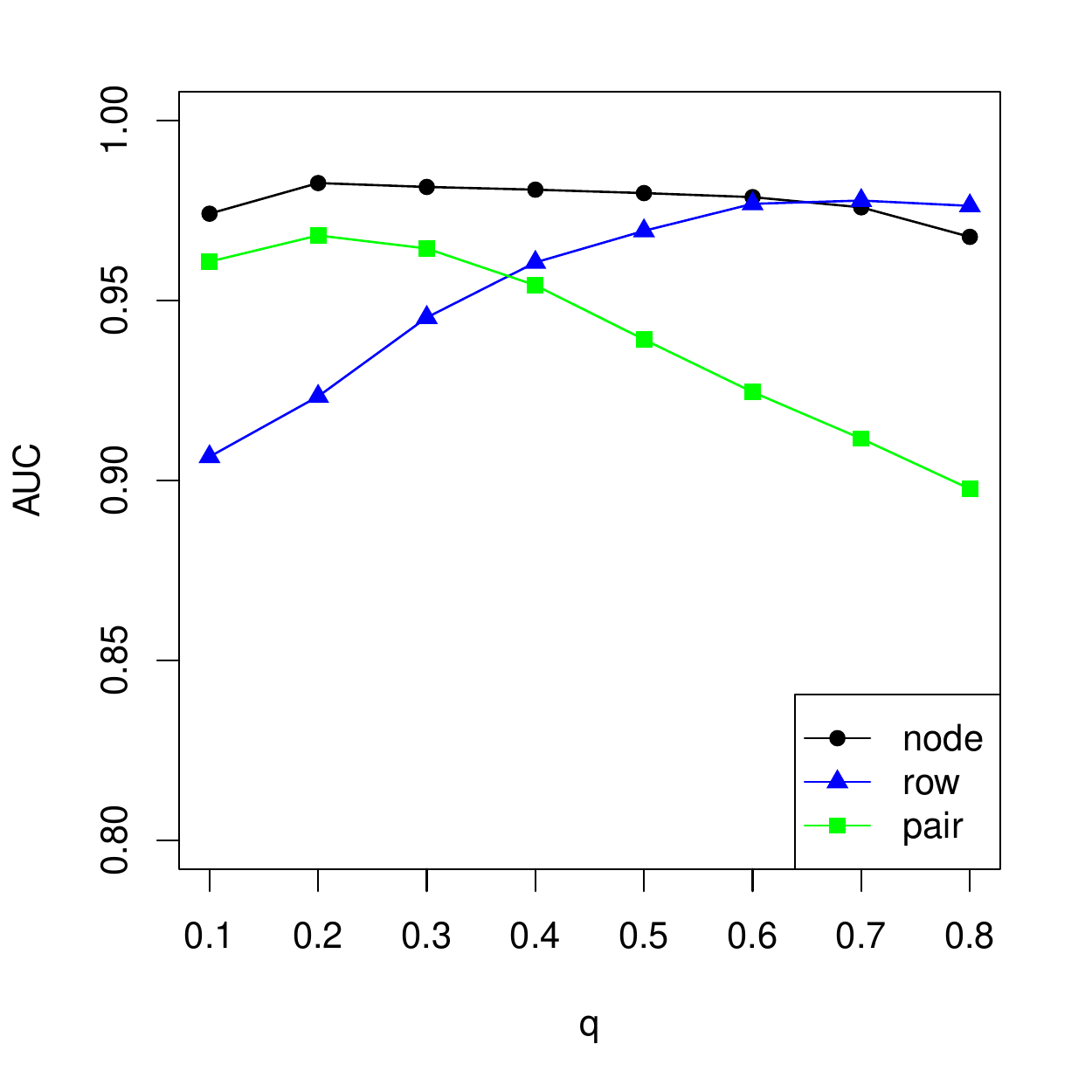}}
\subfloat[$t=10$]{\includegraphics[width=0.3\textwidth]{plot/stabsel/rho20/ratio10/sd1/auc.pdf}}

 \caption{Resampling performance as a function of $t$ with fixed $\rho = 0.2$ and $\sigma_{\alpha}=0.1$.}
	\label{lasso2}
\end{figure}

\begin{figure}[ht!]
\centering
\subfloat[$\rho=0.05$]{\includegraphics[width=0.3\textwidth]{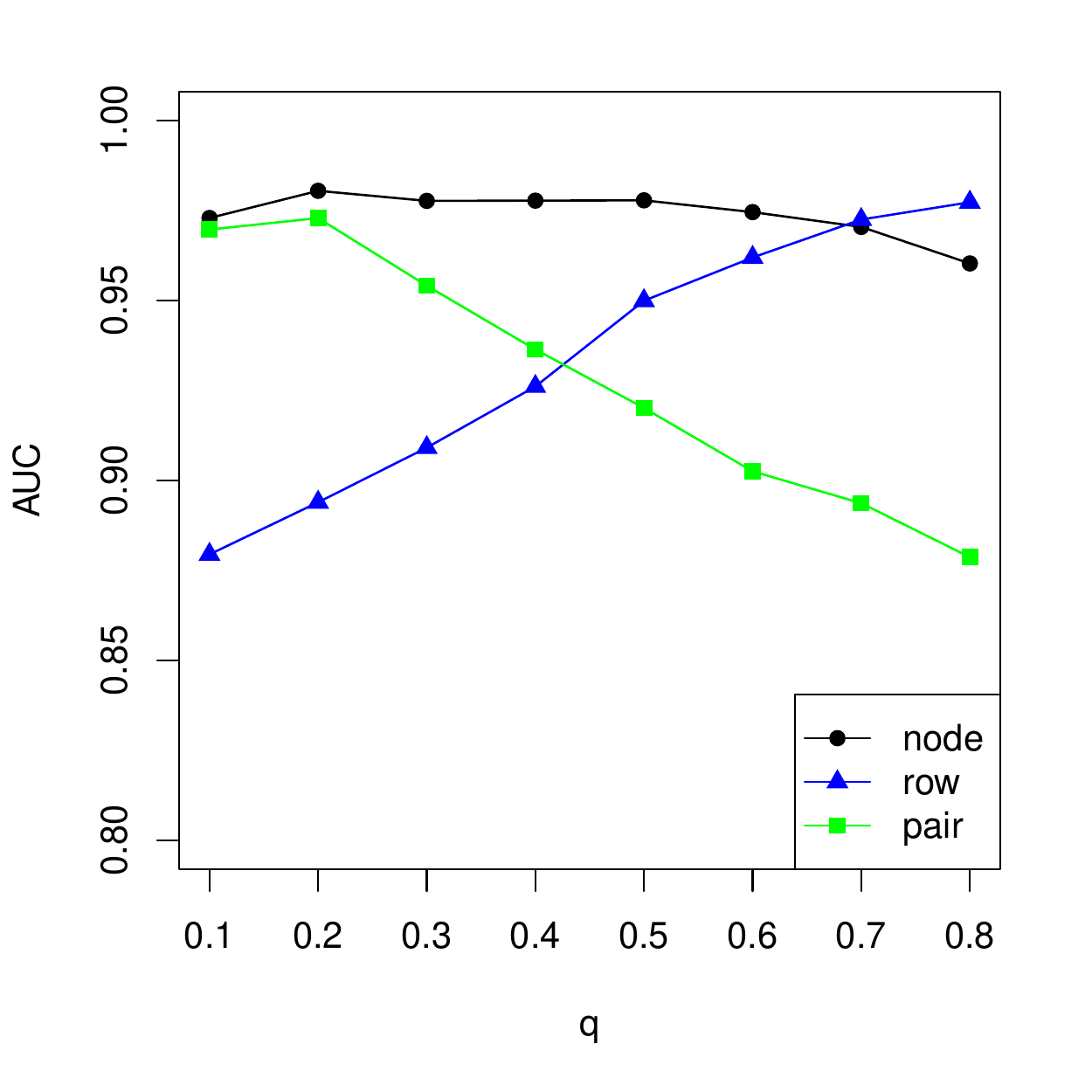}}
\subfloat[$\rho=0.1$]{\includegraphics[width=0.3\textwidth]{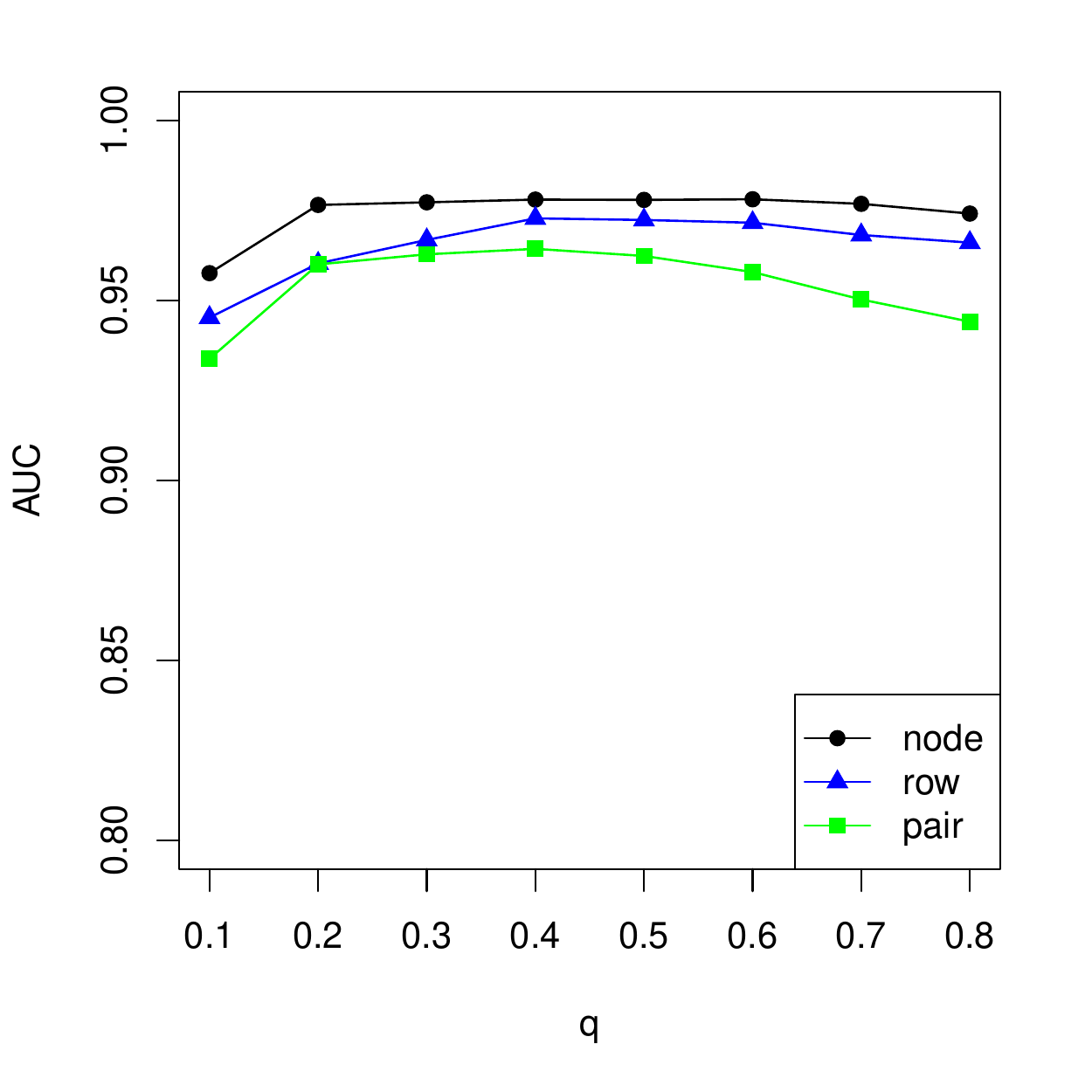}}
\subfloat[$\rho=0.2$]{\includegraphics[width=0.3\textwidth]{plot/stabsel/rho20/ratio10/sd1/auc.pdf}}

 \caption{Resampling performance as a function of $\rho$ with fixed $\sigma_{\alpha}=0.1$ and $t = 10$.}
	\label{lasso3}
\end{figure}

    The simulation setup is going as follows: A graph with 3 blocks with 200 nodes in each community is generated through SBM. Again we vary the ratio of within and between communities edge probability and the edge density of the graph. The individual effect $\alpha$ is generated from $N(c_{k},\sigma_{\alpha})$, where $c_k$ are -1, 0, 1 for nodes from different communities respectively. 25 coefficients $\beta_j$ are drawn from $N(1,1)$ independently, while the rest 75 $\beta_j$ are set to 0. Then $y_i$ is drawn from $N(\alpha_i + \beta^T x_i,1)$. The simulation result is shown in Figure \ref{lasso1}, \ref{lasso2}, \ref{lasso3}. 
    
    The three resampling schemes give different performances in AUC when we vary the subsampling proportion $q$: the performance of $p$ sampling is not sensitive to the choice of $q$, while the performances of row sampling and node pair sampling improve or worsen respectively as $q$ increases. 
    
    Similar to what we observed in the last section, although the subsampling step does bring in uncertainty into the inference, in a regression setting, node covariates often play a more important role and the performance of resampling only the network structure is not as good as expected.
    
%
%
    
\section{A data analysis:  triangle density in Facebook networks}
\label{sec:data}
\begin{figure}[ht!]
\centering
\includegraphics[width=0.9\textwidth]{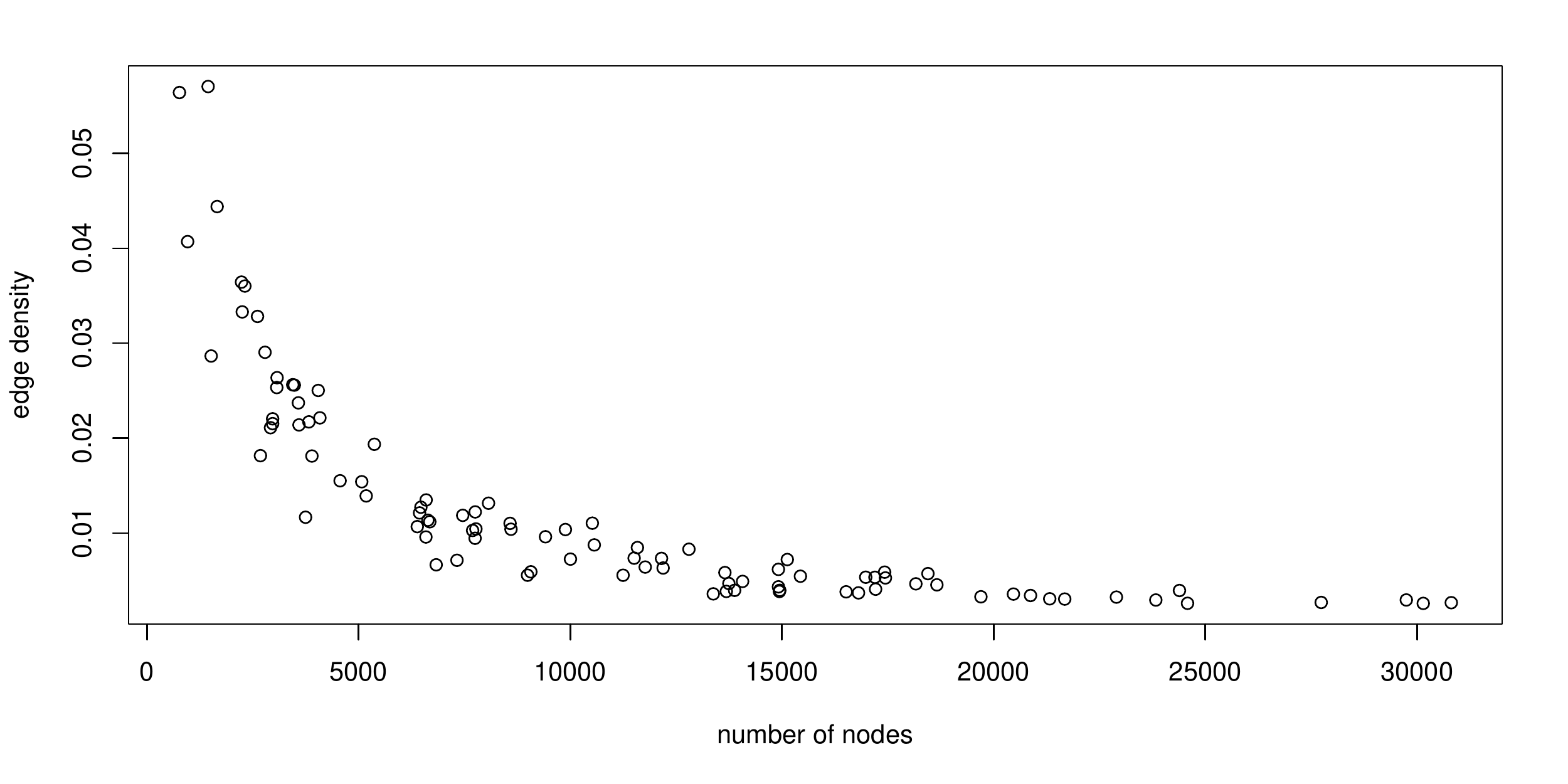}
\caption{Network size and edge density of 95 college Facebook networks.  }
\label{plot:data:rho_vs_n}
\end{figure}

\begin{figure}[ht!]
\centering
\subfloat{\includegraphics[width=0.9\textwidth]{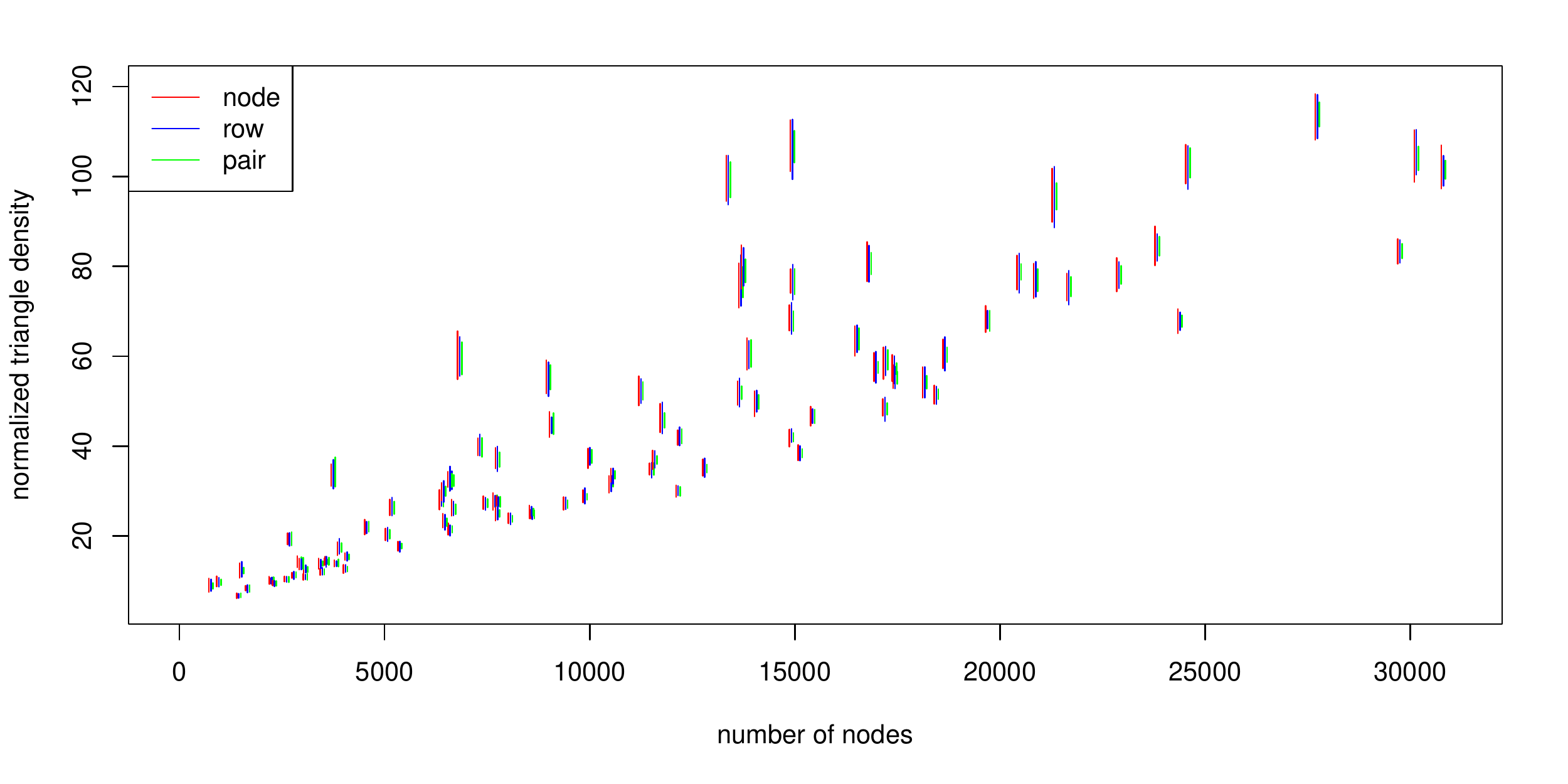}}\\
\subfloat{\includegraphics[width=0.9\textwidth]{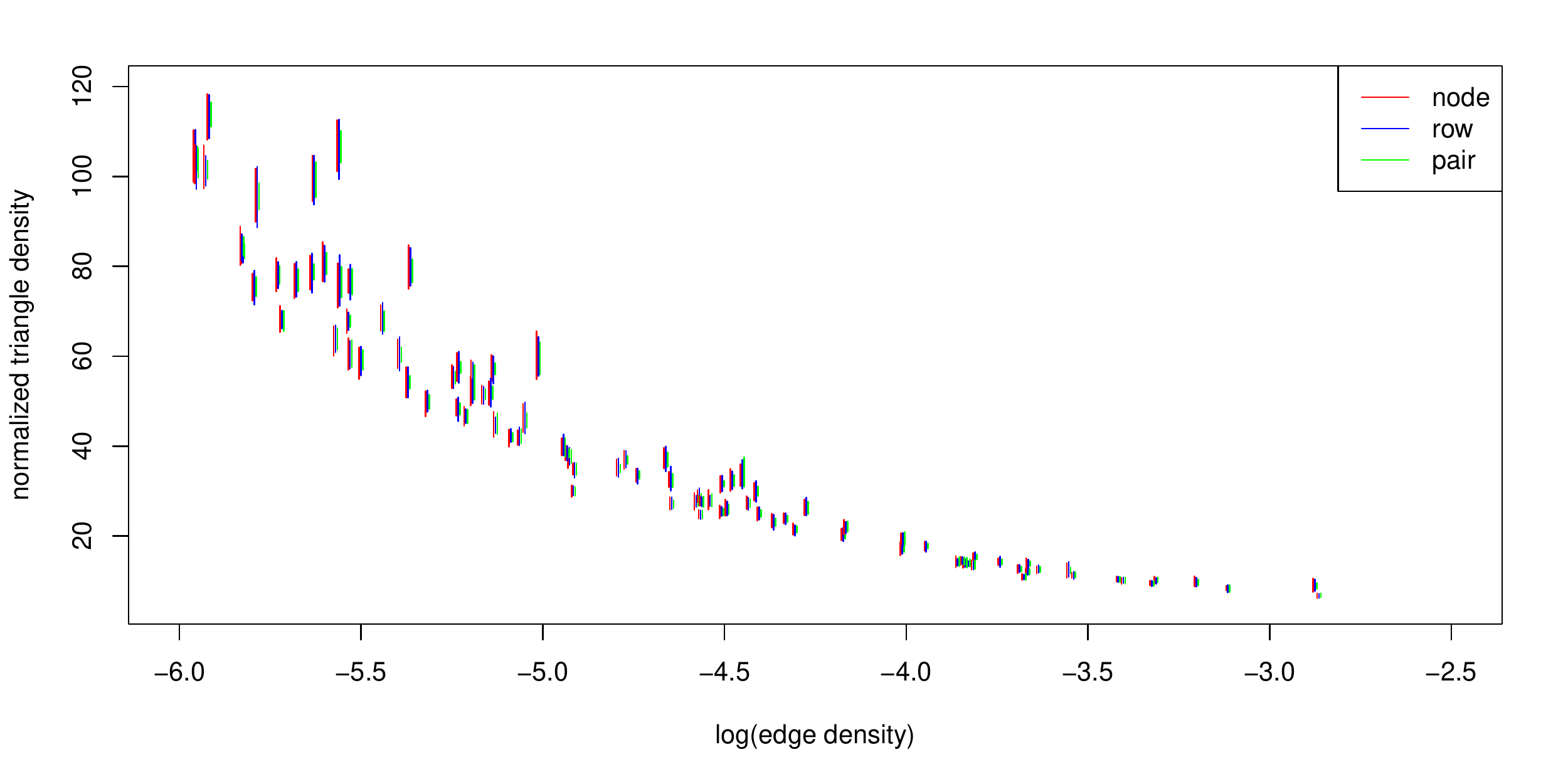}}
\caption{Confidence intervals of normalized triangle density constructed for 95 college Facebook networks.  }
\label{plot:data}

\end{figure}

\begin{figure}[ht!]
\centering
\subfloat{\includegraphics[width=0.9\textwidth]{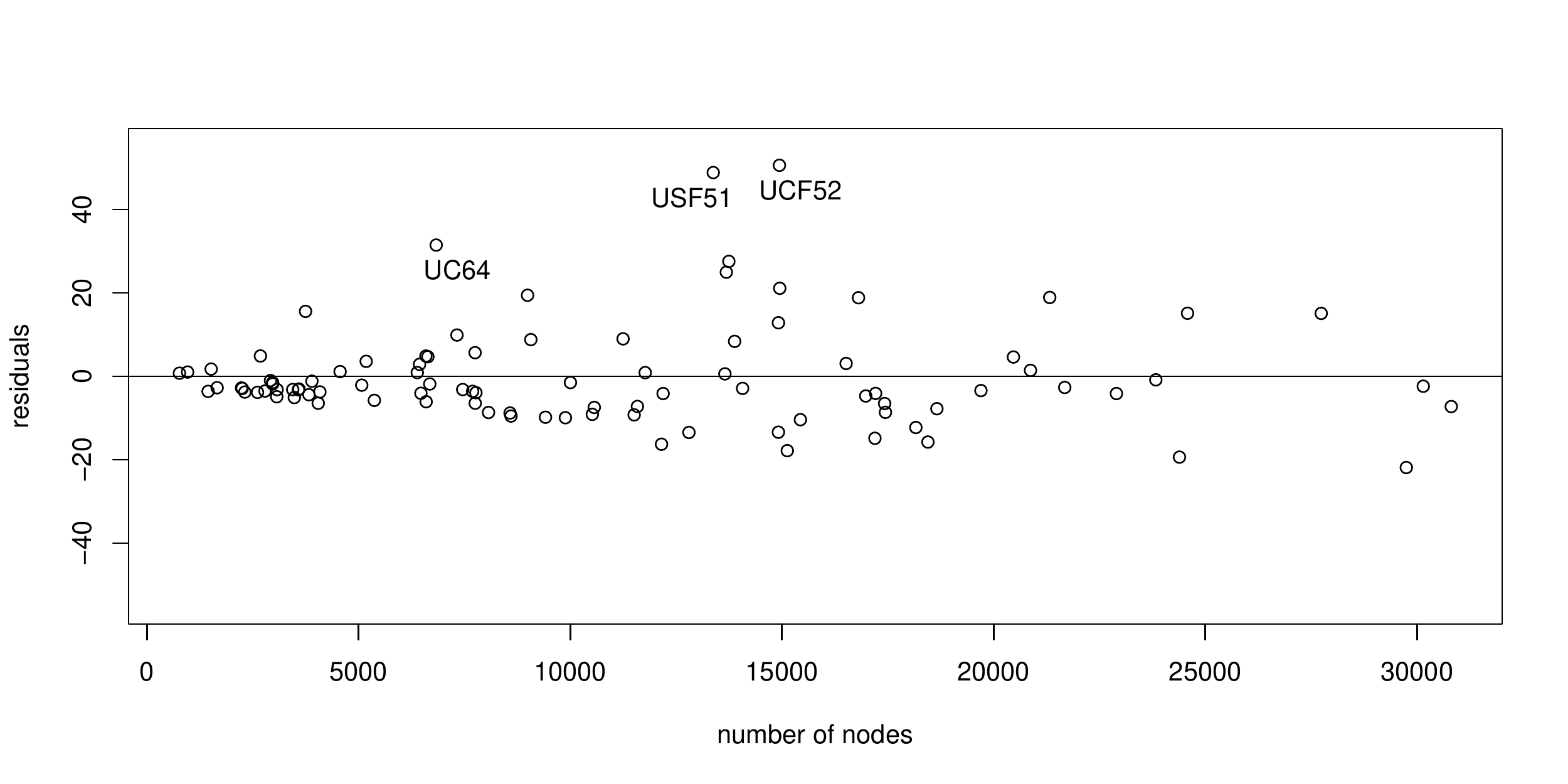}}\\
\subfloat{\includegraphics[width=0.9\textwidth]{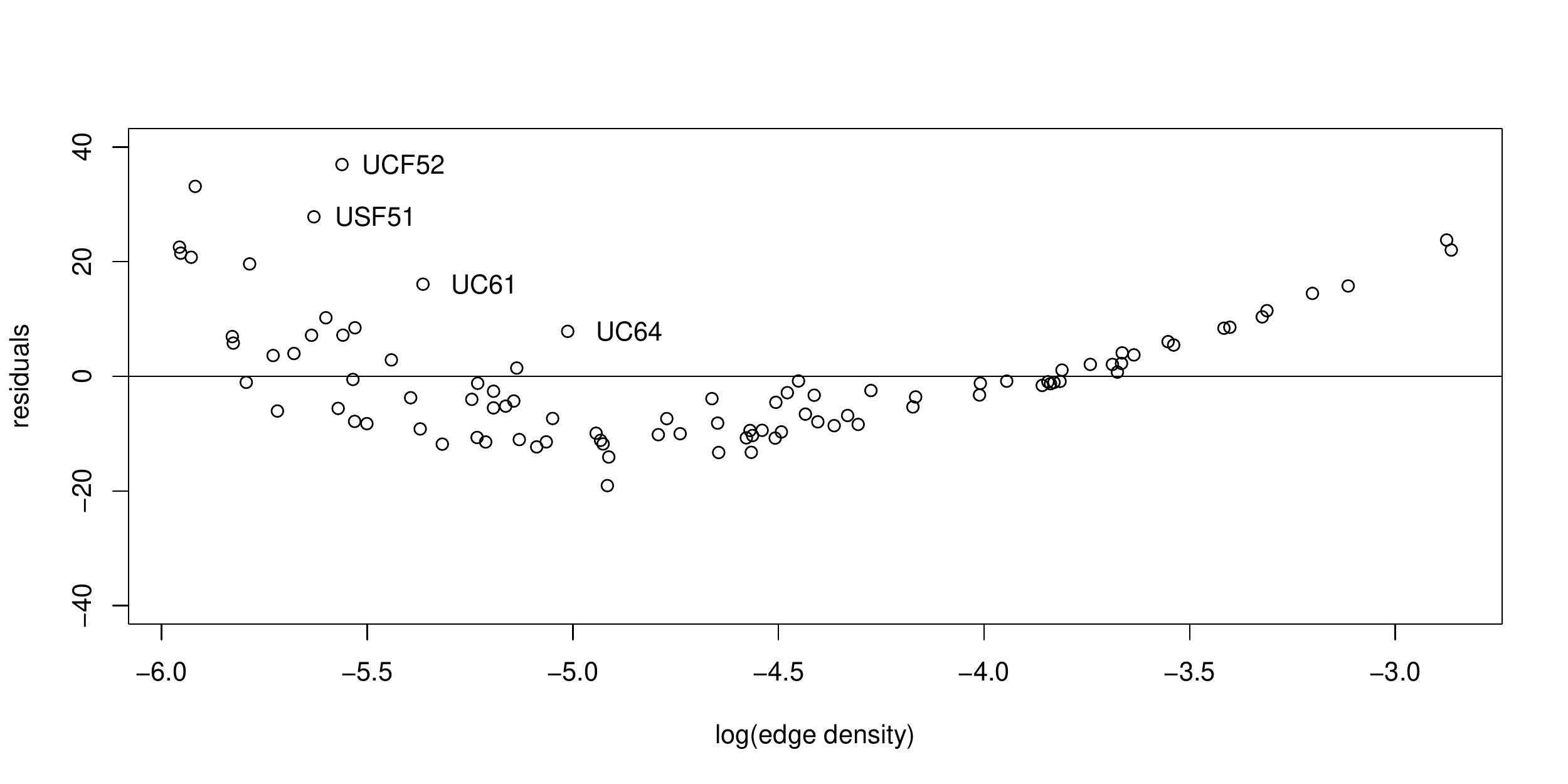}}
\caption{Normalized triangle density residuals after regressing on number of nodes and log edge density respectively.} 
	  \label{plot:data_residuals}
\end{figure}

The data studied are Facebook networks of 95 American colleges and universities from a single-day snapshot taken in September 2005 \cite{ryan_network_2015, traud_2012_social}. Size of these college networks ranges from around 800 to 30000 and their edge density ranges from 0.0025 to 0.06 as shown in Figure~\ref{plot:data:rho_vs_n}. 
Questions of interest include the distribution of normalized triangle density, understanding whether there are significant differences between colleges, and discovering relationships with other important network characteristics, such as size or edge density.  
Answering these and other similar questions requires uncertainty estimates in addition to point estimates.    We construct these uncertainty estimates by applying our Algorithm 1 to obtain 90\% confidence intervals, with subsampling fraction $q$ chosen by Algorithm 2.   The results are shown in Figures \ref{plot:data} and~\ref{plot:data_residuals}.   

There are several conclusions to draw from Figure~\ref{plot:data}.   First, larger networks have higher normalized triangle density, despite normalization, and the confidence intervals are short enough to make these differences very significant.   The same is true for normalized triangle density as a function of edge density:  denser networks have lower normalized triangle density.  This may be due to the fact that the edge density of these social networks goes down as the size of the network goes up, as shown in Figure~\ref{plot:data:rho_vs_n}, since the number of connections to each individual node grows much slower compared to the network size.
Generally, the intervals from all three schemes have similar centers, and those obtained by pair sampling are somewhat shorter than those constructed with node and row sampling.   Algorithm 2 generally chooses small values of subsampling proportion $q$ and Table~\ref{table:data:q} shows the percentage of different $q$ chosen out of the 95 graphs.

\begin{table}[]
\centering
\begin{tabular}{|c|c|c|}
\hline
     & $q=0.1$    & $q=0.2$    \\ \hline
node & 81.1\% & 18.9\% \\ \hline
row  & 54.7\% & 45.3\% \\ \hline
pair & 96.8\% & 3.2\%  \\ \hline
\end{tabular}
\caption{Percentage of different $q$'s chosen for different colleges using different subsampling methods.}
\label{table:data:q}
\end{table}


It is evident from Figure~\ref{plot:data} that there is a strong positive correlation between normalized triangle density and the number of nodes, and so it is instructive to look at triangle density while controlling for network size.   
We show residuals of  normalized triangle density after regressing normalized triangle density on $n$, in Figure \ref{plot:data_residuals}. We observed that residuals mostly centered around zero, we larger variance as $n$ grows,  suggesting there is no longer a dependence on $n$.   However, colleges such as UCF, USF and UC do have larger normalized triangle density compared to colleges of similar size, as seen in both Figure~\ref{plot:data} and \ref{plot:data_residuals}. Similar conclusion can be drawn for edge density.

%
While the example we gave is a simple exploratory analysis, it demonstrates the possibilities of much richer network data analysis  when point estimates of network statistics are accompanied by uncertainty estimates.  

\section{Discussion}
\label{sec:disc}
	  
	  The three different subsampling methods we studied - node, row, and node pair sampling - are all reasonably good at achieving nominal coverage if the sampling proportion $q$ is selected appropriately,  both for estimating uncertainty in network summary statistics and in estimated parameters of a model fitted to network data.     While there is no universally best resampling scheme for all tasks,  node sampling is a safe choice for most, especially compared to row sampling.   The double bootstrap algorithm to choose $q$ we proposed works well empirically over a range of settings, especially for network statistics that are normalized with respect to the size or the edge density of the graph.    Our algorithm is written explicitly for confidence intervals, and would thus need to modified to use with other losses, for instance, prediction intervals, but such a modification would be straightforward.  
	  
An obvious question arises about theoretical guarantees of coverage for bootstrap confidence intervals.   We have intentionally made no assumptions on the underlying network model when deriving these bootstrap methods.    Proving any such guarantee requires assuming an underlying probability model that generated the network.   If such a model is assumed, then, for all common network models, there will be an alternative model-based  bootstrap approach tailor-made for the model.  For example, for latent variable models where network edge probabilities are functions of unobserved latent variables in $R^d$, a bootstrap approach based on estimating latent variables in  $R^d$, bootstrapping those estimated variables, and generating networks from those has been shown to have good asymptotic properties \cite{levin_bootstrapping_2019}.
Similarly, specialized methods for network moments have been proposed in \cite{bhattacharyya_subsampling_2015, lunde_2019_subsampling}.
There is an inevitable tradeoff between keeping a method assumption-free and establishing theoretical guarantees.   In this computational paper, we have focused on the algorithms themselves, and leave establishing guarantees for special cases for future work.  
\FloatBarrier 

\bibliography{draft-bibtex}

\begin{thebibliography}{10}

\bibitem{bera_1987_prepivoting}
R.~Beran.
\newblock Prepivoting to reduce level error of confidence sets.
\newblock {\em Biometrika}, 74(3):457--468, 1987.

\bibitem{bhattacharyya_2014_community}
S.~Bhattacharyya and P.~J. Bickel.
\newblock Community detection in networks using graph distance.
\newblock {\em arXiv preprint arXiv:1401.3915}, 2014.

\bibitem{bhattacharyya_subsampling_2015}
S.~Bhattacharyya and P.~J. Bickel.
\newblock {Subsampling bootstrap of count features of networks}.
\newblock {\em The Annals of Statistics}, 43(6):2384 -- 2411, 2015.

\bibitem{bickel_method_2011}
P.~J. Bickel, A.~Chen, and E.~Levina.
\newblock {The method of moments and degree distributions for network models}.
\newblock {\em The Annals of Statistics}, 39(5):2280 -- 2301, 2011.

\bibitem{bickel_1981_asymptotic}
P.~J. Bickel and D.~A. Freedman.
\newblock {Some Asymptotic Theory for the Bootstrap}.
\newblock {\em The Annals of Statistics}, 9(6):1196 -- 1217, 1981.

\bibitem{bickel_1997_resampling}
P.~J. Bickel, F.~Götze, and W.~R. van Zwet.
\newblock Resampling fewer than n observations: Gains, losses, and remedies for
  losses.
\newblock {\em Statistica Sinica}, 7(1):1--31, 1997.

\bibitem{bickel_2006_covariance}
P.~J. Bickel and E.~Levina.
\newblock {Covariance regularization by thresholding}.
\newblock {\em The Annals of Statistics}, 36(6):2577 -- 2604, 2008.

\bibitem{bickel_hypothesis_2016}
P.~J. Bickel and P.~Sarkar.
\newblock Hypothesis testing for automated community detection in networks.
\newblock {\em Journal of the Royal Statistical Society. Series B (Statistical
  Methodology)}, 78(1):253--273, 2016.

\bibitem{chen_network_2018}
K.~Chen and J.~Lei.
\newblock Network cross-validation for determining the number of communities in
  network data.
\newblock {\em Journal of the American Statistical Association},
  113(521):241--251, 2018.

\bibitem{efron_1986_bootstrap}
B.~Efron and R.~Tibshirani.
\newblock {Bootstrap Methods for Standard Errors, Confidence Intervals, and
  Other Measures of Statistical Accuracy}.
\newblock {\em Statistical Science}, 1(1):54 -- 75, 1986.

\bibitem{efron_1994_introduction}
B.~Efron and R.~J. Tibshirani.
\newblock {\em An introduction to the bootstrap}.
\newblock CRC press, 1994.

\bibitem{fortunato_2010_community}
S.~Fortunato.
\newblock Community detection in graphs.
\newblock {\em Physics reports}, 486(3-5):75--174, 2010.

\bibitem{frank_estimation_1978}
O.~Frank.
\newblock Estimation of the {Number} of {Connected} {Components} in a {Graph}
  by {Using} a {Sampled} {Subgraph}.
\newblock {\em Scandinavian Journal of Statistics}, 5:177--188, 1978.

\bibitem{gel_bootstrap_2017}
Y.~R. Gel, V.~Lyubchich, and L.~L.~R. Ramirez.
\newblock Bootstrap quantification of estimation uncertainties in network
  degree distributions.
\newblock 7(1):1--12, 2017.

\bibitem{green_2017_bootstrapping}
A.~Green and C.~R. Shalizi.
\newblock Bootstrapping exchangeable random graphs.
\newblock {\em arXiv preprint arXiv:1711.00813}, 2017.

\bibitem{hall_1986_bootstrap}
P.~Hall.
\newblock {On the Bootstrap and Confidence Intervals}.
\newblock {\em The Annals of Statistics}, 14(4):1431 -- 1452, 1986.

\bibitem{hall_1988_bootstrap}
P.~Hall and M.~A. Martin.
\newblock On bootstrap resampling and iteration.
\newblock {\em Biometrika}, 75(4):661--671, 1988.

\bibitem{addhealth_2018}
K.~M. Harris.
\newblock The national longitudinal study of adolescent to adult health (add
  health), waves i \& ii, 1994–1996; wave iii, 2001–2002; wave iv,
  2007-2009; wave v, 2016-2018 [machine-readable data file and documentation].
\newblock 2018.

\bibitem{klusowski_estimating_2020}
J.~M. Klusowski and Y.~Wu.
\newblock Estimating the number of connected components in a graph via subgraph
  sampling.
\newblock {\em Bernoulli}, 26(3):1635--1664, Aug. 2020.

\bibitem{kolaczyk_sampling_2009}
E.~D. Kolaczyk.
\newblock {\em Sampling and Estimation in Network Graphs}, pages 1--30.
\newblock Springer New York, New York, NY, 2009.

\bibitem{le_estimating_2015}
C.~M. Le and E.~Levina.
\newblock {Estimating the number of communities by spectral methods}.
\newblock {\em Electronic Journal of Statistics}, 16(1):3315 -- 3342, 2022.

\bibitem{lee_class_1999}
S.~M.~S. Lee.
\newblock On a {Class} of m out of n {Bootstrap} {Confidence} {Intervals}.
\newblock {\em Journal of the Royal Statistical Society. Series B (Statistical
  Methodology)}, 61(4):901--911, 1999.

\bibitem{levin_bootstrapping_2019}
K.~Levin and E.~Levina.
\newblock Bootstrapping networks with latent space structure, 2019.

\bibitem{li_prediction_2019}
T.~Li, E.~Levina, and J.~Zhu.
\newblock {Prediction models for network-linked data}.
\newblock {\em The Annals of Applied Statistics}, 13(1):132 -- 164, 2019.

\bibitem{li_network_2020}
T.~Li, E.~Levina, and J.~Zhu.
\newblock {Network cross-validation by edge sampling}.
\newblock {\em Biometrika}, 107(2):257--276, 04 2020.

\bibitem{lin_2020_on}
Q.~Lin, R.~Lunde, and P.~Sarkar.
\newblock On the theoretical properties of the network jackknife.
\newblock In {\em Proceedings of the 37th International Conference on Machine
  Learning}, volume 119 of {\em Proceedings of Machine Learning Research},
  pages 6105--6115. PMLR, 13--18 Jul 2020.

\bibitem{lunde_2019_subsampling}
R.~Lunde and P.~Sarkar.
\newblock Subsampling sparse graphons under minimal assumptions.
\newblock {\em arXiv preprint arXiv:1907.12528}, 2019.

\bibitem{martin_bootstrap_1990}
M.~A. Martin.
\newblock On {Bootstrap} {Iteration} for {Coverage} {Correction} in
  {Confidence} {Intervals}.
\newblock {\em Journal of the American Statistical Association},
  85(412):1105--1118, 1990.

\bibitem{meinshausen_stability_2010}
N.~Meinshausen and P.~Bühlmann.
\newblock Stability selection.
\newblock {\em Journal of the Royal Statistical Society: Series B (Statistical
  Methodology)}, 72(4):417--473, 2010.

\bibitem{politis_1994_stationary}
D.~N. Politis and J.~P. Romano.
\newblock The stationary bootstrap.
\newblock {\em Journal of the American Statistical Association},
  89(428):1303--1313, 1994.

\bibitem{politis_subsampling_1999}
D.~N. Politis, J.~P. Romano, and M.~Wolf.
\newblock {\em Subsampling}.
\newblock Springer {Series} in {Statistics}. Springer New York, New York, NY,
  1999.

\bibitem{ryan_network_2015}
R.~A. Rossi and N.~K. Ahmed.
\newblock The network data repository with interactive graph analytics and
  visualization.
\newblock 2015.

\bibitem{saade_matrix_2016}
A.~Saade, F.~Krzakala, and L.~Zdeborová.
\newblock Matrix {Completion} from {Fewer} {Entries}: {Spectral}
  {Detectability} and {Rank} {Estimation}.
\newblock {\em arXiv:1506.03498 [cond-mat, stat]}, Jan. 2016.

\bibitem{shao_1989_general}
J.~Shao and C.~F.~J. Wu.
\newblock {A General Theory for Jackknife Variance Estimation}.
\newblock {\em The Annals of Statistics}, 17(3):1176 -- 1197, 1989.

\bibitem{thompson_using_2016}
M.~E. Thompson, L.~L. Ramirez~Ramirez, V.~Lyubchich, and Y.~R. Gel.
\newblock Using the bootstrap for statistical inference on random graphs.
\newblock {\em Canadian Journal of Statistics}, 44(1):3--24, 2016.

\bibitem{tibshirani_regression_1996}
R.~Tibshirani.
\newblock Regression shrinkage and selection via the lasso.
\newblock {\em Journal of the Royal Statistical Society. Series B
  (Methodological)}, 58(1):267--288, 1996.

\bibitem{traud_2012_social}
A.~L. Traud, P.~J. Mucha, and M.~A. Porter.
\newblock Social structure of {Facebook} networks.
\newblock {\em Physica A: Statistical Mechanics and its Applications},
  391(16):4165--4180, Aug. 2012.

\bibitem{zhang_2015_estimating}
Y.~Zhang, E.~D. Kolaczyk, B.~D. Spencer, et~al.
\newblock Estimating network degree distributions under sampling: An inverse
  problem, with applications to monitoring social media networks.
\newblock {\em The Annals of Applied Statistics}, 9(1):166--199, 2015.

\end{thebibliography}


\end{document}